\documentclass[aps,prx,amsmath,superscriptaddress,amssymb,twocolumn]{revtex4-1}
\usepackage[dvipsnames]{xcolor}
\usepackage{graphicx}
\graphicspath{ {images/}}
\usepackage{dcolumn}
\usepackage{bm}
\usepackage[	
citecolor=blue,
colorlinks,
linkcolor=magenta,
urlcolor=blue,
]{hyperref}
\usepackage{braket}
\usepackage{appendix}
\usepackage{wasysym}
\usepackage{slashed}
\usepackage{xcolor}
\usepackage{subfigure}
\usepackage{soul}
\usepackage{dutchcal}
\usepackage{calligra}

\DeclareMathAlphabet{\mathcalligra}{T1}{calligra}{m}{n}
\DeclareFontShape{T1}{calligra}{m}{n}{<->s*[2.2]callig15}{}

\newcommand{\boldscriptr}{\pmb{\mathcalligra{r}}\,}

\newcommand{\zrcl}{ZrCl${}_{3}$}
\def\rim#1{\textcolor{red}{#1}}

\newcommand{\mcl}[1]{\mathcal{#1}}
\newcommand{\mbf}[1]{\mathbf{#1}}
\newcommand{\mbb}[1]{\mathbb{#1}}

\newcommand{\eqnref}[1]{Eq.~(\ref{#1})}
\newcommand{\figref}[1]{Fig.~\ref{#1}}
\newcommand{\bcen}{\begin{center}}
\newcommand{\ecen}{\end{center}}
\newcommand{\btab}{\begin{tabular}}
\newcommand{\etab}{\end{tabular}}
\newcommand{\bdes}{\begin{description}}
\newcommand{\edes}{\end{description}}

\newcommand{\beq}{\begin{equation}}
\newcommand{\eeq}{\end{equation}}
\newcommand{\bea}{\begin{eqnarray}}
\newcommand{\eea}{\end{eqnarray}}

\newcommand{\dou}{\partial}

 \usepackage{pifont}
\begin{document}

\title{Emergent SU(8) Dirac semimetal and novel proximate phases of spin-orbit coupled fermions on a honeycomb lattice}
 \author{Basudeb Mondal}
 \email{basudeb.mondal@icts.res.in}
 \affiliation{International Centre for Theoretical Sciences, Tata Institute of Fundamental Research, Bengaluru 560089, India.}
 \author{Vijay B. Shenoy}
 \email{shenoy@iisc.ac.in}
 \affiliation{Centre for Condensed Matter Theory, Department of Physics, Indian Institute of Science, Bengaluru 560012, India.}
 \author{Subhro Bhattacharjee}
 \email{subhro@icts.res.in}
 \affiliation{International Centre for Theoretical Sciences, Tata Institute of Fundamental Research, Bengaluru 560089, India.}

\begin{abstract}
Emergent Dirac fermions provide the starting point for understanding the plethora of novel condensed matter phases. The nature of the associated phases and phase transitions crucially depends on both the emergent symmetries as well as the implementation of the microscopic ones on the low-energy Dirac fermions. Here, we show that $j=3/2$ electrons in spin-orbit coupled materials on honeycomb lattice can give rise to SU(8) symmetric Dirac semimetals with symmetry implementation very different from that of graphene. This non-trivial embedding of the microscopic symmetries in the low energy is reflected in the nature of phases proximate to the Dirac semimetal. Such phases can arise from finite short-range electron-electron interactions. In particular, we identify 24 such phases -- divided into three classes -- and their low energy properties obtained  by condensing particle-number conserving fermion bilinears that break very different microscopic symmetries and/or are topologically protected by symmetries. The latter includes interesting generalisations of quantum spin-Hall phases. Remarkably some of the resultant phases still support a sub-set of gapless fermions-- protected by a sub-group of SU(8) -- resulting in interesting density wave semimetals. Near the phase transitions to such density wave semimetals, the surviving gapless fermions strongly interact with the bosonic order parameter field and give rise to novel quantum critical points. Our study is applicable to a wide class of $d^1$ and $d^3$ transition metals with strong spin-orbit coupling and predicts that such materials can harbour a very rich interplay of symmetries and competing interactions in the intermediate correlation regime.

\end{abstract}

\maketitle

\section{Introduction}

Massless Dirac fermions arise in a variety of condensed matter systems~\cite{vafek2014dirac,goerbig2017dirac}. Perhaps the most well-known is the recently studied -- both experimentally and theoretically -- example of monolayer graphene~\cite{PhysRev.71.622,novoselov2005two,zhang2005experimental,geim2009graphene,RevModPhys.81.109,goerbig2011electronic,RevModPhys.83.407,RevModPhys.83.837,RevModPhys.83.851} where such Dirac fermions arise as a low energy limit of electrons hopping on the honeycomb lattice. More generally such Dirac fermions may arise in a variety of other two and three  dimensional lattices~\cite{wehling2014dirac,vafek2014dirac,PhysRevLett.115.126803,hirata2021interacting} relevant for several materials including organic semiconductors like $\alpha$-(BEDT-TTF)$_2$I$_3$~\cite{kobayashi2007massless,PhysRevB.78.045415,hirata2021interacting}, the d$_{x^2-y^2}$-wave superconductor in cuprates~\cite{PhysRevLett.74.797,kirtley1995symmetry,volovik1993superconductivity,PhysRevB.52.R3876,balents1998nodal}, Dirac and Weyl semimetals~\cite{wan2011topological,PhysRevLett.108.140405} and surface of 3D topological insulators~\cite{PhysRevB.75.121306,PhysRevB.79.195322,PhysRevLett.98.106803,RevModPhys.82.3045,RevModPhys.83.1057,chen2009experimental}. These low energy Dirac fermions have indelible signatures in a plethora of low energy experiments of these candidate materials as is evident in the integer quantum Hall effect ~\cite{pacile2008two,zhang2005experimental} as well as other spectroscopic and transport properties~\cite{li2009scanning,nair2008fine,li2008dirac,levy2010strain,RevModPhys.83.407,PhysRevB.79.115434} of monolayer graphene, surface transport of 3D topological insulators~\cite{bardarson2013quantum} or spectroscopy of $d$-wave superconductors~\cite{PhysRevLett.74.797,kirtley1995symmetry}.

Dirac fermions also arise in a somewhat different context as low energy theories of certain quantum spin liquids (QSL). Indeed in U(1) Dirac QSLs, low energy fermionic spinons-- minimally coupled to an emergent U(1) gauge field-- have free Dirac dispersion with enhanced symmetries at low energies within parton mean-field theories~\cite{PhysRevB.72.104404}. While in this case of QED$_3$, the fluctuations of the gauge field ultimately lead to the destruction of the quasi-particles ~\cite{affleck1988large,marston1989large,rantner2001electron,wen2002quantum,hermele2004stability,PhysRevB.72.104404,PhysRevLett.87.257003,PhysRevB.66.144501,PhysRevB.72.104404}, the proximate ordered phases can be obtained by condensing appropriate spinon bilinears that gap out the Dirac spinons. Also, such effective Dirac theories, with or without dynamic gauge fields have also been recently discussed in context of $(2+1)$ dimensional fermionic particle-vortex dualities emanating out of conjectures of Dirac composite fermions in half-filled Landau level in quantum Hall systems~\cite{PhysRevX.5.031027}. 

An equally important question pertains to the nature of the different phases obtained~\cite{boyack2021quantum} upon gapping out the Dirac fermions via short-ranged four-fermion interactions/other bosonic fields or via external perturbations such as originating from substrate effects in graphene~\cite{zhou2007substrate}. For the former, a typical effect of such interactions is to condense a fermion-bilinear that dynamically generates mass for the gapless Dirac fermions for a finite strength of the interaction. The nature of the resultant gapped phases~\footnote{This phase can have other gapless mode such as Goldstone boson if a continuous symmetry is broken.}~\cite{boyack2021quantum,ryu2009masses,PhysRevB.79.085116,PhysRevLett.95.036402,PhysRevB.72.104404} as well as the theory of the associated phase transition from the proximate Dirac semimetal via Gross-Neveu-Yukawa~\cite{PhysRevD.10.3235,zinn1991four,sachdev1999quantum,boyack2021quantum,PhysRevLett.97.146401} field theories have received considerable attention in a wide array of condensed matter settings and allows for systematic understanding of novel quantum phase transitions including Landau forbidden {deconfined quantum criticality}~\cite{senthil2004deconfined,PhysRevB.74.064405}.

A central aspect of the above diverse physics is the implementation of microscopic symmetries on the low-energy Dirac fermions. These {\it ultraviolet} (UV) symmetries typically consist of lattice symmetries, time reversal as well as possible spin-rotation symmetries of the electrons occupying the underlying atomic orbitals. Material-dependent microscopic energetics allow for different implementations of these UV symmetries on the underlying low energy Dirac fermions opening up avenues to probe the Dirac semimetal~\cite{RevModPhys.83.407} as well as stabilize novel proximate phases. In regards to the latter, masses that preserve microscopic symmetries, but gap out the fermions,  have recently provided crucial  insights in the development of the theory of symmetry-protected topological phases (SPTs)~\cite{PhysRevLett.95.226801,PhysRevLett.61.2015}. In a large class of fermionic SPTs, this has been made possible via the interplay of atomic spin-orbit coupling~\cite{witczak2013correlated} and electron correlations that allow for such non-trivial implementation of usual condensed matter symmetries on the low energy degrees of freedom. Central to our interest here is the possibility of realising a new class of {\it spin-orbitally coupled Dirac fermions} such that the UV symmetries are implemented in a novel way allowing for new material platforms for the interplay of symmetries and interactions.  What, then, are the nature of the gapped phases in these spin-orbit coupled two-dimensional Dirac materials?

\begin{figure*}
	\centering
	\subfigure[]{
	\includegraphics[scale=0.6]{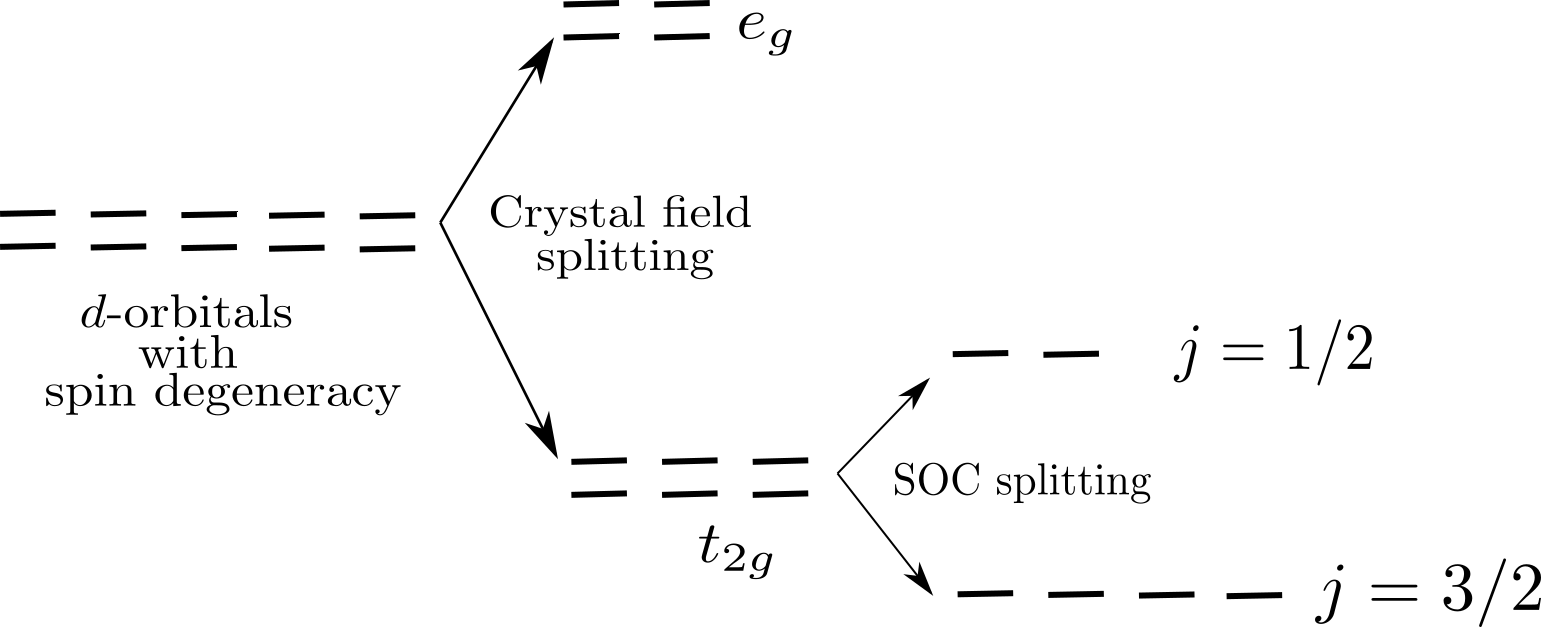}
	\label{fig_SOC_splitting}
	}
	\subfigure[]{
	\includegraphics[scale=0.3]{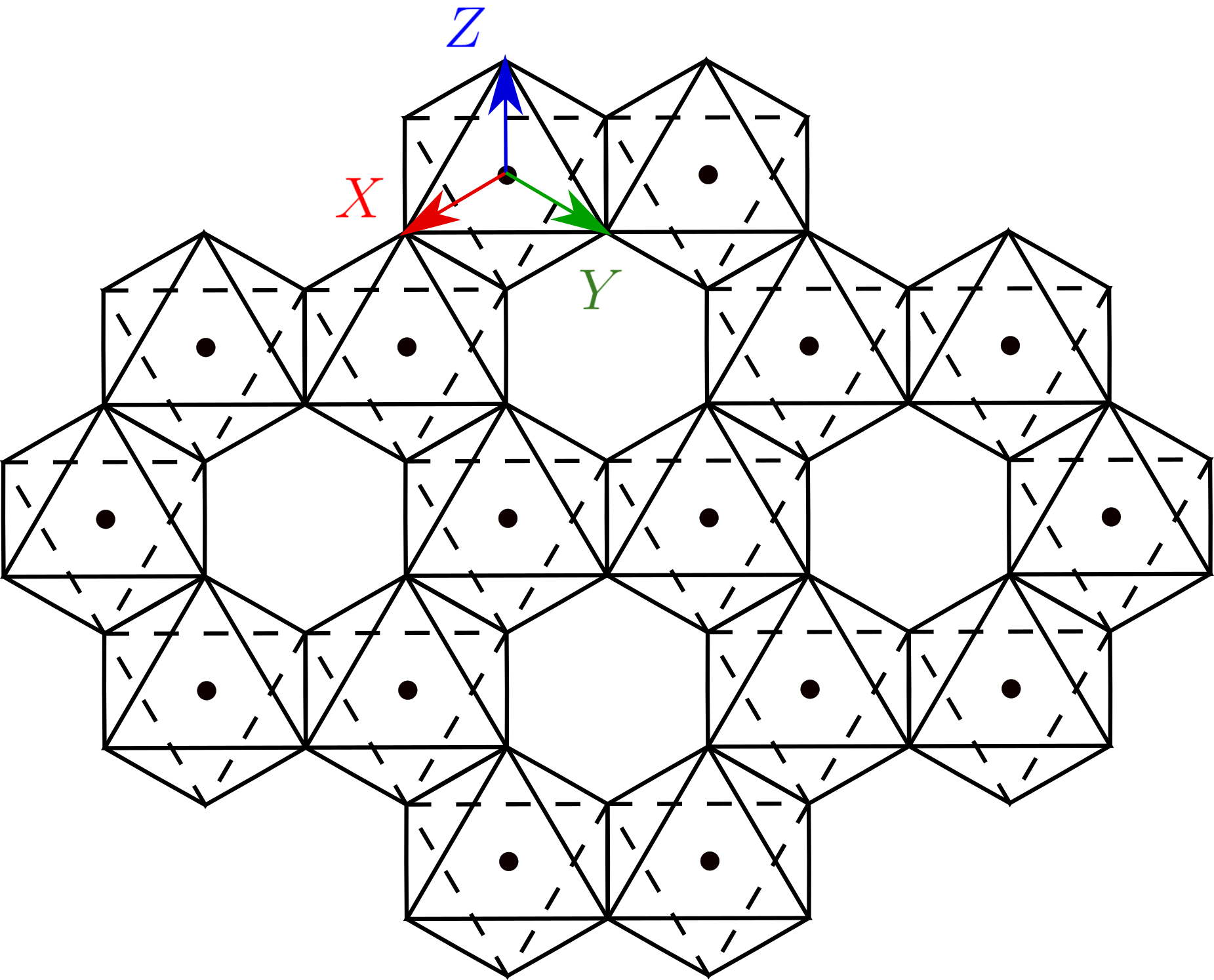}
	\label{fig:3d crystal structure of ZrCl3}
	}
	\caption{(a) Splitting of the $d$-orbitals due to crystal field and SOC. (b) Edge-sharing octahedra forming a honeycomb lattice. As elaborated in the main text and in Appendix \ref{appen_sub_lattice}, the honeycomb lattice lies in the [111] plane of the Cartesian coordinate system whose projections are denoted by $X, Y$, and $Z$.}
\end{figure*}

In this paper, we present a new material-relevant platform for realizing spin-orbit (SO)-assisted SU(8) Dirac fermions in two spatial dimensions and discuss a plethora of rich phases  proximate to such a Dirac semimetal. Somewhat counterintuitively, the underlying microscopics involving strong spin-orbit coupling (SOC),  octahedral crystal field effect on $d$-orbitals, and hopping pathways can lead to a large symmetry enhancement resulting in gapless Dirac semimetal for non-interacting electrons with an emergent SU(8) symmetry at low energy or {\it infrared} (IR). This can be realized, for example, in systems containing transition metal ions  with strong SOC in $d^1$ electronic configuration on a honeycomb lattice in an edge-sharing octahedral motif (see Fig. \ref{fig:3d crystal structure of ZrCl3}), which leads to a quarter filled $j=3/2$ atomic orbitals as low energy electronic degree of freedom. This lattice is fairly common in the context of honeycomb Iridates as well as $\alpha$-RuCl$_3$ where the transition metal ion has a $d^5$ configuration leading to $j=1/2$ moments with possible relevance to Kitaev QSLs. A much more relevant material in the same class, for the present purpose, is \zrcl{} where Zr$^{3+}$ is in a $4d^1$ configuration. Our calculations, however, are quite general and show that generally interesting low-energy physics can emerge in a relevant parameter regime for $d^1$ systems and its particle-hole conjugate $d^3$, which we expect will be synthesized in the near future.

The low energy SU(8) Dirac fermions -- described by a free Dirac action of $N_F=4$ flavours of $4$-component Dirac fermions (Eq. \ref{eq_dirac_l})-- obtained here fundamentally differ from a large flavour/spin generalisation of graphene in terms of the implementation of lattice transformations which, due to strong SOC, is intertwined with the spin-rotations. This is reflected in the low energy theory via the transformation of the Dirac modes under various lattice symmetries. A direct fallout of this non-trivial symmetry implementation is observed in the phases that are proximate to the SU(8) Dirac semimetal and can be  obtained from it for finite electron-electron  interactions and/or additional hopping perturbations via breaking of the SU(8) symmetry by condensing various fermion bilinears. In this work, we study the 64 particle-hole bilinears ({\it i.e.} bilinears that do not carry a net electronic charge) consisting of the SU(8) singlet and the 63 dimensional adjoint multiplet. This adjoint multiplet is further broken down into singlets, doublets, and triplets by the UV symmetry group and corresponds to different ordered phases. For the present symmetry realisation we find that the 64 bilinears result in 24 different phases proximate to the SU(8) Dirac semimetal. 

Typically condensation of the bilinears gaps out the single-fermion spectrum resulting in a broken symmetry phase with a single-fermion gap. We show several examples of such gapped broken symmetry  insulators, typically with different spin/charge modulations. In addition, this work shows interesting instances where the condensate leaves intact a sub-set of the gapless Dirac fermions which describe various types of spin density wave semimetals. Notably, while the bilinears necessarily break the IR symmetry of the free Dirac theory-- the SU(8) and parity or time reversal, they may still be invariant under all or some microscopic/UV symmetry transformations. Indeed, the classification of such masses that are allowed by microscopic symmetries and gap out the fermions, results in SPTs with gapped fermionic spectrum in the bulk and gapless edges whose low energy bulk theory is given by various Chern-Simons actions that characterise the appropriate quantized Hall response of such insulators. In the present context, we find several such interaction-driven SPTs, variously dubbed as topological Mott insulators~\cite{PhysRevLett.100.156401,PhysRevLett.100.156804} both in the presence and absence of time reversal symmetry. These include interesting generalisations of the quantum spin-Hall phases as well as newer ones such as the quantum spin-quadrupole Hall phase, with the latter being odd under time reversal, unlike the former. A more subtle version realised in the present case includes a ferro spin-quadrupole ordered insulator with spin-octupole filtered edge currents (Sec. \ref{sec:ferro-spinQuadru-insulator}) in addition to a ferro spin-octupole ordered insulator with anomalous charge quantum Hall effect (Sec. \ref{subsubsec:IsingFerroSpinOctopolar}). These {\it anomalous} fermionic masses, along with the conventional (microscopic) symmetry-breaking ones -- made possible by non-trivial implementation of UV symmetries on the Dirac fermions -- provide a rich phase diagram for the phases and associated phase transitions proximate to the SO coupled SU(8) Dirac semimetal.

The classification of the bilinears naturally opens the floodgate of questions regarding the nature of concomitant quantum transitions primarily out of the Dirac semimetal into one of the 24 symmetry-broken phases. As mentioned above, these transitions occur at a finite value of the short-range four fermion interactions (Eq. \ref{eq_interactions}). For continuous phase transitions, the resultant Gross-Neveu-Yukawa type critical theories are obtained by decoupling the four-fermion interaction along a particular bilinear channel resulting in coupling between the Dirac fermions with the  bosonic order parameter corresponding to the fermion bilinear which, in turn, gain dynamics on integrating out the higher energy fermions~\cite{PhysRevB.80.075432,vafek2014dirac}. The properties of such critical points when the fermions are gapped out across the transition can be understood within various renormalisation group schemes such as $3-\epsilon$ expansions~\cite{PhysRevB.80.075432} or Large $N_f$ expansions~\cite{PhysRevLett.97.146401}. We shall encounter several possible examples of such transitions and mention some of the expected fallouts in respective places while the details will be taken up elsewhere. A particular class of transition worth mentioning involves the possibility of gapping out only a subset of fermions across the transition, typically resulting in density-wave semimetals of various types. Such transitions between two different semimetals serve as examples of (semi)metallic criticality, which have been of recent interest~\cite{PhysRevLett.125.257202,PhysRevLett.128.087201}.

A more subtle structure in the phase diagram arises in the form of unnecessary phase transitions~\cite{PhysRevX.9.021034} or more precisely, unnecessary multi-critical points. These are observed in two or more lattice triplets whose components are made up of incompatible mass matrices ({\it i.e.}, they do not mutually anticommute). For such a triplet, isolated points of gaplessness (See Fig. \ref{fig_spectrum_on_sphere}) occur on a $2$-sphere denoting the mass manifold for the triplet. These isolated gapless points denote unnecessary multi-critical points since any two generic gapped points on this sphere can also be connected entirely by avoiding these gapless isolated points and hence avoiding the transition altogether. Such unnecessary multi-critical points can be understood as a fallout of the particular embedding of the microscopic symmetries in the emergent SU(8) (see Fig. \ref{fig_projection_from_S5_to_S2}). 

Finally, the bilinears corresponding to different order parameters may carry fermionic modes at their topological defects~\cite{abanov2000theta,PhysRevLett.100.156804,ryu2009masses}. The simplest is the gapless chiral edge fermions associated with the domain walls of the Integer Chern insulator or the anomalous Hall insulator. These lead to Chern-Simons terms in the action once the fermions are integrated out in the gapped phase and account for the gapless edge modes. In addition, we also find a slew of generalised Quantum spin-Hall insulators with edge/domain walls carrying gapless fermion modes captured by mutual Chern-Simons terms. The above analysis is easily extended to other topological defects including vortices and skyrmions. In particular, for the Skyrmions of a triplet quantum spin-octupole order parameter, we find the corresponding skyrmions are bosonic and carry {\it four} units of electronic charge. Thus, condensing such skyrmions naturally gives rise to a charge-$4e$ superconductor.

Considering the length and the  number of results that we present in this work, we  start with an overview of the results that summarises the work and helps the reader navigate the text.

\section{Overview of the results}

The focus of our work is crystalline systems on a honeycomb lattice formed out of edge-sharing octahedra (Fig. \ref{fig:3d crystal structure of ZrCl3}) where the electronically active transition metal ions, with strong atomic SOC, sit at the centres of such octahedra. Such structures are quite common and occur in a several stacked SOC magnets of recent interest such as the honeycomb Iridates A$_2$IrO$_3$ (A=Na, Li)~\cite{PhysRevB.82.064412,PhysRevB.83.220403,PhysRevLett.108.127203}, and ruthenates $\alpha$-RuCl$_3$~\cite{banerjee2016proximate,sandilands2015scattering,plumb2014alpha,choi2012spin}. In this work, we focus on such honeycomb system where the $j=3/2$ electronic orbitals (Fig. \ref{fig_SOC_splitting}) form the low energy manifold and in particular with a single electron ($d^1$ configuration) in the $j=3/2$ orbitals. An example of such a situation is $\alpha$-ZrCl$_3$~\cite{swaroop1964crystal,b1964synthesis,yamada_2018_emergent_SU(4)} where Cl$^{-}$ forms the edge sharing octahedral network with a Zr$^{3+}$-- in $d^1$ configuration -- sits and gives rise to a quarter filled $j=3/2$ system. While the fate of the low energy electronic phase of ZrCl$_3$ is still to be settled~\cite{yamada_2018_emergent_SU(4),ushakov2020can}, we expect that a large number of such materials exist  whose physics is governed by various regimes of electronic correlations. Our analysis will be applicable to the whole class of such materials with $j=3/2$ orbitals in $d^1$ (and $d^3$, by particle-hole symmetry) configuration.

\subsection{Spin-orbit coupled SU(8) Dirac semimetal}

The material set-up consists of the honeycomb system with electrons occupying the $j=3/2$ atomic orbitals (Eq. \ref{eq:4-component psi}) and at quarter filling ({\it i.e.}, $d^1$ configuration) as detailed in Section \ref{sec_microscopics}. The hopping Hamiltonian accounting for the electron hopping via the ligands that form the octahedra (Eq. \ref{eq:hamiltonian in j=3/2 orbitals}), introduced in Ref. \cite{yamada_2018_emergent_SU(4)}, is -- in an appropriately rotated {\it local} basis (Eq. \ref{eq:definition of phi})-- nothing but four copies of nearest neighbour hopping model on a honeycomb lattice in $\pi$-flux at $1/4$th filling (Eq. \ref{eq:Hamiltonian in SU(4) diagonal form}). The four copies stem from the four $j=3/2$ atomic orbitals, although the symmetry is enhanced to SU(4). On solving this Hamiltonian we obtain two valleys (Eq. \ref{eq:position of dirc points}) of Dirac band-touching at the $1/4$th filling (Fig. \ref{fig:band structure}). Note that these Dirac points, unlike graphene, are not at the Brillouin zone (BZ) corners, but at locations inside the magnetic BZ as shown in Fig. \ref{fig:band structure} with momenta given by Eq. \ref{eq:position of dirc points}.

The low energy theory takes the canonical Dirac form with the (Euclidean) Lagrangian density~\cite{kaplan2009chiral}
\begin{align}
    \mathcal{L}_0=v_F \sum_{f=1}^{N_F}\bar{\chi}_f({\bf r})(-i\slashed{\partial})\chi_f({\bf r})
	\label{eq_dirac_l}
\end{align}
and the corresponding Hamiltonian is given by Eq. \ref{eq_freediracham}. Here, $\slashed{\partial}\equiv\sum_{i=0}^2\gamma_i\partial_i$  with $\gamma^0, \gamma^1$ and $\gamma^2$ being $4\times 4$ matrices (see Eq. \ref{eq_gamma}) that generate an Euclidean Clifford algebra, {\it i.e.}, $\{\gamma_\mu,\gamma_\nu\}=2\delta_{\mu\nu}$ and $v_F$ is the velocity of the Dirac fermions. $\chi_f$ denotes $4-$component Dirac fermions (Eq. \ref{eq_16spinor}) with $N_F=4$ flavours ({\it i.e.}, $f=1,2,\cdots, N_F$) with $\bar{\chi}_f=i\chi^\dagger_f\gamma_0$. The four components of each $\chi_f$ are given by
\begin{align}
\chi_{f\sigma\tau}
\end{align}
where $\sigma=1,2$ corresponds to the two bands that touch at the Dirac points and $\tau=\pm$ are the two Dirac valleys/nodes. 

The free Dirac theory (Eq. \ref{eq_dirac_l}) has a very large internal SU(2$N_{F}$) symmetry in addition to the usual symmetries of the Dirac theory-- emergent Lorentz transformation and continuous translations as well as discrete charge conjugation, parity (reflection in two dimension~\cite{PhysRevLett.53.2449}) and time reversal. In the present case, the emergent SU(8) symmetry is a combination of the SU(4) flavour symmetry of the hopping model and the emergent chiral SU(2) symmetry with the latter being similar to spinless graphene~\cite{ryu2009masses}.

This IR symmetry is, of course,  emergent and much larger than the microscopic/UV symmetries of the lattice theory. In various candidate materials, different energy scales of the atomic orbitals provide for different microscopic symmetry groups which are then embedded within the larger IR symmetry group of the Dirac theory. Indeed, in the present case, the strong SOC resulting in the $j=3/2$ atomic orbitals result in an implementation of the honeycomb lattice symmetries on the resultant Dirac fermions, distinct from graphene. In particular, in the present case, the effect of SOC is manifested in the symmetry transformation of the $\chi$ spinors via mixing of the SU(4) flavours under lattice transformation (Table  \ref{tab:lat_sym} and Eq. \ref{eq: microscopic time reversal}). This difference in embedding, coupled with the difference in the location of the Dirac points, reflects in the properties of the present SO-coupled Dirac semimetal  and its proximate phases that are much richer than a larger $N_F$ flavour mono-layer graphene.

 
\subsection{Phases proximate to the SU(8) Dirac semimetal}

The emergent SU(8) and relativistic invariance of the low energy theory indicates that a large number of correlation functions decay identically at long distances. These correlation functions correspond to a wide set of very different phases as far as the microscopic symmetry breaking is concerned~\cite{PhysRevB.72.104404} and adding four-fermion interactions can lead to spontaneously symmetry broken and/or symmetry-protected topological phases proximate to the Dirac semimetal by favouring one of the channels.  

The Dirac semimetal is perturbatively stable to short-range four-fermion interactions (Eq. \ref{eq_interactions}). However, as the strength of the four-fermion term is increased it can favour the condensation of a particle-hole fermion bilinear (Eq. \ref{eq_massexp})
\begin{align}
    \langle\bar\chi_{f\sigma\tau}\chi_{f'\sigma'\tau'}\rangle\neq 0
\end{align}
that breaks the SU(8) symmetry spontaneously. There are 64 such bilinears divided into a SU(8) scalar
\begin{align}
    -i\langle \bar{\chi}\chi\rangle
    \label{eq_su8scalar}
\end{align}
and the 63 fold adjoint multiplet of SU(8) (also given in Eq. \ref{eq_su8adjoint})
\begin{align}
    -i\langle \bar{\chi}\mathcal{P}_a\chi\rangle,~~~~~a=1,\cdots,63.
\end{align}
and $\mathcal{P}_a$ being the SU(8) generators (Eq. \ref{eq_su8_generators}) obtained by combining the 15 generators of flavour SU(4), $\Sigma_i$ (defined below Eq. \ref{eq_umatrices}), and 3 generators of chiral SU(2), $\zeta_j$ (Eq. \ref{eq_zetasu2}), along with their respective identities.

Mean-field decomposition (exact in the limit $N_F\rightarrow\infty$~\cite{moshe2003quantum,sachdev1999quantum}) of the four-fermion interactions into the bilinears  (Eq. \ref{eq_dirac_massiveaction}) generically lead to fully gapping of the Dirac fermions resulting in a SU(8) symmetry broken insulator. However, we also find situations where only a subset of fermions are gapped out leading, typically to a density wave semimetal with a lower (than SU(8)) symmetry (see below). 

\begin{table*}
	\centering
	\caption{The SU(4) invariant chiral phases. The microscopic symmetry elements mentioned in this and the subsequent tables are defined in Table~\ref{tab:lat_sym}. Also, the $\zeta_i$ are generators of the chiral SU(2) and $\Sigma_i$ are generators of flavor SU(4) which are defined in Eq.~\ref{eq_zetasu2} and Appendix~\ref{appen_gamma} respectively.}
	\label{tab_chiral_masses}
	\begin{tabular}{|c|c|c|c|c|c|}\hline
		\# 	&   The Phase        	& Irrep         	&  Broken    		& Fermion            	& Comments \\ 
		&		         	&       			&  Microscopic      &  Bilinear          	&  \\ 
		&                 	&               	& symmetries        &       				& \\ \hline\hline
		1. &   Integer Chern    & $\mcl{A}^o_{2g}$    &   $\mbf{C_2'}$, $\mbf{\sigma_d}$, TR               & $-i\langle\bar\chi\chi\rangle$ & Fully gapped. \\  
		& insulator  (ICI).  &               &                       &                                & Charge Hall response\\
		&   (Sec.~\ref{sec:ICI-phase})              &               &                       &                                & given by $N_F=4$ \\
		&                  &               &                       &                                & CS theory (Eq. \ref{eq_icilag}) \\ \hline 	
		2.	&  Stripy     &  $\mcl{T}^e_{1g}$   &   $\mbf{T_1},\mbf{T_2},$   			& $-i \langle \bar{\chi} \zeta_1 \chi\rangle,   $       													& Fully gapped. \\    
		&charge density  &  &  {$\mbf{C_3,S_6,C_2',\sigma_d}$}                     	&  $-i \langle\bar{\chi} \zeta_2 \chi\rangle,$                  	&    Stripy modulation        \\      
		& wave (StCDW).&&& $-i\langle \bar{\chi} \zeta_3\chi\rangle$ & of electronic charge \\ 
            &(Sec.~\ref{sec:chiral-CDW-phase})&&&& density (Fig. \ref{fig:stripy_CDW})\\\hline 		
	\end{tabular}
\end{table*}

A natural question, then, pertains to the nature of such phases obtained by fermion-bilinear condensation that lie proximate to the Dirac semimetal. To answer this question, the actual embedding of the microscopic/UV symmetries in the emergent low-energy symmetries becomes important. We find that the 64 masses, break up into 27 irreducible representations representing 24 different phases (Eqs. \ref{eq_chiraldecomp_1}-\ref{eq:T2u_T1g_direct_product}). The microscopic symmetries only allow for singlet, doublet and triple irreducible representations (Irreps), and the 24 phases are made up of six singlets, five  doublets, and  sixteen triplets.

The 24 phases can be subdivided into three groups depending on the participation of the SU(4) flavour and the SU(2) chiral sectors in the fermion bilinear (Eq. \ref{eq_massexp}) which dictate their transformation properties under the microscopic symmetries. These are~:

\textit{Group-1 : Chiral masses : }The chiral masses (Eqs. \ref{eq_chiraldecomp_1}-\ref{eq_chiraldecomp_2}) are composed of SU(4) flavour singlets. These four bilinears have a structure of $-i\langle\bar\chi\zeta_j\chi\rangle$ where $j=0,1,2,3$ and we have included the SU(8) scalar (Eq. \ref{eq_su8scalar}) corresponding to the case of $\zeta^0$. This singlet corresponds to ($N_F=4$) Integer Chern Insulator (ICI) and is a SU(4) generalisation of the Haldane mass~\cite{PhysRevLett.61.2015} as given by Eq. \ref{eq_chern_insulator}. The other three masses (Eq. \ref{eq_su4_cdw}) form a triplet under lattice symmetries that correspond to three stripy charge-density-wave insulators (StCDW) (Fig. \ref{fig:stripy_CDW}). Notably, the presence of the $\pi$-flux breaks up the chiral masses as ${\bf 4} = {\bf 1} \oplus {\bf 3}$, as opposed to graphene where the four chiral masses are decomposed as ${\bf 4} = {\bf 1} \oplus {\bf 1} \oplus {\bf 2}$ corresponding to $N_F=2$ Integer Chern insulator, staggered charge density waves and two Kekule orders respectively~\cite{ryu2009masses}. The summary of these four masses is given in Table \ref{tab_chiral_masses} and the details are discussed in Section \ref{subsec_group1}. 

 \begin{table*}
	\centering
	\caption{The flavor phases}
	\label{tab_flavor_masses}
	\begin{tabular}{|c|c|c|c|c|c|}\hline
		\# 	&   The Phase        	& Irrep         	&  Broken    		& Fermion            	& Comments \\ 
		&		         	&       			&  Microscopic      & Bilinear           	& \\ 
		&                 	&               	& symmetries        &       				& \\ \hline\hline
		3. 	& Singlet        & $\mcl{A}^e_{1g}$   & None  & $-i\langle\bar{\chi} \Sigma_{45}\chi\rangle$  & Fully Gapped. \\ 
		& Quantum &               &       &                                               & Spin-octupole Hall response in  \\ 
		& spin-octupole     &               &       &                                               & presence of electric field (Eq. \ref{eq:octupolar_hall_action}) via \\
		&  Hall insulator.             &               &       &                                               & spin-octupole filtered edges  \\
		&    (Sec.~\ref{sec_45,33_mass})           &               &       &                                               & protected by U(1)$\rtimes$ Z$_{\rm 2}^{TR}$. \\ \hline		
		4. 	& Triplet  	& $\mcl{T}^e_{1g}$ 	&  $\mbf{T_1},\mbf{T_2},$ 					& $-i \langle \bar{\chi} \Sigma_{23} \chi\rangle,$  & Fully gapped. Spin-octupole  \\
		& Quantum  				& 				&  $\mbf{C_3}$, $\mbf{S_6}$, 						& $-i \langle\bar{\chi} \Sigma_{13} \chi\rangle,$  	& filtered edge currents. \\
		& spin-octupole  							&  				& $\mbf{C_2'}$, $\mbf{\sigma_d}$  	&  $-i\langle \bar{\chi} \Sigma_{12}\chi\rangle$  	& The Skyrmion configurations of \\
		&  Hall insulator.							&  				&  									&  													& the triplet order parameter carry $4$\\
		&  (Inversion even)							&  				&  									&  													&  units of electronic charge. Such \\
		&  	(Sec.~\ref{subsec_te1gsop})						&  				&  									&  													& skyrmion condensation leads \\ 
		& 							& 				& 									& 													& to a novel $4e$ superconductor. \\ \hline 		
		5. 	& Triplet   	& $\mcl{T}^e_{1u}$    & $\mbf{T_1},\mbf{T_2},$              & $-i\langle\bar\chi\Sigma_{34}\chi\rangle,$       													 & Non-compatible masses.  \\ 
		& Quantum                 &            & $\mbf{C_3}$, $\mbf{S_6}$,        & $-i\langle\bar\chi\left(-\frac{1}{2}\Sigma_{14}-\frac{\sqrt{3}}{2}\Sigma_{15}\right)\chi\rangle,$     & Generally fully gapped  \\
		& spin-octupole                          &     &   $\mbf{C_2'},\mbf{\sigma_d},\mbf{I}$                      & $-i\langle\bar\chi\left(-\frac{\sqrt{3}}{2}\Sigma_{25}+\frac{1}{2}\Sigma_{24}\right)\chi\rangle.$     & except for isolated points  \\
		&  Hall insulators.                         &               &                                         &                                                                                                   & protected by symmetries.   \\
		& (Inversion odd)                          &               &                                         &                                                             & Spin-octupole \\ 
		&  (Sec.~\ref{subsec_flavour_special})                      &               &                                         &     & filtered edge.\\ \hline
        6. & Triplet                          & $\mcl{T}^e_{2u}$ & $\mbf{T_1},\mbf{T_2},$                                        &$-i\langle\bar\chi\Sigma_{35}\chi\rangle,$                                                             & Similar to entry No.5 \\ 
		& Quantum                          &               &  $\mbf{C_3}$, $\mbf{S_6}$,                                        & $-i\langle\bar\chi\left(\frac{\sqrt{3}}{2}\Sigma_{24} + \frac{1}{2}\Sigma_{25}\right)\chi\rangle,$    &  of this table but with different \\
		& spin-octupole                    &               &   $\mbf{C_2'},\mbf{\sigma_d},\mbf{I}$                  & $ -i\langle\bar\chi\left(\frac{\sqrt{3}}{2}\Sigma_{14} - \frac{1}{2}\Sigma_{15}\right)\chi\rangle.$   &   spin-octupole current at the edges \\
        &Hall insulator &&&& protected by different set of symmetries.\\ 
        &(Inversion odd) &&&& \\
        & (Sec.~\ref{subsec_flavour_special}) &&&&\\\hline

        7.	& Doublet        & $\mcl{E}^o_u$       & $\mbf{C_2'}$, $\mbf{I}$,    & $-i\langle\bar\chi\Sigma_4\chi\rangle, -i\langle\bar\chi\Sigma_{5}\chi\rangle$        & Fully gapped.\\
		& Quantum                &               &  $\mbf{C_3}$, $\mbf{S_6},$                                      &                                                                                       & Vortices carry zero\\
		& spin-quadrupole Hall                           &               &  $\mbf{\sigma_d},$ TR       &                                                                                       & modes with charge and \\
		&insulator.                           &               &                                       &                                                                                       & quadrupole quantum numbers.\\ 
  &(Sec.~\ref{subsec_qsqp})&&&&\\\hline
		8. 	& Triplet 	   	& $\mcl{T}^o_{2g}$ 	& $\mbf{T_1},\mbf{T_2},$                            & $-i\langle\bar\chi\Sigma_1\chi\rangle,$   & Fully gapped.\\
		& Quantum         	&		        & $\mbf{C_3}$, $\mbf{S_6}$, 			                & $-i\langle\bar\chi\Sigma_2\chi\rangle,$   & Quadrupole filtered edge \\
		&spin-quadrupole Hall                         	&               &    $\mbf{C_2'},\mbf{\sigma_d}$, TR                           &$-i\langle\bar\chi\Sigma_3\chi\rangle$     & modes protected by U(1)  \\
		& insulator.                          &               &                                                   &                                           & symmetry. TR broken. \\
        &(Sec.~\ref{subsec_qsqp})&&&&\\\hline	
	\end{tabular}
\end{table*}

{\it Group-2 : Flavour masses :} The flavour masses are composed of chiral singlets. There are fifteen such bilinears of the form $-i\langle\bar\chi\Sigma_j\chi\rangle$ where $j$ runs over the fifteen generators of SU(4) (listed in Appendix \ref{appen_gamma}). Ten of them are time reversal (TR) even (Eq. \ref{eq_so5_generator_index}) and transform into each other under an adjoint representation of an SO(5) sub-group of the SU(4) flavour group while the other five are TR odd (Eq. \ref{eq_so5-vector}) and transform under a vector representation of the same SO(5). Under lattice symmetries, the ten TR even masses break up into one singlet and three triplets (Eqs. \ref{eq:decomposition_21} and \ref{eq:decomposition_22}) while the five TR odd masses break up into a doublet and a triplet (Eq. \ref{eq:decomposition_3}).  All these 15 flavour masses correspond to six generalised quantum spin-Hall phases~\cite{PhysRevLett.95.226801}. In particular, they result in four TR even quantum spin-octupole Hall phases (Sec. \ref{subsec_qsop}) and two TR odd quantum spin-quadrupole Hall phases (Sec. \ref{subsec_qsqp}) with the latter being an interesting analog of symmetry protected topological phases without time-reversal symmetry. These masses are summarised in Table \ref{tab_flavor_masses} while the details are given in Sec. \ref{subsec_group2}.

{\it Group-3 : Mixed masses :} The mixed masses are composed of non-trivial combinations of flavour and chiral sectors. There are forty-five such bilinears of the form $-i\langle\bar\chi\Sigma_j\zeta_k\chi\rangle$ where $j (k)$ runs over the 15(3) SU(4) flavour (SU(2) chiral) generators. These masses, under lattice symmetries, are divided into 11 triplets, 4 doublets and 4 singlets as shown in Eq. \ref{eq:decomposition_4}. The nature of the resultant phases is quite rich and  corresponds to various types of dipolar, quadrupolar and octupolar density waves that break both SU(4) flavour as well as lattice symmetries. In particular, these masses describe four density wave patterns for various types of spin-quadrupole and spin-octupole order parameters. These are -- (1) Uniform (ferro) order, (2) Staggered (``Neel") order (Fig. \ref{fig_Neel_pattern}), (3) Stripy order (Fig. \ref{fig:stripy_CDW}), and, (4) Zig-zag order (Fig. \ref{fig:zig_zag_density_wave}). Moreover, depending on the fate of the fermions in these phases, they are divided into two classes : (a) insulators, where all the Dirac fermions are gapped out (summarised in Table \ref{tab_mixed_insulators}), and, (b) semimetals, where a subset of Dirac fermions remain gapless even after the condensation of the bilinear (summarised in Table \ref{tab_mixed_semimetals}). We discuss them in detail in Section \ref{subsec_group3}.  It is useful to note that amongst the insulators, there are two singlets, $\mathcal{A}^e_{1g}$ and $\mathcal{A}_{2g}^o$ (see Table \ref{tab_mixed_insulators}; Secs. \ref{sec:ferro-spinQuadru-insulator} and \ref{subsubsec:IsingFerroSpinOctopolar}) that have non-trivial gapless edge modes. While the former is time reversal even corresponds to an Ising ferro spin-quadrupolar insulator with quantized spin-octupole filtered edge modes that are captured by a non-trivial mutual CS term (Eq. \ref{eq_cs_mutualising}), the latter is time reversal odd and represents an anomalous Hall insulator with quantized charge Hall response (Eq. \ref{eq_cs_anomaloushall}). 

{ In section~\ref{sec:Discussion} that follows the main body of work presented in sections~\ref{sec_microscopics}-\ref{subsec_group3},  we discuss the implications of the above results in a broader context bringing forth the unique features and opportunities brought out by the specific spin-orbit coupled system discussed in this paper. Various technical details are summarised in different appendices.
 }

\begin{table*}
	\centering
	\caption{The mixed phases (insulators)}
		\label{tab_mixed_insulators}
	\begin{tabular}{|c|c|c|c|c|c|}\hline
		\# 	&   The Phase        	& Irrep         	&  Broken    		& Fermion            	& Comments \\ 
		&		         	&       			&  Microscopic      & Bilinear           	& \\ 
		&                 	&               	& symmetries        &       				& \\ \hline\hline
		9.  &Singlet            & $\mcl{A}_{1g}^e$    & None      & $-i\Braket{\bar{\chi} \left(\Sigma_{3}\zeta_{1} - \Sigma_{1}\zeta_{3} - \Sigma_{2}\zeta_{2} \right)\chi}/{\sqrt{3}}$    & Fully gapped.\\
		&  Spin-quadrupolar	    &               &           &                                                                                                                               & Octupole ($\Sigma_{45}$)\\
  		&  ferro Hall 	    &               &           &                                                                                                                               & filtered edge modes.\\ 
        & insulator.  &&&&\\
            & (Sec.~\ref{sec:ferro-spinQuadru-insulator})&&&&\\\hline
		10.  &Singlet               & $\mcl{A}_{2g}^o$    & $\mbf{C_2'}$, $\mbf{\sigma_d}$,TR        & $-i\Braket{\bar{\chi} \left(\Sigma_{12}\zeta_{1} - \Sigma_{23}\zeta_{3} + \Sigma_{13}\zeta_{2} \right)\chi}/{\sqrt{3}}$    & Fully gapped.\\
		&  Spin-octupolar				      &               &                   &                                                                                                                                   & Gapless edges carry\\
  	    &  Anomalous ferro				      &               &                   &                                                                                                                                   & qunatized charge current\\
        & Hall Insulator.&&&& via CS term.\\
        & (Sec.~\ref{subsubsec:IsingFerroSpinOctopolar}) &&&&\\ \hline
		11. & Doublet   & $\mcl{E}_u^o$       & $\mbf{C_2'}$, $\mbf{\sigma_d}$, $\mbf{I}$, $\mbf{C_3}$, $\mbf{S_6}$   & See Eq.~\ref{eq_euo_mass_2}                            & Fully gapped.  \\
		& Spin-octupolar   &  (2 doublets) & TR                                      & and Eq.~\ref{eq_euo_mass_3}						     & Both doublets correspond \\ 
		& N\'eel insulator.							&				&										& 									 &to the same phase.\\
        &(Sec.~\ref{sec:Neel-spinOctupole-insulator})&&&  	& Vortices can carry  \\
        &&&&& non-trivial quantum \\
        &&&&&number.\\\hline	

		12.	& Triplet                      &$\mcl{T}_{2g}^o$             & $\mbf{T_1},\mbf{T_2},$        &  $-i\Braket{\bar{\chi}\Sigma_{45}\zeta_1\chi}$,        & Fully gapped.        \\ 
		& Spin-octupolar                     &                       &  $\mbf{C_3}$, $\mbf{S_6}$, $\mbf{C_2'},\mbf{\sigma_d},$      &  $-i\Braket{\bar{\chi}\Sigma_{45}\zeta_2\chi}$,         &          \\
		& stripy density wave                                          &                       &   TR                            &  $-i\Braket{\bar{\chi}\Sigma_{45}\zeta_3\chi}$         &           \\ 
        & (StDW) insulator. &&&&\\
        & (Sec.~\ref{sec:stripy-spinOctupole-insulator}) &&&& \\\hline		
		13. 	& Triplet 			&$\mcl{T}_{1u}^e$      &   $\mbf{T_1},\mbf{T_2},$             &  $-i\Braket{\bar{\chi} \Sigma_5 \zeta_1 \chi}$    									&  Fully gapped,       \\ 
		& Spin-quadrupolar	         &                           &   $\mbf{C_3}$, $\mbf{S_6}$, $\mbf{C_2'},\mbf{\sigma_d},\mbf{I}$       & $-\frac{i}{2}\Braket{\bar{\chi} (-\sqrt{3}\Sigma_4+\Sigma_5)\zeta_3 \chi}$,   		&  Masses are \\
		& zig-zag density wave                        &                           &                                      & $-\frac{i}{2}\Braket{\bar{\chi} (-\sqrt{3}\Sigma_4-\Sigma_5)\zeta_2\chi}$        	&  non-comaptible, \\ 
		& (ZDW) insulator.		&							&                                      &  								                                                &  Gapless modes \\
		&	(Sec.~\ref{sec:zigzag-spinQuadru-insulator})						&							&                                      & 										& for some special linear \\
		&						&							&                                      & 			& combinations.  \\\hline	
        14.  &Triplet&$\mcl{T}_{2u}^e$&$\mbf{T_1},\mbf{T_2},$&$-i\Braket{\bar{\chi} \Sigma_4 \zeta_1 \chi}$,& Similar to entry No.13 \\
            &Spin-quadrupolar&&$\mbf{C_3}$, $\mbf{S_6}$, $\mbf{C_2'},\mbf{\sigma_d},\mbf{I}$&$-\frac{i}{2}\Braket{\bar{\chi} (\sqrt{3}\Sigma_5+\Sigma_4)\zeta_3 \chi}$,&  in this table, but \\
            &ZDW insulator&&&$-\frac{i}{2}\Braket{\bar{\chi} (\sqrt{3}\Sigma_5-\Sigma_4)\zeta_2\chi}$ & different quadrupole \\
            & (Sec.~\ref{sec:zigzag-spinQuadru-insulator})&&&& operators are  \\ 
            &&&&& ordered.\\\hline
		15.	& Triplet			     &  $\mcl{T}_{1u}^o$       & $\mbf{T_1},\mbf{T_2},$                    &  See                                          	               & Both triplets correspond \\
		& Spin-octupolar      		&  (2 triplets)     & $\mbf{C_3}$, $\mbf{S_6}$, $\mbf{C_2'},\mbf{\sigma_d},\mbf{I},$          &  Eq.~\ref{eq_t1u0_masses_1}, \ref{eq_t1uo_masses_2}	           & to the same phase.  \\
		& ZDW insulator.                          &         &   TR                                        &   										            	       & Fully gapped,  \\
		&  (Sec.~\ref{sec:zigzag-spinOctupole-insulator})                 &     &                                           &           		& Masses are non-compatible. \\
		&							&					&                                            &																	& Gapless modes appear \\ 
            &&&&& for special linear\\
            &&&&&  combinations. \\\hline		
        16.  & Triplet & $\mcl{T}_{2u}^o$ &$\mbf{T_1},\mbf{T_2},$& See& Similar to entry No. 15 \\
            & Spin-octupolar & (2 triplets)   & $\mbf{C_3}$, $\mbf{S_6}$, $\mbf{C_2'}, \mbf{\sigma_d}, \mbf{I},$ & Eq.\ref{eq_t2uo_masses_1}, \ref{eq_t2uo_masses_2} & in this table, but \\
            & ZDW insulator && TR && different spin-octupole \\
            & (Sec.~\ref{sec:zigzag-spinOctupole-insulator}) &&&& operators are ordered.\\ \hline
	\end{tabular}
\end{table*}

\begin{table*}
	\centering
	\caption{The mixed phases (semimetals)}
		\label{tab_mixed_semimetals}
	\begin{tabular}{|c|c|c|c|c|c|}\hline
		\# 	&   The Phase        	& Irrep         	&  Broken    		& Fermion            	& Comments \\ 
		&		         	&       			&  Microscopic      & Bilinear           	& \\ 
		&                 	&               	& symmetries        &       				& \\ \hline\hline
		17. 	& Singlet    &  $\mcl{A}_{1u}^o$   & TR, $\mbf{I},\mbf{\sigma_d}$             & See Eq.~\ref{eq_neel_singlet1}					& Have semimetallic features.    \\
		& Spin-octupolar    &            &                            & 					&  Can give rise to integer \\
		&   N\'eel semimetal.             &     &               &   											    &    QH phase which is different \\
		& (Reflection odd)                      &               &                   			&                                                    &   from the ICI phase. \\
            &(Sec.~\ref{sec:Neel-spinOctupole-semimetal})&&&&\\\hline
        18.  &Singlet& $\mcl{A}_{2u}^o$ & TR, $\mbf{I}$, $\mbf{C_2'}$ & See Eq.~\ref{eq_neel_singlet2} & Similar to entry No.17\\
            &Spin-octupolar&&&&  in this table, but different\\
            &N\'eel semimetal&&&& spin-octupole operator is ordered.\\
            & (Reflection even) &&&& \\ 
            &(Sec.~\ref{sec:Neel-spinOctupole-semimetal})&&&& \\\hline

		19.		& Triplet     		&$\mcl{T}_{1g}^o$     & $\mbf{T_1},\mbf{T_2},$      &$-i\Braket{\bar{\chi}\left(-\Sigma_{13}\zeta_3 - \Sigma_{23}\zeta_2 \right)\chi}/\sqrt{2}$        &Masses are non-compatible, \\
		& Spin-octupolar        	&            &  $\mbf{C_3}$, $\mbf{S_6}$, $\mbf{C_2'},\mbf{\sigma_d},$        &$-i\Braket{\bar{\chi}\left(\Sigma_{12}\zeta_2 - \Sigma_{13}\zeta_1 \right)\chi}/\sqrt{2}$      & 8 fermionic modes are\\
		& StDW semimetal            			&     &   TR                         &$-i\Braket{\bar{\chi}\left(\Sigma_{23}\zeta_1 + \Sigma_{12}\zeta_3 \right)\chi}/\sqrt{2}$      & gapless, number of \\
		&    (Sec.~\ref{sec:stripy-spinOctupole-semimetal})                        &               &       &                                                                                            &gapless modes are  same for \\
		&            &               &   	&              & all linear combinations of \\ 
		&							&				&                               &            & the masses within a given \\
		& 							&				&                               &            & triplet. \\\hline 
		20.  	& Triplet      		&$\mcl{T}_{2g}^o$     & $\mbf{T_1},\mbf{T_2},$      &$-i\Braket{\bar{\chi}\left(\Sigma_{13}\zeta_3 - \Sigma_{23}\zeta_2 \right)\chi}/\sqrt{2}$        &Masses are non-compatible, \\
		& Spin-octupolar        	&            &  $\mbf{C_3}$, $\mbf{S_6}$, $\mbf{C_2'},\mbf{\sigma_d},$    &$-i\Braket{\bar{\chi}\left(-\Sigma_{12}\zeta_2 - \Sigma_{13}\zeta_1 \right)\chi}/\sqrt{2}$      & 4 fermionic modes are always\\
		& StDW semimetal             			&     &     TR                       &$-i\Braket{\bar{\chi}\left(-\Sigma_{23}\zeta_1 + \Sigma_{12}\zeta_3 \right)\chi}/\sqrt{2}$      & gapless, \\
		&   (Sec.~\ref{sec:stripy-spinOctupole-semimetal})                        &               &       &                                                                                            & this number changes depending \\
		&            	&               &  		&         	& on the linear combinations of \\ 
		& 							&				&                               &           & the masses.  \\
		& 							&				&                               &            &  \\\hline   
        21.  &Triplet &$\mcl{T}_{2g}^e$& $\mbf{T_1},\mbf{T_2},$ & $-i\Braket{\bar{\chi}\left(-\Sigma_1\zeta_2 + \Sigma_2\zeta_3 \right)\chi}/\sqrt{2}$ & Similar to entry No.19\\
            & Spin-quadrupolar && $\mbf{C_3}$, $\mbf{S_6}$, $\mbf{C_2'},\mbf{\sigma_d},$ & $-i\Braket{\bar{\chi}\left(\Sigma_3\zeta_2 + \Sigma_2\zeta_1 \right)\chi}/\sqrt{2}$ & in this table, but different\\
            & StDW semimetal. &&& $-i\Braket{\bar{\chi}\left(\Sigma_3\zeta_3 + \Sigma_1\zeta_1 \right)\chi}/\sqrt{2}$ & spin-quadrupole operators \\
            & (Sec.~\ref{sec:stripy-spinQuadru-semimetal}) &&&& are ordered.\\ \hline
        22.  &Triplet & $\mcl{T}_{1g}^e$ & $\mbf{T_1},\mbf{T_2},$ & $-i\Braket{\bar{\chi}\left(-\Sigma_1\zeta_2 - \Sigma_2\zeta_3 \right)\chi}/\sqrt{2}$ & Similar to entry No.20\\
            & Spin-quadrupolar && $\mbf{C_3}$, $\mbf{S_6}$, $\mbf{C_2'},\mbf{\sigma_d}$ & $-i\Braket{\bar{\chi}\left(-\Sigma_3\zeta_2 + \Sigma_2\zeta_1 \right)\chi}/\sqrt{2}$ & in this table, but different\\
            & StDW semimetal &&& $-i\Braket{\bar{\chi}\left(\Sigma_3\zeta_3 - \Sigma_1\zeta_1 \right)\chi}/\sqrt{2}$ & spin-octupole operators are ordered.\\
            & (Sec.~\ref{sec:stripy-spinQuadru-semimetal}) &&&&\\ \hline
          23.  & Doublet          & $\mcl{E}_g^e$       & $\mbf{C_2'},\mbf{\sigma_d}$, $\mbf{C_3}$, $\mbf{S_6}$                  &   See Eq.~\ref{eq_ege_doublet_1}                                                 & Some bands remain gapless. \\
		& spin-quadrupolar  	 &               &                                       &                                                                                       & Can give rise to integer QH    \\
		& ferro semimetal.                          &               &                                       &                                                                                       & phase different the ICI phase \\ 
            & (Sec.~\ref{sec:ferro-spinQuadru-semimetal}) &&&&\\\hline
	24.    	& Doublet       & $\mcl{E}_g^o$       & $\mbf{C_2'},\mbf{\sigma_d}$, $\mbf{C_3}$, $\mbf{S_6},$              &   See Eq.~\ref{eq_ego_doublet}                                                   & Some bands remain gapless \\
		& Spin-octupolar 	   &               &TR                                       &                                                                                       &                               \\ 
        & ferro  semimetal. &&&&\\
        & (Sec.~\ref{subsubsec:FSOPSWSM}) &&&&\\\hline			
	\end{tabular}
\end{table*}

\section{The Dirac theory for the \texorpdfstring{$d^1$}{} systems on honeycomb lattice}
\label{sec_microscopics}

The starting point of our analysis is the low energy single electron atomic orbitals of  the $d^1$ ion in an octahedral crystal field with strong atomic SOC~\cite{PhysRevLett.118.217202,yamada_2018_emergent_SU(4)} as shown in Fig.~\ref{fig_SOC_splitting}. In the absence of SOC, the single electron occupies the six-fold degenerate $t_{2g}$ atomic orbitals $|d_{XY},\sigma\rangle, |d_{YZ},\sigma\rangle$ and $|d_{ZX},\sigma\rangle$ with $\sigma=\uparrow,\downarrow$ while the high energy $e_g$ orbitals remain empty and are projected out. The SOC, projected on the $t_{2g}$ manifold gives $H_{\rm SOC}^{t_{2g}}=-\lambda~{\bf l\cdot s}$ ($\lambda>0$), where $l=1$ is the effective orbital angular momentum of the $t_{2g}$ orbitals\cite{PhysRevLett.118.217202,PhysRevLett.101.076402} (see Appendix \ref{appen_sub_j3/2} for further details).

The strong SOC selects the low energy orbitals by splitting the six-fold degeneracy into $4\oplus 2$ which corresponds to the lower $j=3/2$ and higher $j=1/2$  (${\bf j=l+s}$) orbitals respectively with the splitting being $3\lambda/2$~\cite{PhysRevLett.101.076402}. For the $d^1$ configuration, the low energy physics is therefore of a 1/4th filled $j=3/2$ orbitals (Fig.~\ref{fig_SOC_splitting}) with single electron creation operators at the lattice site are  given by~\cite{PhysRevLett.118.217202,yamada_2018_emergent_SU(4)} 
\begin{align}\label{eq:4-component psi}
	\psi^{\dagger} = \left( \psi^{\dagger}_{1/2}, \psi^{\dagger}_{-1/2}, \psi^{\dagger}_{3/2},\psi^{\dagger}_{-3/2} \right).
	\end{align}
The interplay of hopping and interaction of electrons occupying these four orbitals then determine the low energy electronic properties of the system.	

\subsection{The lattice and microscopic symmetries}

Consider such $j=3/2$ orbitals on a honeycomb geometry with edge-sharing octahedra. In this geometry, the active atoms sit at the centre of each octahedron and form the honeycomb lattice. It is useful to consider the honeycomb lattice to lie in a plane perpendicular to the Cartesian $[111]$ direction (details in Appendix \ref{appen_sub_lattice}) such that the three nearest neighbour bonds are parallel to the three Cartesian planes shown in Fig. \ref{fig:3d crystal structure of ZrCl3}. Correspondingly we denote these bonds as $x$, $y$, and $z$ bonds if they are parallel to the $YZ$, $ZX$, and $XY$ planes respectively, following the, by now popular nomenclature in the context of the Kitaev spin model~\cite{kitaev2006anyons} on the honeycomb lattice~\cite{PhysRevB.102.235124}.

\begin{figure}
\centering
	\includegraphics[scale=0.4]{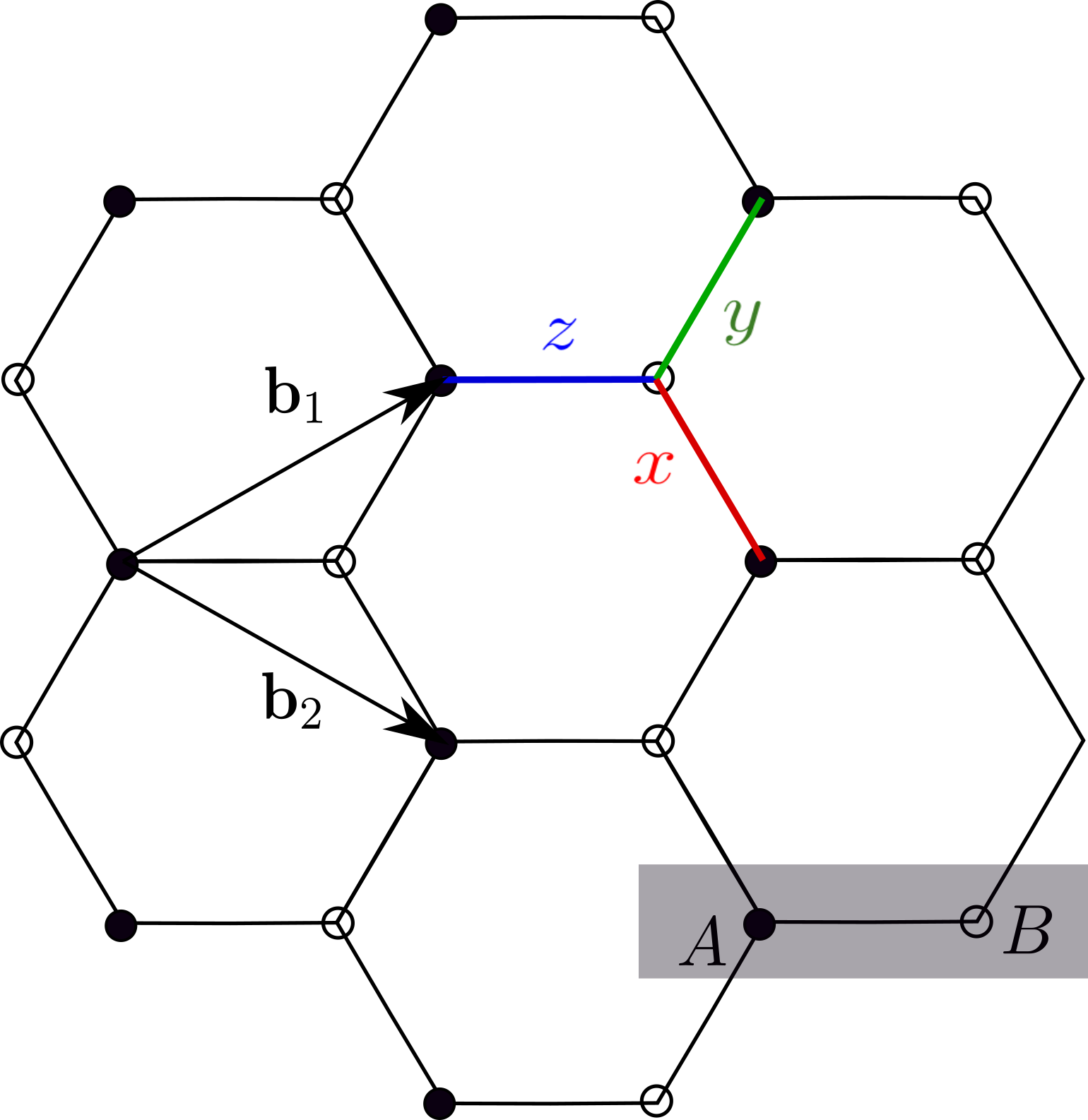}
	\caption{The filled circles are points of A sub-lattice and the hollow circles are of B sub-lattice. The gray shaded area shows the two points of a single unit cell with $\mathbf{b}_1$ and $\mathbf{b}_2$ being unit lattice vectors (see Eq. \ref{eq_lattice_vectors}).}
	\label{fig:2-point honeycomb}
\end{figure}

The honeycomb net has a triangular Bravais lattice and a two-site unit cell with two sub-lattices, $\mathcal{s}=A,B$ as shown in Fig. \ref{fig:2-point honeycomb}.  Each point on the honeycomb lattice is labeled by $(\mathcal{u}_1,\mathcal{u}_2,\mathcal{s})$ where $\mathcal{u}_1,\mathcal{u}_2\in \mathbb{Z}$ denote the position of the unit cell via 
\begin{align}
{\bf r}=\mathcal{u}_1{\bf b}_1+\mathcal{u}_2{\bf b}_2
\end{align}
with ${\bf b}_1$ and ${\bf b}_2$ are unit lattice vectors shown in Fig. \ref{fig:2-point honeycomb}.

\begin{table}
    \centering
    \begin{tabular}{|c|c|c|}\hline 
     & ${\bf r}(\mathcal{u}_1,\mathcal{u}_2,\mathcal{s})\rightarrow {\bf r}'(\mathcal{u}_1',\mathcal{u}_2',\mathcal{s}')$ & $\psi({\bf r})\to\psi'({\bf r}')$\\\hline\hline
     ${\bf T}_1$ & $(\mathcal{u}_1,\mathcal{u}_2,\mathcal{s})\rightarrow (\mathcal{u}_1+1,\mathcal{u}_2,\mathcal{s})$ &$\psi \to \mcl{U}_{\mbf{T}_1}\psi$\\ \hline
     ${\bf T}_2$ & $(\mathcal{u}_1,\mathcal{u}_2,\mathcal{s})\rightarrow (\mathcal{u}_1,\mathcal{u}_2+1,\mathcal{s})$ &$\psi \to \mcl{U}_{\mbf{T}_2}\psi$\\ \hline
     ${\bf C_3}$ & $(\mathcal{u}_1, \mathcal{u}_2, \mathcal{s}) \to (\mathcal{u}_2-1, -\mathcal{u}_1-\mathcal{u}_2+ \delta_{\mathcal{s},A}, \mathcal{s})$ &$\psi \to \mcl{U}_{\mbf{C_3}}\psi$\\ \hline
     ${\bf S_6}$ & $(\mathcal{u}_1, \mathcal{u}_2, \mathcal{s}) \to (\mathcal{u}_1+\mathcal{u}_2- \delta_{\mathcal{s},A},- \mathcal{u}_1, \bar{\mathcal{s}})$ &$\psi \to \mcl{U}_{\mbf{S_6}}\psi$\\ \hline
     ${\bf C_2'}$ & $(\mathcal{u}_1,\mathcal{u}_2, \mathcal{s}) \to (\mathcal{u}_2-1, \mathcal{u}_1+1 ,\mathcal{s})$ &$\psi \to \mcl{U}_{\mbf{C_2}}\psi$\\ \hline
     ${\boldsymbol{\sigma_d}}$ & $(\mathcal{u}_1, \mathcal{u}_2, \mathcal{s}) \to (-\mathcal{u}_2, -\mathcal{u}_1, \bar{\mathcal{s}})$ &$\psi \to \mcl{U}_{\mbf{\sigma_d}}\psi$\\ \hline
     ${\bf I}$ & $(\mathcal{u}_1, \mathcal{u}_2, \mathcal{s}) \to (-\mathcal{u}_1-1, -\mathcal{u}_2+1, \bar{\mathcal{s}})$ &$\psi \to \mcl{U}_{\mbf{I}}\psi$\\ \hline
\end{tabular}
    \caption{The action of the microscopic symmetries on the $j=3/2$ orbitals (Eq. \ref{eq:4-component psi}). Note $\bar{\mathcal{s}}=B(A)$ for $\mathcal{s}=A(B)$. The $\mcl{U}_\mbb{S}$ ($\mbb{S}$ represents some lattice symmetry) are $4\times 4$ unitary matrices which are given in Appendix~\ref{appen_symm} (Eqs \ref{eq:transformation of psi under translation} - \ref{eq:transformation of psi under C2}).}
    \label{tab:lat_sym}
\end{table}

To understand the lattice symmetries, we take the ideal honeycomb structure of $\alpha$-ZrCl$_3$ as a prototypical example as it has all the representative symmetries. The point group of the $\alpha$-ZrCl$_3$ lattice is $D_{3d}$ which has 12 elements. Keeping in mind the geometry of the edge-sharing ligand octahedra surrounding the active ion (Fig. \ref{fig:3d crystal structure of ZrCl3}), the generators of the lattice symmetries of the system are as follows and their action on the lattice coordinates are given in Table \ref{tab:lat_sym}.
\begin{itemize}
\item{${\bf T}_1, {\bf T}_2 :$ Two-dimensional lattice translations} of the honeycomb lattice by ${\bf b}_1$ and ${\bf b}_2$ respectively.
\item ${\bf C_3 :}$ {Rotations by angle $\frac{2\pi}{3}$} about the center of a honeycomb plaquette.
\item ${\bf S_6 :}$ Rotations about  the center of a honeycomb plaquette by angle \texorpdfstring{$\frac{\pi}{3}$}{} followed by a reflection about the honeycomb plane.
\item ${\bf C'_2} :$ Rotations by angle $\pi$ about the axes lying on the honeycomb plane and passing through two opposite vertices of a honeycomb plaquette. There are three of such axes. One of the ${\bf C_2'}$ axes is {parallel to the $z$-bonds} (see Fig.~\ref{fig:2-point honeycomb}).
\item ${\boldsymbol{\sigma}_d} :$ {Reflections about planes which are perpendicular to the honeycomb plane and bisect the angle between two consecutive ${\bf C'_2}$ axes.} There are three such planes. One of the planes is the perpendicular bisector of one of the $z$-bonds in Fig~\ref{fig:2-point honeycomb}.
\item ${\bf I :}$ Inversion about the center of a honeycomb plaquette.
\end{itemize}

In addition, the system also has time reversal (TR) symmetry $\mathbb{T}$, with
\begin{align}
\mathbb{T}^2=-1.
\label{eq:microtr}
\end{align}
In appendix \ref{appen_symm}, we provide the details of the transformation of the $t_{2g}$ and the $j=3/2$ orbitals under the above symmetries.


\subsection{The tight binding Hamiltonian for indirect hopping}

Starting with the hopping Hamiltonian for the $t_{2g}$ orbitals and taking into account the indirect hopping amplitudes via the ligand in the edge-sharing geometry (Fig. \ref{fig:3d crystal structure of ZrCl3}), the effective Hamiltonian is obtained by projecting it to the $j=3/2$ manifold using Eq. \ref{eq:projection from t2g to j=3/2}. This minimal hopping Hamiltonian for the $j=3/2$ orbitals (Eq.~\ref{eq:4-component psi}) is given by 
\begin{eqnarray}\label{eq:hamiltonian in j=3/2 orbitals}
	H = -\frac{t}{\sqrt{3}} \sum_{\braket{\mathbf{ r,\mathcal{s};r^{\prime},\mathcal{s}'}}} \psi^{\dagger}(\mathbf{r},\mathcal{s}) U^{\mathcal{s}\mathcal{s}'}_{\mathbf{ rr^{\prime}}} \psi(\mathbf{r^{\prime}},\mathcal{s}') + {\rm h.c.}
\end{eqnarray}
where  $U^{\mathcal{s}\mathcal{s}'}_{\mathbf{rr'}}$ are hopping amplitudes of overall strength $t$, on nearest neighbour bonds (hence only between different sublattices)  that are given by $4 \times 4$ Hermitian matrices which depend on the type ($x,y$ or $z$, see Fig. \ref{fig:2-point honeycomb})  of the $\braket{\mbf{r},\mcl{s};\mbf{r}',\mcl{s}'}$ bond~\cite{yamada_2018_emergent_SU(4)}  given by 
\begin{eqnarray}
	&U^{AB}_{\mathbf{rr^{\prime}}} &\equiv U_x = - \Sigma_1 \text{ , if $\braket{\mbf{r},A;\mbf{r}',B}=x$  } \nonumber \\  
	&& \equiv U_y = -\Sigma_2  \text{ , if $\braket{\mbf{r},A;\mbf{r}',B}=y$  }\nonumber\\
	&& \equiv U_z = -\Sigma_3 \text{ , if $\braket{\mbf{r},A;\mbf{r}',B}=z$  }
	\label{eq_umatrices}
\end{eqnarray}
Here, $\Sigma_{i}$ are sixteen $4\times 4$ traceless Hermitian matrices with $\Sigma_0$ being the identity matrix and the rest being generators of SU(4). They can be obtained by using $j=3/2$ matrices as shown in Appendix \ref{appen_gamma}.

The three $U_\alpha$(for $\alpha = x,y,z$) matrices square up to identity and mutually anti-commute, i.e.,
\begin{align}
	U_{\alpha}^2=\mathbb{I}_4\equiv\Sigma_0,~~~~\{U_{\alpha} ,U_{\beta} \}= 2\delta_{\alpha\beta}\Sigma_0~~~~\forall~\alpha,\beta=x,y,z.
\end{align}

Before proceeding to diagonalize Eq.~\ref{eq:hamiltonian in j=3/2 orbitals} to obtain the electron band structure, we identify the generic nature of the electron dispersion.

\subsubsection{The  \texorpdfstring{SU(4)}{} Symmetry and \texorpdfstring{$\pi$}{}-flux}

As a first step, it is useful to consider the phase picked up by the electron on encircling any closed loop of the lattice. Such loops are formed out of the honeycomb plaquette consisting of six sites. The phase is given by the product of the $U_\alpha$ matrices around a honeycomb plaquette and is generically given by
\begin{align}
    \prod_{\braket{\mathbf{r,\mathcal{s};r^{\prime},\mathcal{s}'}}\in \hexagon} U^{\mathcal{s}\mathcal{s}'}_{\mathbf{rr^{\prime}}} =\sum_i W_i~\Sigma_i
    \label{eq_flux}
\end{align}
where $W_i$ are the respective coefficients that denote a generic direction in the SU(4) space. In the above sum, $i$ runs over the 16 indices of the $\Sigma_i$ matrices defined in Appendix~\ref{appen_gamma}.

Crucially, however, it was noticed in Ref.~\cite{yamada_2018_emergent_SU(4)} that the explicit form of $U$-matrices (Eq. \ref{eq_umatrices}) give
\begin{eqnarray}\label{eq:pi flux statement in U}
	\prod_{\braket{\mathbf{r,\mathcal{s};r^{\prime},\mathcal{s}'}}\in \hexagon} U^{\mathcal{s}\mathcal{s}'}_{\mathbf{rr^{\prime}}} = -\Sigma_0
\end{eqnarray}
such that no direction in SU(4) space is favoured and the system has an underlying SU(4) symmetry. This SU(4) symmetry can be made manifest by suitable site-dependent unitary rotations of the $\psi$ fermions (see below, Eq. \ref{eq:definition of phi}) \cite{yamada_2018_emergent_SU(4)}. 
 
An equally important feature is the negative sign in Eq. \ref{eq:pi flux statement in U} which shows that such SU(4) fermions experience a $\pi$-flux through every hexagon. Thus the above problem of $d^1$ fermions is that of SU(4) symmetric fermions hopping on a honeycomb lattice with $\pi$-flux per plaquette. The non-trivial effect of $\pi$-flux at $1/4$-th filling is already apparent by considering the simpler case of spinless fermions on honeycomb lattice with $\pi$-flux at quarter-filling. This, as we discuss below, leads to Dirac fermions at low energy whose properties are quite different from those in graphene.

In the rest of this work, we uncover the interplay of SU(4) symmetry and the $\pi$-flux that, along with electron-electron interactions, leads to rich low-energy electronic properties of $d^1$ systems. 

\subsubsection{SU(4) diagonalization and the local basis}
Following Ref. \cite{yamada_2018_emergent_SU(4)}, the SU(4) symmetry of the hopping Hamiltonian in Eq.~\ref{eq:hamiltonian in j=3/2 orbitals} can be made manifest by performing site-dependent (local) unitary transformations on the fermions. 

\begin{figure}
\centering
\includegraphics[scale=0.4]{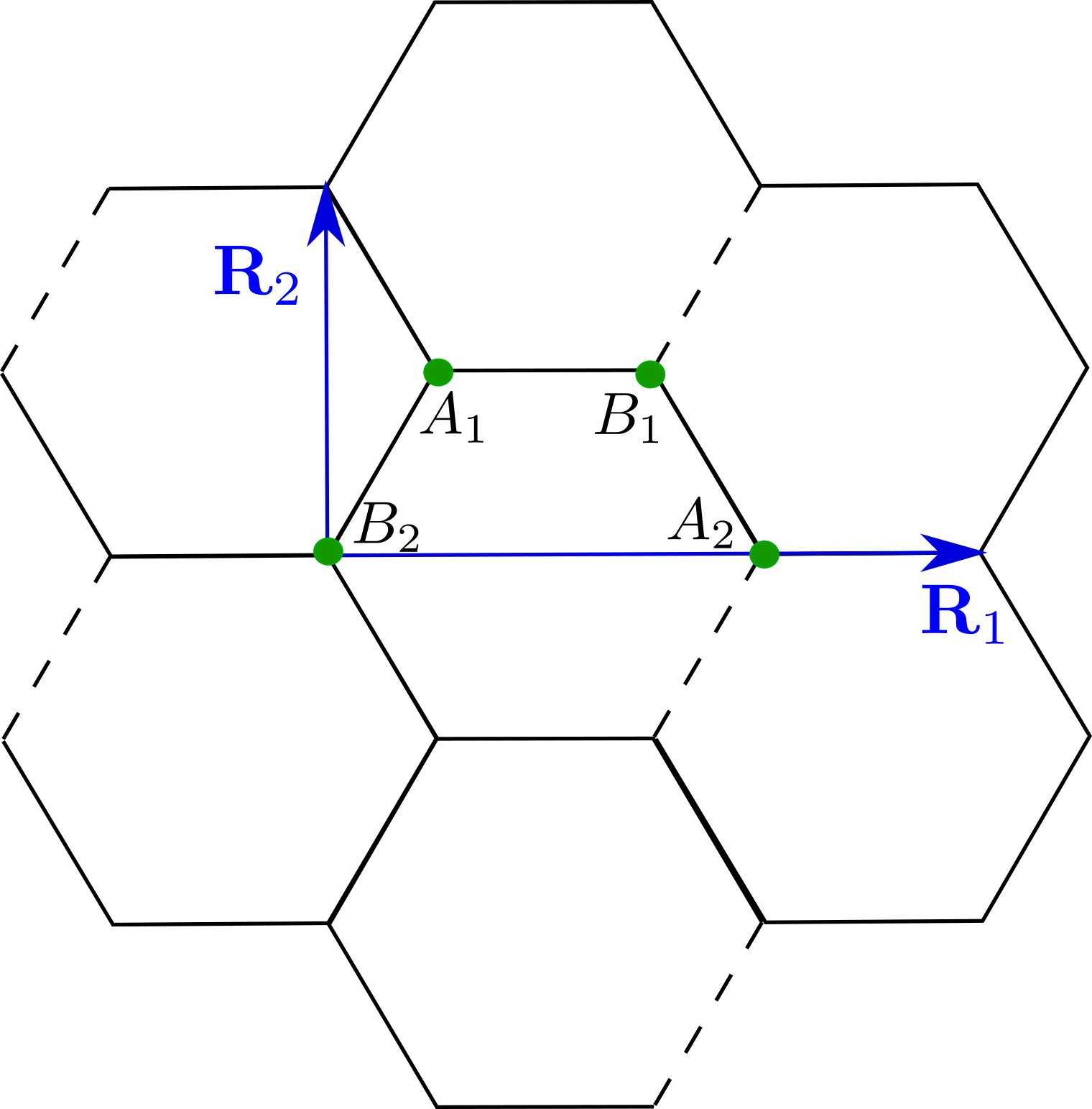}
	\caption{Honeycomb lattice with $\pi$-flux and the four-point magnetic unit-cell in the gauge choice (Eq.~\ref{eq_etaexp}): the dashed (continuous) bonds have $\eta({\mathbf{\boldscriptr_\mathcal{S},\boldscriptr_{\mathcal{S}'}^{\prime}}})=-1(+1)$ (see Eq. \ref{eq:pi flux statement in eta}).}
	\label{fig:unit cell}
\end{figure}

To obtain this manifestly SU(4) invariant basis, and also due to the $\pi$-flux, it is useful to consider a four site {\it magnetic unit cell} as shown in Fig. \ref{fig:unit cell}. The four sites, $A_1, A_2, B_1, B_2$, in the magnetic unit cell comprise two sites each of $A$ and $B$ sub-lattices of the underlying honeycomb net. The lattice translation vectors for this magnetic unit-cell, as shown in Fig. \ref{fig:unit cell}, are given in terms of the underlying honeycomb lattice primitive vectors as
\begin{align}
    {\bf R}_1={\bf b}_1+{\bf b}_2;~~~~~~{\bf R}_2={\bf b}_1-{\bf b}_2
\end{align}
such that the sites with reference to the enlarged unit cell are given by
\begin{align}\label{eq:definition of r_mu}
\boldscriptr_{\mathcal{S}} = n_x\mathbf{ R}_1 + n_y \mathbf{R}_2+ \mathbf{d}_{\mathcal{S}}\equiv\boldscriptr+{\bf d}_\mathcal{S}.
\end{align}
with $\mathcal{S} \in \{A_1,A_2,B_1,B_2\}$ denotes the four sites in the magnetic unit-cell, ${\bf d}_{\mathcal{S}}$ are the position vectors of the $\mcl{S}$-sublattice site w.r.t. the $B_2$ site (see Fig.~\ref{fig:unit cell}) of the same magnetic unit cell, $\boldscriptr$ labelled by integers $n_x$ and $n_y$. 

With this, we can now define new fermion annihilation operators given by 
\begin{align}
\phi(\boldscriptr_\mathcal{S})=[\phi_1(\boldscriptr_\mathcal{S}),\phi_2(\boldscriptr_\mathcal{S}),\phi_3(\boldscriptr_\mathcal{S}),\phi_4(\boldscriptr_\mathcal{S})]^T
\label{eq:localbasis}
\end{align}
as
\begin{eqnarray}\label{eq:definition of phi}
\phi(\boldscriptr_\mathcal{S}) = \mathcal{G}(\boldscriptr_\mathcal{S})^{\dagger} \psi(\boldscriptr_\mathcal{S})
\end{eqnarray}
where $\mathcal{G}(\boldscriptr_\mathcal{S})$ are $4\times 4$ unitary matrices whose explicit forms are given in Appendix \ref{sec:appen_g_matrices}. Any many-body operator can be expressed either in the $\phi$ basis or $\psi$. In this article, we use the terms ``local basis" and ``global basis" respectively to refer to these two ways.

The Hamiltonian (Eq. \ref{eq:hamiltonian in j=3/2 orbitals}) written in the local basis is
\begin{eqnarray}\label{eq:Hamiltonian in SU(4) diagonal form}
	&H &=  -\frac{t}{\sqrt{3}}\sum_{\braket{\boldscriptr_{\mathcal{S}},\boldscriptr^\prime_{\mathcal{S}'}}}\eta({\mathbf{\boldscriptr_\mathcal{S},\boldscriptr_{\mathcal{S}'}^{\prime}}})\phi^{\dagger}(\boldscriptr_\mathcal{S}) \phi(\boldscriptr^\prime_{\mathcal{S}'}) + \text{ h.c. } 
\end{eqnarray}
which is manifestly SU(4) invariant and $\eta({\mathbf{\boldscriptr_\mathcal{S},\boldscriptr_{\mathcal{S}'}^{\prime}}})=\pm 1$ implementing the $\pi$-flux constraint of Eq. \ref{eq:pi flux statement in U}, via
\begin{eqnarray}\label{eq:pi flux statement in eta}
\prod_{\braket{\mathbf{\boldscriptr_\mathcal{S},\boldscriptr_{\mathcal{S}'}^{\prime}}}\in \hexagon} \eta({\mathbf{\boldscriptr_\mathcal{S},\boldscriptr_{\mathcal{S}'}^{\prime}}}) = -1
\end{eqnarray}  

Fig. \ref{fig:unit cell} shows a choice for $\eta({\mathbf{\boldscriptr_\mathcal{S},\boldscriptr_{\mathcal{S}'}^{\prime}}})$ which is given by 
\begin{align}
    &\eta({\mathbf{\boldscriptr_\mathcal{S},\boldscriptr_{\mathcal{S}'}^{\prime}}})=
    \left\{\begin{array}{cl}
    -1 &   \text{ if }\mcl{S}=B_1, \mcl{S}'=A_2 \text{ and } \boldscriptr' = \boldscriptr+\mbf{R}_2\\
     +1 & \text{ (otherwise)}\\
     \end{array}\right.
    \label{eq_etaexp}
\end{align}

Obviously, there are many other choices for $\eta({\mathbf{\boldscriptr_\mathcal{S},\boldscriptr_{\mathcal{S}'}^{\prime}}})$ which are related to each other through gauge transformations which correspond to different signs of the $\mathcal{G}(\boldscriptr_\mathcal{S})$ matrices with respect to the ones introduced in Appendix \ref{sec:appen_g_matrices}. An alternate choice for $\eta({\mathbf{\boldscriptr_\mathcal{S},\boldscriptr_{\mathcal{S}'}^{\prime}}})$ and indeed the magnetic unit cell is shown in Fig. \ref{fig:piFluxUnitCell}. For the rest of our calculation in the main text, we choose $\eta({\mathbf{\boldscriptr_\mathcal{S},\boldscriptr_{\mathcal{S}'}^{\prime}}})$ as given by Eq. \ref{eq_etaexp}.

\subsubsection{The band structure}

\begin{figure*}
    \centering
        \subfigure[]{
    \includegraphics[scale=0.5]{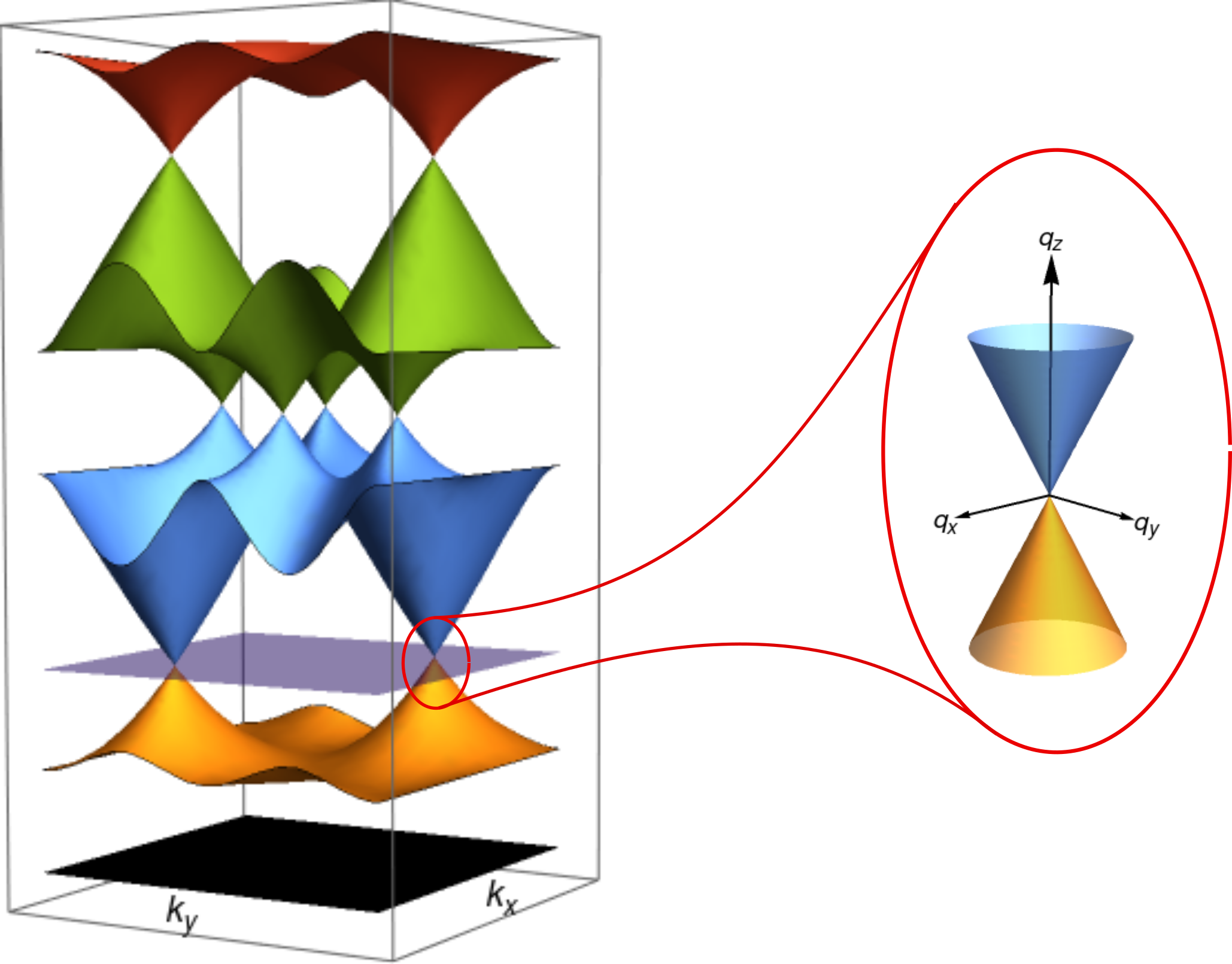}
    }\hfill    
    \subfigure[]{
    \includegraphics[scale=0.4]{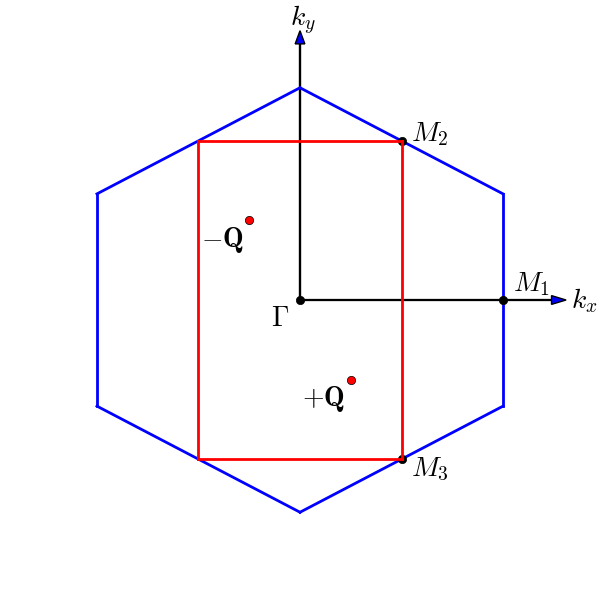}
    } 
    \caption{(a) Band structure for the Hamiltonian in Eq. \ref{eq:Hamiltonian in SU(4) diagonal form} for the magnetic unit cell consisting of four sites (Fig. \ref{fig:unit cell}). Each band is four-fold degenerate. For $d^1$ system, the lowest four bands are occupied with the chemical potential crossing the two Dirac points as shown. (b) Position of the two Dirac points (Eq.\ref{eq:position of dirc points}) in the chosen gauge (Fig. \ref{fig:unit cell} and Eq. \ref{eq_etaexp}) in the magnetic Brillouin zone (in red). The primitive Brillouin zone of the hexagonal lattice is also drawn in blue.}
    \label{fig:band structure}
\end{figure*}

Eq. \ref{eq:Hamiltonian in SU(4) diagonal form} represents four copies of nearest neighbour hopping model on honeycomb lattice in presence of $\pi$-flux. A single copy of such model at half filling was studied in Ref. \cite{ryu2009masses,PhysRevB.90.075140}. However, we shall find that the underlying SU(4) symmetry in the present case and the $1/4$th filling for $d^1$ materials (see below) along with SOC open up a new regime of possibilities for the resultant system at low energies.

To disentangle the role of the SU(4) and the $\pi$-flux, it is useful to consider a single flavour ``spinless" version of Eq. \ref{eq:Hamiltonian in SU(4) diagonal form} with $\phi(\boldscriptr_\mathcal{S})$ being a single component fermion.  This is worked out in Appendices \ref{appen_band} and \ref{appen:spinlessPiFluxHex}. The resultant band structure is shown in Fig. \ref{fig:band structure} and consists, for quarter filling, two linearly dispersing band-touching points--Dirac cones-- at 
\begin{align}\label{eq:position of dirc points}
  \pm\mathbf{ Q}=\pm \left[\frac{\pi}{6},-\frac{\pi}{2\sqrt{3}} \right].  
\end{align}
We label the two Dirac points (valleys) by the Ising variable $\tau=\pm$. Similar Dirac points also occur at 3/4th filling by particle-hole symmetry of the microscopic problem. Also note that there are four Dirac points at half filling~\cite{PhysRevB.90.075140} as is shown in Fig. \ref{fig:band structure}. In the rest of the discussion though we shall consider exclusively the vicinity of quarter filling and the nature of the low energy Dirac fermions at the two valleys at $\pm {\bf Q}$ given by Eq. \ref{eq:position of dirc points}.

Turning back to the case of $j=3/2$ orbitals in $d^1$ configuration (Eq. \ref{eq:Hamiltonian in SU(4) diagonal form}), the band structure is now four-fold degenerate due to the SU(4) symmetry such that the lower four bands are completely filled with the chemical potential again at the two Dirac cones given by Eq. \ref{eq:position of dirc points}. As remarked above, the similar Dirac cones are also present for three-quarter filling and hence the rest of our discussion is also applicable to materials with $d^3$ electronic configuration.

At this point, we would like to take a small detour by discussing the above band structure in the {\it global basis} (Eq. \ref{eq:4-component psi}) which provides interesting complementary insights into the results that follow in the rest of this paper. This alternate insight arises from the observation that while for a single flavour $\pi$-flux problem we are forced to use the magnetic unit cell (Fig. \ref{fig:unit cell}), for the four flavour version relevant to $d^1$ or $d^3$ systems, it is possible to use the two-site primitive honeycomb unit-cell (Fig. \ref{fig:2-point honeycomb}) by diagonalizing the Hamiltonian in Eq. \ref{eq:hamiltonian in j=3/2 orbitals}. However, in this {\it global basis} neither the SU(4), nor the $\pi$-flux is manifest but are mixed together non-trivially. As a result, while all the lattice symmetries (Table \ref{tab:lat_sym}), act in a linear fashion, i.~e., they are {\em non-projective} despite the fact that each hexagonal plaquette hosts a $\pi$-flux as shown in \eqnref{eq:pi flux statement in U}, to overcome the slightly more involved nature (arising explicit mixing of the lattice symmetries and the SU(4) in the global basis) we use the {\it local basis} for most of our discussion in the rest of the main-texts. Notwithstanding, the {\it global basis} is useful to understand certain structures in our calculations which we refer to at relevant places throughout the rest of the paper. The considerations in the global basis are presented in Appendix~\ref{appen:globalbasis}. Briefly, Bloch diagonalizing the Hamiltonian in global basis (Eq. \ref{eq:hamiltonian in j=3/2 orbitals}) obtains four bands arising from the four $j=3/2$ orbitals (Eq. \ref{eq:4-component psi}) and each two-fold degenerate due to inversion symmetry. The first set of bands touches the second set of bands at {\em four} distinct points with Dirac cone structure, see \figref{fig:globalBZBands}. With the quarter filling of the bands the chemical potential is tuned to the Dirac points at the four $\mathbf{Q}_g$ vectors, $\Gamma$, M$_1$, M$_2$, M$_3$ termed as valleys, in the original honeycomb Brillouin zone. This is to be contrasted with the local basis where one obtains two valleys due to the doubling of the unit cell, and the concomitant folding of the bands. One of the central insights of the global basis is that under lattice symmetries such as ${\bf S}_6$ (Table \ref{tab:lat_sym}), {\it only} three of the Dirac cones sitting at the three $M$ points (Fig. \ref{fig:globalBZBands}) mix amongst themselves while the Dirac cone at the BZ center, {\it i.e.} at $\Gamma$-point,  remains isolated. This naturally distinguishes the different valleys into two groups-- one containing only the $\Gamma$ point cone and the other containing the other three at the three nonequivalent $M$ points. As we shall see later, the above grouping is a fallout of the fact that the microscopic lattice symmetries get embedded in a larger low energy {\it IR space group} (see Section~\ref{sec:phases-and-phase-transitions}) that allows up to three-dimensional representations such that the above grouping is a block diagonalization of a reducible representation, {\it i.e.}, ${\bf 4=1\oplus 3}$. This insight will be important in understanding a subset of {\it partially gapless masses} discussed in section~\ref{subsec_semimetals}.

\subsection{Low-energy Dirac theory}

Turning back to the local basis (Eq. \ref{eq:localbasis} and Fig. \ref{fig:band structure}), for $1/4$th filling, the low energy theory is obtained by expanding the lattice fermions, $\phi({\boldscriptr_\mathcal{S}})$, in terms of the soft-modes, around the two Dirac points, $\pm {\bf Q}$, as 
\begin{align}\label{eq:expansion of phi in soft modes}
    \phi_f({\boldscriptr_\mathcal{S}})\sim\mathcal{W}^{(+)}_{\mathcal{S}\sigma}\chi_{f\sigma+}(\boldscriptr) e^{i\bf Q\cdot \boldscriptr}+\mathcal{W}^{(-)}_{\mathcal{S}\sigma}\chi_{f\sigma-}(\boldscriptr) e^{-i\bf Q\cdot \boldscriptr}
\end{align}
where $\chi_{f\sigma\tau}({\bf x})$ are the soft modes in the continuum evaluated at ${\bf x}=\boldscriptr$ with $f=1,\cdots,4$  denote the SU(4) flavour index, $\sigma=1,2$ is the particle or hole-like band index  and $\tau=\pm$ is  the valley index coming from the two Dirac nodes at $\pm {\bf Q}$. $\mathcal{W}^{(\pm)}_{\mathcal{S}\sigma}$ are two $4\times 2$ matrices (one at each valley, $\tau=\pm$) in the (magnetic) unit-cell $(\mathcal{S})$-particle-hole ($\sigma$) space. The details are given in  Appendix \ref{appen_band}. 

In terms of the soft modes, the low energy Hamiltonian takes the canonical Dirac form in two spatial dimensions and is given by 
	\begin{align}
	H_D = -iv_F \sum_{f=1}^4\int d^2{\bf x}~ \chi_f^{\dagger}(\mathbf{x}) (\alpha_1 \partial_1 +\alpha_2 \partial_2) \chi_f(\mathbf{x})
	\label{eq_freediracham}
	\end{align}
where $v_F = \frac{t\mcl{l}}{\sqrt{2}}$ is the fermi velocity, $\mathcal{l}$ is the length of each side of the hexagon and $\partial_i=\partial/\partial x_i$ ($i=1,2$),  with
\begin{eqnarray}
    \chi_f(\mathbf{x}) &&= \left( \chi_{f1+}, \chi_{f2+}, \chi_{f1-}, \chi_{f2-} \right)^T
    \label{eq_16spinor}
\end{eqnarray}
a 4-component spinor, one for each SU(4) flavour $f=1,2,3,4$, which can be further stacked up to form a $16$-component spinor, and 
	\begin{eqnarray}
	\alpha_1 = \tau_3 \sigma_1,~~~~~~~~~~~~~~\alpha_2 = \tau_0 \sigma_2
	\label{eq_diracmatrices}
	\end{eqnarray}
are the two Dirac Matrices. Here $\tau_{\mu}$ and $\sigma_{\mu}$ ($\mu=0,1,2,3$) are Pauli matrices that act in the valley space and band/particle-hole space respectively. 
 
The corresponding Euclidean action is given by
	\begin{align}
	{S}_0=\int d^2{\bf x}d\tau~\mathcal{L}_0
	\label{eq_dirac_action}
	\end{align}
	where, $\mathcal{L}_0$ is given by Eq. \ref{eq_dirac_l} with $N_F=4$ and  $\bar{\chi}_f=i\chi^\dagger_f\gamma_0$ and
	\begin{align}
	\gamma_0=\tau_3\sigma_3,~~~~~~~\gamma_1=\tau_0\sigma_2~~~~~~~\gamma_2=-\tau_3\sigma_1
	\label{eq_gamma}
	\end{align}
	such that $\alpha_1=i\gamma_0\gamma_1$ and $\alpha_1=i\gamma_0\gamma_2$. Here $\gamma_0,\gamma_1$ and $\gamma_2$ generate the Euclidean Clifford Algebra that satisfy $\left\{\gamma_{\mu},\gamma_{\nu} \right\} = 2\delta_{\mu \nu}$ with $\mu,\nu=0,1,2$~\cite{kaplan2009chiral}.

The above low-energy free Dirac theory has a much larger symmetry compared to the microscopic system. Firstly, Eq.~\ref{eq_dirac_action} is invariant under SU(2) transformations on each flavour of $\chi_f$ generated by
\begin{align}
\{\tau_3\sigma_0/2,\tau_1\sigma_2/2,\tau_2\sigma_2/2\}\equiv \{\zeta_1,\zeta_2,\zeta_3\}/2.
\label{eq_zetasu2}
\end{align}

 This denotes rotation in the valley and band space similar to Dirac fermions in graphene~\cite{PhysRevB.79.085116} which we refer to as chiral symmetry~\cite{ryu2009masses}. This, along with the manifest invariance under the SU(4) {\it flavour} symmetry generated by $\Sigma_i$ (defined in Appendix.~\ref{appen_gamma}), nominally gives rise to an internal symmetry of SU(4) $\otimes$ SU(2). However, the emergent internal symmetry is SU(8) which is generated by 63 traceless Hermitian matrices, $\mathcal{P}_b$, that are obtained as  
\begin{eqnarray}\label{eq_su8_generators}
    \mcl{P}_b = \Sigma_i\zeta_j
\end{eqnarray}
where on the LHS, $b=1,2,\cdots, 63$ which are made up of the fifteen SU(4) generators, $\Sigma_i$, given in Appendix~\ref{appen_gamma} and three SU(2) generators, $\zeta_j$, defined in Eq.~\ref{eq_zetasu2} along with the identities in the two spaces $\Sigma_0$ and $\zeta_0$ respectively.

  Under the SU(8) generated by the $16\times 16$ traceless Hermitian matrices, $\mathcal{P}_b$, the spinors $\chi$ transform as
\begin{align}\label{eq_transformation under SU8}
    \chi\rightarrow \exp(i\xi_b \mathcal{P}_b)\chi
\end{align}
where $\chi=(\chi_1^T,\chi_2^T,\chi_3^T,\chi_4^T)^T$ is the 16-component spinor with each $\chi_f~(f=1,2,3,4)$ given by Eq. \ref{eq_16spinor}. This leads to the conservation of the SU(8) flavour current
\begin{align}
    J_{\mu,b}=-i\bar{\chi}\gamma_\mu \mathcal{P}_b\chi
    \label{eq_flavourcurrent}
\end{align}
{\it i.e.} $\partial_\mu J_{\mu,b}=0$ $\forall~b=1,\cdots,63$. This is to be contrasted with SU(8) Dirac fermions realised in the $\pi$-flux phase on a square lattice~\cite{PhysRevB.72.104404} for fermionic spinons in a class of quantum spin-liquids, where the resultant implementation of the symmetries on the low energy fermions are very different. Crucially, in the present case, the non-trivial SOC of the underlying orbitals results in the mixing of the spin and the real spaces, under various lattice symmetries and time reversal which leads to important observable consequences which are reflected in the nature of the phases proximate to the semimetal, as we show below. In addition, in the square lattice spin liquid problem, the spinons couple to an emergent dynamic SU(2) gauge field which is absent in the present case. 

In addition to the above internal SU(8), the free Dirac action of Eq. \ref{eq_dirac_action} has a usual set of emergent space-time symmetries that include :

\paragraph{Emergent Lorentz symmetry}  generated by  the three matrices
    \begin{align}
    \gamma_{\mu\nu}=-\frac{i}{4}[\gamma_\mu,\gamma_\nu]
\end{align}
along with simultaneous rotations of the Euclidean space-time coordinates. Under Lorentz transformation, the spinors transform as $\chi\rightarrow \exp(i\Omega_{\mu\nu}\gamma_{\mu\nu})\chi$. Note that $\gamma_\mu=\epsilon^{\mu\nu\lambda}\gamma_{\nu\lambda}$ where $\mu,\nu,\lambda=0,1,2$.

\paragraph{Continuous spatial translation symmetry} under which the soft modes at the two valleys transform as
\begin{align}
    {\bf T}_{\bf x_0}^{\rm cont} :\left\{\begin{array}{c}
    \chi_+({\bf x})\rightarrow\chi'_+({\bf x})=e^{i{\bf Q\cdot x_0}}~\chi_+({\bf x-x_0})\\
    \chi_-({\bf x})\rightarrow\chi'_-({\bf x})=e^{-i{\bf Q\cdot x_0}}~\chi_-({\bf x-x_0})\\
    \end{array}\right.
    \label{eq_cont_trans}
\end{align}
where 
\begin{align}
    \chi_\pm({\bf x})=\frac{1}{2}(1\pm \zeta_1)\chi({\bf x})
    \label{eq_chipm}
\end{align}
are the two spinors associated with the two valleys respectively located at ${\bf \pm Q}$. Using Eq. \ref{eq_chipm}, we can re-write the free Dirac action (Eq. \ref{eq_dirac_action}) as ${S}_0={S}_0^++{S}_0^-$ where ${S}_0^\pm=v_F\int d^2{\bf x}d\tau~\bar{\chi}_\pm({\bf x})(-i\slashed{\partial})\chi_\pm({\bf x})$ are the actions at the two valleys.

\paragraph{Emergent CPT symmetries :}  The free Dirac action ${S}_0$ is also invariant under emergent charge conjugation($\mcl{C}$), emergent parity($\mcl{P}$) and emergent time reversal($\mcl{T}'$) symmetries. These symmetries act on the spinors in the following way: 
\begin{subequations}
\begin{align}
    \mcl{C}:& \quad \chi(\mbf{x},t) \to -i\gamma_2\gamma_0\bar{\chi}^T(\mbf{x},t),\\
    \mcl{P}:& \quad \chi(x_1,x_2,t) \to -i\gamma_1\chi(-x_1,x_2,t),\\    
    \mcl{T}':& \quad \chi(\mbf{x},t) \to -i\gamma_2K\chi(\mbf{x},-t),
    \label{eq_ftcpt}
\end{align}
\end{subequations}
with $K$ being the complex conjugation operator. Here we denote the emergent time-reversal by $\mcl{T}'$ to distinguish it from the microscopic time reversal operation defined in Eq. \ref{eq: microscopic time reversal} (which we denote with $\mbb{T}$ in Eq. \ref{eq:microtr}). 

\subsection{Microscopic symmetries in the low energy theory}\label{sec:action_of_microscopic_symmetries_on_chi}

The enhanced IR symmetries provide important insights into the low-energy physics  including the properties of the Dirac semimetal and associated quantum phase transitions into proximate symmetry broken phases. The latter is determined by the underlying UV/microscopic symmetries. These UV symmetries are embedded as a subgroup of the emergent (larger) IR symmetry group and are implemented as a combination of the IR symmetry transformations (see, for example, the discussion below Eq. \ref{eq: microscopic time reversal} for the time-reversal symmetry). This is particularly interesting in the present case where the underlying SOC mixes the lattice and the $j=3/2$ flavour  space such that the embedding of the microscopic symmetries in the IR symmetry group can be rather intricate.   It is, therefore, useful to list the symmetry transformation of the low energy Dirac fermions, $\chi$, under various microscopic symmetries discussed above.

The total electronic charge is conserved in the microscopic system. This U(1) electronic charge conservation leads to the conservation of a current
\begin{align}
    J^{charge}_\mu=-i\bar\chi\gamma_\mu\chi
    \label{eq_u1charge}
\end{align}
in the low energy Dirac theory, {\it i.e.}, $\partial_\mu J_\mu^{charge}=0$. 

On the other hand, the transformation of the low energy Dirac fermions, $\chi$, under the discrete lattice symmetries (Table \ref{tab:lat_sym}) as well as microscopic time reversal (Eq. \ref{eq:microtr}) have the generic form (see Appendix \ref{sec:Symmetry_transformation_calculations} for details)
\begin{align}
\chi({\bf x})\xrightarrow{\mathbb{S}}\chi'({\bf x}')&= \left( \Omega^f_{\mathbb{S}} \otimes~\Omega^c_{\mathbb{S}}  \right)\chi(\mathbb{S}^{-1}\mbf{x})
\label{eq_pointgr}
\end{align}
where $\mathbb{S}(={\bf T_1, T_2, C_3, S_6, C_2'},\boldsymbol{\sigma_d}, {\bf I})$ stands for the generators of the lattice symmetries listed in Table \ref{tab:lat_sym} and $\Omega_{\mathbb{S}}^f, \Omega_{\mathbb{S}}^c$  both are $4\times 4$ unitary matrices that act on the SU(4) flavor space and the chiral space respectively. The explicit form of these matrices is given in Appendix~\ref{appen_softmodetrans}. A central aspect of Eq. \ref{eq_pointgr} is the fact that because of underlying SOC, both $\Omega^f_\mathbb{S}$ and $\Omega^c_\mathbb{S}$ are non-trivial matrices for all the lattice symmetries.

Finally, under the microscopic time-reversal symmetry (Eq. \ref{eq:microtr}), we have
\begin{align}\label{eq: microscopic time reversal}
    \mathbb{T}~:~\chi({\bf x},t)\rightarrow\chi'({\bf x},t)&=i\gamma^1~\Sigma_{13}\zeta_2~K~\chi({\bf x},-t)
\end{align}
such that $\mathbb{T}^2=-\mathbb{I}_{16}$ and thereby accounting for the Kramers' degeneracy for the $j=3/2$ orbitals. Notably, this transformation is proportional to a simultaneous emergent time reversal, $\mathcal{T}'$ (Eq. \ref{eq_ftcpt}) combined with a SU(8) rotation by $\Sigma_{13}\zeta_2$ and a Lorentz boost.

\section{Short range Interactions}

Having described the free low energy theory for the electrons and its enhanced IR symmetries, we now turn to the effect of interactions on them. More precisely we consider the effect of short-range four-fermion interactions. A generic form of such interaction Hamiltonian obtained from an underlying multi-orbital Hubbard-type model for the lattice fermions is given by
\begin{align}
    H_{\rm int}=&\int d^2{\bf x}d^2{\bf x'}~V_{ijkl}({\bf x-x'}) \chi_i({\bf x})^\dagger\chi_j({\bf x'})^\dagger\chi_k({\bf x})\chi_l({\bf x'})\nonumber\\
    &~~~~~~~~~~~~~~~~~~~~~~~~~~~~~~~~~~~~~~~~~~~~~+\cdots
    \label{eq_interactions}
\end{align}
where $V_{ijkl}$ denotes potential ($i,j,k,l$ collectively spans over the different indices) and $\cdots$ corresponds to more irrelevant higher fermion interactions. We assume that the interactions are short-ranged in the sense $V_{ijkl}$ is only appreciable for ${\bf x}$ and ${\bf x'}$ being proximate with a suitable UV regulation. Further, we assume that the form of $V$ is constrained enough that at low energy it has the full SU(8) symmetry. Long-ranged Coulomb interactions as well as on-site Hubbard repulsion, for example, have such SU(8), symmetry. This immediately means that even in the presence of these short-range interactions, all the flavour currents $J_{\mu,a}$ (Eq. \ref{eq_flavourcurrent}), in addition to the electronic current $J^{charge}_\mu$ (Eq. \ref{eq_u1charge}) remain conserved unless spontaneously broken.

While short-range quartic interactions are perturbatively irrelevant at the free Dirac fixed point~\cite{PhysRevLett.97.146401,vafek2014dirac}, on cranking them up they lead to phase transitions possibly gapping out the fermions and associated spontaneous breaking of the symmetries of the free Dirac theory. In the rest of this paper, we provide an understanding of the phases that are obtained by condensing various  fermion bilinears which do not carry a net electronic charge, {\it i.e.} {invariant under the U(1) described in Eq. \ref{eq_u1charge} and therefore have the form}
\begin{align}
    &\Delta^{a}=-i\langle\bar{\chi} \mathcal{M}_a\chi\rangle\neq 0
    \label{eq_massexp}
\end{align}
where $\mathcal{M}_a$ are $16\times 16$ {\it mass matrices} such that $\gamma_0\mathcal{M}_a$ anti-commutes with both the Dirac matrices-- $\alpha_1$ and $\alpha_2$ -- given by Eq. \ref{eq_diracmatrices}. . {This leaves out another important class of fermion bilinears symbolically of the form $\langle \chi~\chi\rangle$ that describes different types of superconductors which will be taken up elsewhere~\cite{ankushsc}.}

For such a U(1) invariant massive phase, the mean-field Hamiltonian is given by
\begin{align}
&{S}_{MF}={S}_0+{S}_m=-i\int d^2{\bf r}d\tau~\bar{\chi}({\bf r})\left[v_F \slashed{\partial}-\Delta^{a}\mathcal{M}_{a}\right]\chi({\bf r}).
\label{eq_dirac_massiveaction}
\end{align}

There are 64 such linearly independent $\mathcal{M}_a$ matrices that can be broadly classified into two classes according to their transformation properties under SU(8). The first class contains a single SU(8) singlet given by Eq. \ref{eq_su8scalar} while the second class corresponds to 63 SU(8) adjoint multiplet 
\begin{align}
    -i\langle \bar{\chi}\mathcal{P}_a\chi\rangle,~~~~~a=1\cdots,63.
    \label{eq_su8adjoint}
\end{align}
and $\mathcal{P}_a$ being the SU(8) generators (Eq. \ref{eq_su8_generators}).

The microscopic operators that characterise the same broken symmetry -- hence can serve as valid order-parameters for appropriate symmetry broken phases-- have the same transformation properties as the bilinear and hence are proportional to each other. In principle, the microscopic operators can also get contributions from the conserved currents of the same symmetry, but usually, such currents decay faster than the bilinears, and hence at long distances, the correlation function of the microscopic fields is determined by that of the field theory bilinear~\cite{PhysRevB.72.104404}. 

The correspondence between the microscopic operators and the low-energy fermion bilinears is obtained by comparing their symmetry transformations. In particular, the transformation of the low-energy fermion bilinears under microscopic symmetries can be used to systematically uncover the nature of the phases proximate to the Dirac semimetal. Bilinears that are related by microscopic symmetries together constitute a single phase. This leads to the classification of the fermion masses in terms of broken microscopic symmetries and/or anomalies. In the present case, we find that the 64 masses group together to give rise to 24 phases which we now turn to understand in detail.

 
\section{Classification of the fermionic bilinears : phases and transitions}\label{sec:phases-and-phase-transitions}

This task of classifying the bilinears according to their microscopic symmetries (and hence identifying the phases) is much more involved compared to the same problem in graphene~\cite{ryu2009masses} since the  SOC mixes the $j=3/2$ flavour and the real spaces in a non-trivial way. As a result, the lattice translations, ${\bf T_1}, {\bf T_2}$ (Table \ref{tab:lat_sym}) do not necessarily commute with the point group symmetries such as ${\bf C_3}, {\bf S_6}, {\bf C'_2}, \boldsymbol{\sigma_d},{\bf I}$ (Table \ref{tab:lat_sym}) and microscopic time reversal, $\mathbb{T}$, (Eq. \ref{eq:microtr}). This is clear by looking at the transformations of the Dirac spinor, $\chi$, under the above lattice symmetries (Eq. \ref{eq_pointgr} with the detailed forms given by Eqs. \ref{eq:chitr} - \ref{eq:chisigd}). Hence we need to analyse the action of the entire set of transformations generated by the space group and microscopic time reversal on $\chi$ to understand the transformation of the fermion bilinears in Eq. \ref{eq_massexp}. The resultant symmetry group, we dub as {\it IR space group}.

\paragraph*{IR Space group :} 
To understand the structure of this {\it IR space group}, we note that the $j=3/2$ electron states transform under a double group representation of $D_{3d}$, which has 24 elements. Since the translations do not commute with point group transformations, corresponding to any element (say, $\mathbb{S}$) of the double group of $D_{3d}$, there are four elements in the {\it IR space group} which can be constructed as (say) $\mathbb{S}, \mbf{T_1}\mathbb{S},\mathbb{S}\mbf{T_2}$ and $\mbf{T_1}\mathbb{S}\mbf{T_2}$ by composing it with translations, ${\bf T_1}$ and $\mbf{T_2}$ (Table \ref{tab:lat_sym}).  So, the group of microscopic symmetries that act on the spinors has 96 elements in total. These elements can be divided into 20 conjugacy classes and hence there are 20 different irreducible representations of the IR space group. Among these 20, only 10 has $+$ve character for $2\pi$ rotation. Since the fermion bilinears are always invariant under a $2\pi$ rotation, we consider only these irreps for the classification of the masses. 

Among these 10 irreps of the IR space group, four are 1-dimensional ($\mcl{A}_{1g}, \mcl{A}_{2g}, \mcl{A}_{1u}, \mcl{A}_{2u}$), two are 2-dimensional ($\mcl{E}_g,\mcl{E}_u$) and four irreps are 3-dimensional ($\mcl{T}_{1g},\mcl{T}_{2g},\mcl{T}_{1u}, \mcl{T}_{2u}$). Following conventional notation, the subscripts $1(2)$ and $g(u)$ denote that the irrep is  even (odd) under rotation, $\mbf{C_2'}$ and inversion, ${\bf I}$, respectively (Table. \ref{tab:lat_sym}). Further, to incorporate microscopic time reversal, $\mathbb{T}$ (Eq. \ref{eq:microtr}), we will add a superscript $e(o)$ (e.g., $\mcl{A}_{1u}^{e(o)}$) to denote the particular irrep is even (odd) under time-reversal. The details of these irreps are given in Appendix~\ref{appen_definition_of_space_group_representaiton}.

The central question we now turn to investigate in the rest of the paper are the nature of the phases obtained by condensing the fermion bilinears $\langle\bar{\chi}\mcl{P}_a\chi\rangle$. Since this is decided by the microscopic symmetries, we decompose the above bilinears in terms of the irreducible representation of the microscopic symmetries~\cite{PhysRevB.72.104404}. We find that these 64 bilinears break up into 27 irreps of the space group among which there are six 1-dimensional representations, five 2-dimensional representations and sixteen 3-dimensional representations.  This can be obtained as follows.

Starting with Eqs. \ref{eq_pointgr} and \ref{eq: microscopic time reversal}, we can derive the action of the microscopic symmetries on the members of the 64 fermion bilinears, which leads to the following structure. Under the action of a lattice symmetry transformation (say $\mathbb{S}$, corresponding to Table \ref{tab:lat_sym} and Eq.~\ref{eq_pointgr}), a fermion bilinear of the form $\bar{\chi}\mcl{P}_a\chi$ (with $\mcl{P}_a$ given by Eq.~\ref{eq_su8_generators}) transforms as
\begin{align}\label{eq_transformation of mass under lattice symmetries}
-i\bar{\chi}\mcl{P}_a\chi = \chi^{\dagger}\gamma_0 \Sigma_i \zeta_j\chi \xrightarrow{\mathbb{S}} \chi^{\dagger} (\Omega_\mathbb{S}^{f\dagger}\Sigma_i\Omega_\mathbb{S}^{f})\otimes(\Omega_\mathbb{S}^{c \dagger} \gamma_0\zeta_j \Omega_\mathbb{S}^{c})\chi
\end{align}  
and under the action of microscopic time reversal (Eqs. \ref{eq:microtr} and \ref{eq: microscopic time reversal}), we get
\begin{align}\label{eq_transformation of mass under TR}
	\chi^{\dagger}\gamma_0 \Sigma_i \zeta_j\chi\xrightarrow{\mbb{T}} \chi^{\dagger} (\Sigma_{13}\Sigma_i^*\Sigma_{13})\otimes(\tau_1\sigma_0 ~\gamma_0^*\zeta_j^* ~\tau_1\sigma_0)\chi
\end{align}  

The above structure shows that the action of each symmetry is implemented as a product of the transformations in the flavour and chiral spaces, {\it i.e.}, for lattice symmetries,
	\begin{eqnarray}\label{eq_Sigma_under_lattice_symmetries}
	\Sigma_i &\xrightarrow{\mbb{S}} \Omega_\mbb{S}^{f\dagger}\Sigma_i\Omega_\mbb{S}^{f};\quad
	\quad\gamma_0\zeta_j &\xrightarrow{S} \Omega_S^{c\dagger}\gamma_0\zeta_j\Omega_S^{c}\label{eq:transformation_of_zeta}
	\end{eqnarray} 
 and for time reversal
 \begin{eqnarray}\label{eq_transformation of sigma under TR}
    \Sigma_i &\xrightarrow{\mbb{T}} \Sigma_{13}\Sigma_i^*\Sigma_{13};\quad\quad\gamma_0\zeta_j &\xrightarrow{\mbb{T}} \tau_1\sigma_0 ~\gamma_0^*\zeta_j^* ~\tau_1\sigma_0.\label{eq_transformation of zeta under TR}
\end{eqnarray}
However, due to the SOC, the real space transformations get non-trivially coupled to the flavour space and this resultant inter-locking is reflected in the form of the transformation matrices, particularly in Eq. \ref{eq_Sigma_under_lattice_symmetries} where the $\Omega_S^f$ reflects the degree of interlocking between real and flavour spaces. In fact, it is this non-trivial structure that distinguishes the spin-orbit coupled Dirac fermions-- the topic of the present work-- with multi-flavour (larger $N_F$) generalisation of graphene where such SOC is usually neglected (except for the spin-Hall effect~\cite{PhysRevLett.95.226801} and related phases where SOC is essential).

	 \begin{table}
	     \centering
	 	\begin{tabular}{c| c| c| c}
	 		&Irrep& $\Sigma_p$ & $\mbb{T}$ \\\hline
	 		Singlets&$\mcl{A}_{1g}^e$& $\Sigma_0=\mbf{1}_4$&    	even\\
	 		&$\mcl{A}_{2g}^o$& $\Sigma_{45}$&	odd\\ \hline 
	 		Doublet&$\mcl{E}_u^e$ & $\{\Sigma_4,\Sigma_5\}$ & even\\ \hline			
	 		Triplets&$\mcl{T}_{1g}^e$& $\{\Sigma_3,\Sigma_1,-\Sigma_2 \}$& even \\
	 		&$\mcl{T}_{2g}^o$	& $\{\Sigma_{12},\Sigma_{23},\Sigma_{13} \}$& odd\\
	 		&$\mcl{T}_{1u}^o$& $ \{\Sigma_{35}, \frac{\sqrt{3}\Sigma_{14}}{2}   - \frac{\Sigma_{15}}{2}, \frac{\sqrt{3}\Sigma_{24}}{2}+\frac{\Sigma_{25}}{2}  \}$ & odd \\ 			
	 		&$\mcl{T}_{2u}^o$& $\{ \Sigma_{34},-\frac{\Sigma_{14}}{2}   -  \frac{\sqrt{3}\Sigma_{15}}{2}, \frac{\Sigma_{24}}{2}   - \frac{\sqrt{3}\Sigma_{25}}{2} \}$& odd\\\hline
	 	\end{tabular}	
		     \caption{Irreps of the matrices SU(4) flavour space.}
	     \label{tab:SU4_space_irreps}
	 \end{table}

 Due to the direct product structure of the above transformations, we can analyze the action of the microscopic symmetries on the flavor and the chiral spaces separately and then use Clebsch-Gordon decomposition, e.g., if the matrix $\Sigma_i$ transforms in some irreducible representation (say, $\mcl{D}_1$) and $\gamma_0\zeta_j$ transforms in some other irrep (say, $\mcl{D}_2$), then the bilinear $\bar{\chi}\Sigma_i\zeta_j\chi$ transforms in the product representation $\mcl{D}_1\otimes \mcl{D}_2$. This direct product representation is reducible in general which then is reduced into a direct sum representation.

Table \ref{tab:SU4_space_irreps} shows the $\Sigma_i$ matrices in the SU(4) flavour space and their irreducible representations under the IR space-group transformations as well as TR. The transformations of the $\gamma_0\zeta_j$ (Eq.~\ref{eq_zetasu2}) matrices in the SU(2) chiral space are written in Table~\ref{tab:real_space_irreps}. Details of their symmetry transformations are given in Appendix \ref{appen_definition_of_space_group_representaiton}.


	 \begin{table}
	     \centering
	 	\begin{tabular}{c| c| c| c}
	 		&Irrep& $\gamma_0\zeta_j$ &$\mbb{T}$\\\hline
	 		Singlet&$\mcl{A}_{2g}^o$& $\gamma_0$ & odd\\
	 		Triplet&$\mcl{T}_{1g}^e$& $\{ \gamma_0\zeta_1, \gamma_0\zeta_2, \gamma_0\zeta_3  \}$& even \\
	 	\end{tabular}	
		     \caption{Irreps for the matrices in the SU(2) chiral space.}
	     \label{tab:real_space_irreps}
	 \end{table}


The 64 fermion bilinears are subdivided into three groups depending on the participation of the flavour, $\Sigma_i$ and the chiral elements, $\zeta_j$, in the fermion bilinear (Eq. \ref{eq_massexp}) which, in turn, dictate their transformation properties under the microscopic symmetries. These are -- (1)~Group-1~: the chiral masses composed of flavour singlets, (2)~Group-2~: the flavour masses composed of chiral singlets, and, (3)~Group-3~: the mixed masses which are composed of non-trivial combinations of both the flavour and chiral sectors. Here we list  the masses in the groups mentioned above. The following sections contain a detailed discussion of their physics.

\subsection{Group-1 : The chiral masses} There are four masses of the form $-i\langle\bar\chi\chi\rangle$ and $-i\langle\bar\chi\zeta_i\chi\rangle$ ($i=1,2,3)$ that are invariant under the SU(4) flavour symmetry and charge conservation which are  broken down by the lattice symmetries and TR as ${\bf 4}={\bf 1}\oplus {\bf 3}$, {\it i.e.},
	\begin{subequations}
	\begin{eqnarray}
	&&\left[\mcl{A}_{1g}^e\right]^\Sigma \otimes \left[\mcl{A}_{2g}^o\right]^\zeta = \mcl{A}_{2g}^o \label{eq_chiraldecomp_1}\\
	   &&\left[\mcl{A}_{1g}^e\right]^\Sigma \otimes \left[\mcl{T}_{1g}^e\right]^\zeta = \mcl{T}_{1g}^e
	    \label{eq_chiraldecomp_2}
	\end{eqnarray}
	\end{subequations}
where $\left[\cdots\right]^\Sigma$ and $\left[\cdots\right]^\zeta$ denote the two irreducible representations taken from Tables \ref{tab:SU4_space_irreps} and \ref{tab:real_space_irreps} respectively.

As mentioned in the overview (Table \ref{tab_chiral_masses}), the singlet represents an Integer Chern insulator (ICI) phase, the triplet corresponds to the three stripy charge density waves (CDW) (Fig.~\ref{fig:stripy_CDW}).  Since the flavour index plays no role, we can quantitatively compare the spinless version of the present problem (Appendix \ref{appen:spinlessPiFluxHex}) with spinless electrons in graphene \cite{ryu2009masses,PhysRevLett.98.186809}. In the case of graphene, the irreducible representation splits up into ${\bf 4}={\bf 1}\oplus {\bf 1}\oplus {\bf 2}$ where the two singlets represent the ICI phase\cite{PhysRevLett.61.2015,PhysRevB.74.085308} and staggered (N\'eel) CDW, and the doublet corresponds to the two Kekule patterns \cite{ryu2009masses,PhysRevLett.98.186809}. This is very different from the present case and this provides a startling example where the microscopic SOC changes the low energy symmetry implementation. We discuss these masses in more detail in Sec.~\ref{subsec_group1}.

\subsection{Group-2 : The flavour masses} There are 15 masses of the form $i\langle\bar\chi\Sigma_j\chi\rangle$ where $\Sigma_j$ are the 15 generators of SU(4) as given by Eq. \ref{eq:Sigma_ij_matrices} in Appendix \ref{appen_gamma}. Under microscopic symmetries, they break up into six different irreps, {\it i.e.,} ${\bf 15=1\oplus3\oplus3\oplus3\oplus2\oplus3}$ given by Eqs. \ref{eq:decomposition_21}, \ref{eq:decomposition_22} and \ref{eq:decomposition_3}. As pointed out in the overview section (Table \ref{tab_flavor_masses}) these correspond to six generalised spin-Hall phases that are summarised in Table \ref{tab_flavor_masses} while the details are given in Sec. \ref{subsec_group2}.  

In order to explore the nature of the resultant phases, it is useful to understand in detail the mathematical structure of the implementation of the various microscopic symmetries that break up the 15 flavour masses further into different irreducible representations. Starting with TR, ten of the flavour masses are TR even and are of the form
\begin{align}\label{eq_so5_generator_index}
    -i\langle \bar\chi\Sigma_{j}\chi\rangle,~~~~~j=12, 13, 14, 15, 23, 24, 25, 34, 35~{\rm and}~45.
\end{align}
These transform into each other under an adjoint representation of an SO(5) sub-group (generated by themselves, Eq. \ref{eq_so5_generator_index}) of the SU(4) flavour group. 

The other five are TR odd and transform under a vector representation of the same SO(5) and are given by
\begin{align}
-i\langle \bar\chi\Sigma_{j}\chi\rangle~~\text{with}~~    j=1,2,3,4~{\rm and}~5
    \label{eq_so5-vector}
\end{align}

Next, the lattice inversion (Table \ref{tab:lat_sym}), $\mathbf{I}$, breaks each of the above two sets further. Out of the 10 adjoint ones (Eq. \ref{eq_so5_generator_index}), four 
\begin{align}
   -i\langle\bar\chi\Sigma_{j}\chi\rangle~~{\rm with}~~j=45, 12, 13~{\rm and}~23  
   \label{eq_so5-4}
\end{align}
are even under inversion, $\mathbf{I}$, while the other six
\begin{align}
   -i\langle\bar\chi\Sigma_{j}\chi\rangle~~{\rm with}~~j=14, 15, 24, 25, 34~{\rm and}~35 
   \label{eq_so5-6}
\end{align}
are odd under it. For the 5 vector masses (Eq. \ref{eq_so5-vector}), two
\begin{align}
   -i\langle\bar\chi\Sigma_{j}\chi\rangle~~{\rm with}~~j=4~{\rm and}~5
   \label{eq_so5-2}
\end{align}
are inversion odd, while three 
\begin{align}
   -i\langle\bar\chi\Sigma_{j}\chi\rangle~~{\rm with}~~j=1, 2~{\rm and}~3
   \label{eq_so5-3}
\end{align}
are even.

Each of the above four subsets (Eqs. \ref{eq_so5-4}-\ref{eq_so5-3}) is individually closed under a U(1) $\otimes$ SU(2) sub-group of the SO(5) (eq. \ref{eq_so5_generator_index}) that is generated respectively by 
\begin{align}
\Sigma_{45}
\label{eq_u145}
\end{align}
and 
\begin{align}
    \{ \Sigma_{12}, \Sigma_{13},\Sigma_{23}\}.
    \label{eq_so5-su2}
\end{align}

In particular, in the first subset (Eq.~\ref{eq_so5-4}), the first mass is a U(1) $\otimes$ SU(2) singlet while the rest are only U(1) singlets that transform as spin-1 under the SU(2). The three masses in Eq.~\ref{eq_so5-3} are U(1) singlets and an SU(2) triplet, while the two masses in Eq.~\ref{eq_so5-2} are SU(2) singlets and transform into each other under the U(1). Finally the six masses in Eq.~\ref{eq_so5-6}, decompose into two SU(2) triplets~:
\begin{align}
    &\{-i\langle\bar\chi\Sigma_{14}\chi\rangle, -i\langle\bar\chi\Sigma_{24}\chi\rangle,-i\langle\bar\chi\Sigma_{34}\chi\rangle\}
    \label{eq_so5-6-31}   
\end{align}
and
\begin{align}
    \{-i\langle\bar\chi\Sigma_{15}\chi\rangle, -i\langle\bar\chi\Sigma_{25}\chi\rangle,-i\langle\bar\chi\Sigma_{35}\chi\rangle\}.
    \label{eq_so5-6-32}
\end{align}
The three components of the first triplet mix with their corresponding components of the second triplet under the U(1) generated by Eq. \ref{eq_u145}. 

Now, considering the other lattice symmetries the above four subsets (Eqs.~\ref{eq_so5-4}-\ref{eq_so5-3}) break up further into the irreps of the microscopic symmetry group as follows. 

The four masses in Eq. \ref{eq_so5-4} break up into a singlet and a triplet as 
	\begin{subequations}\label{eq:decomposition_21}
		\begin{eqnarray}
        \label{eq:decomp_21 singlet}
		&&\left[\mcl{A}_{2g}^o\right]^\Sigma \otimes \left[\mcl{A}_{2g}^o\right]^\zeta = \mcl{A}_{1g}^e\\
        \label{eq:decomp_21 triplet}
		&&\left[\mcl{T}_{2g}^o\right]^\Sigma \otimes \left[\mcl{A}_{2g}^o\right]^\zeta = \mcl{T}_{1g}^e	
		\end{eqnarray}
\end{subequations}
whose explicit forms are given in Eqs.~\ref{eq_a1ge_mass} and \ref{eq_flavour-1} respectively and correspond to two different kinds of quantum spin-octupole phases discussed in Sec. \ref{sec_45,33_mass} and \ref{subsec_te1gsop}.

The six inversion-odd masses (Eq. \ref{eq_so5-6}), on the other hand, break up into two triplets 
	\begin{subequations}\label{eq:decomposition_22}
		\begin{eqnarray}
		&&\left[\mcl{T}_{1u}^o\right]^\Sigma \otimes \left[\mcl{A}_{2g}^o\right]^\zeta = \mcl{T}_{2u}^e\\		
		&&\left[\mcl{T}_{2u}^o\right]^\Sigma \otimes \left[\mcl{A}_{2g}^o\right]^\zeta = \mcl{T}_{1u}^e.
		\end{eqnarray}
\end{subequations}
given by a linear combination of the two triplets in Eq. \ref{eq_so5-6-31} and \ref{eq_so5-6-32} as given by Eqs. \ref{eq_flavour-21} and \ref{eq_flavour-22} respectively. These too correspond to spin-octupole Hall phases, albeit with interesting fine-tuned gapless points for a special combination of the three components of the triplets as discussed in Sec. \ref{subsec_flavour_special}.

The doublet and the triplets in Eqs. \ref{eq_so5-2} and \ref{eq_so5-3} remain intact and result in 
	\begin{subequations}\label{eq:decomposition_3}
		\begin{eqnarray}
		&&\left[\mcl{E}_{u}^e\right]^\Sigma \otimes \left[\mcl{A}_{2g}^o\right]^\zeta = \mcl{E}_u^o\\	
		&&\left[\mcl{T}_{1g}^e\right]^\Sigma \otimes \left[\mcl{A}_{2g}^o\right]^\zeta = \mcl{T}_{2g}^o
		\end{eqnarray}
\end{subequations}
with explicit forms being given by Eqs. \ref{eq_flavour-3} and \ref{eq_flavour-4} respectively. These phases break time-reversal symmetry and describe quantum spin-quadrupole Hall phases as described in Sec. \ref{subsec_qsqp}.

\subsection{Group 3 : The mixed masses} Finally, the largest set of masses is obtained by taking the direct product of the flavour multiplets and the chiral multiplets. There are 45 such masses of the form $-i\langle\bar\chi\Sigma_i\zeta_j\chi\rangle$ where $\Sigma_i$ are the fifteen generators of flavor SU(4) (see Appendix~\ref{appen_gamma}) and $j=1,2,3$. Therefore these masses transform into each other under transformations of the SU(4)~$\otimes$~SU(2) subgroup of SU(8) of the free Dirac theory. Their group decomposition to irreducible representations under the microscopic symmetries is given by
\begin{subequations}\label{eq:decomposition_4}		
		\begin{eqnarray}	
	\label{eq:A2g_T1g_direct_product}	&&\left[\mcl{A}_{2g}^o\right]^\Sigma \otimes \left[\mcl{T}_{1g}^e\right]^\zeta = \rim{\mcl{T}_{2g}^o}\\
	\label{eq:Eu_T1g_direct_product}	&&\left[\mcl{E}_{u}^e\right]^\Sigma \otimes \left[\mcl{T}_{1g}^e\right]^\zeta = \rim{\mcl{T}_{1u}^e} \oplus \rim{\mcl{T}_{2u}^e}\\
	\label{eq:T1g_T1g_direct_product}	&&\left[\mcl{T}_{1g}^e\right]^{\Sigma} \otimes \left[\mcl{T}_{1g}^e\right]^{\zeta} = \mcl{T}_{1g}^e \oplus \mcl{T}_{2g}^e \oplus \mcl{E}_g^e \oplus \rim{\mcl{A}_{1g}^e}\\
	\label{eq:T2g_T1g_direct_product}	&&\left[\mcl{T}_{2g}^o\right]^\Sigma \otimes \left[\mcl{T}_{1g}^e\right]^\zeta = \mcl{T}_{1g}^o \oplus \mcl{T}_{2g}^o \oplus \mcl{E}_g^o \oplus \rim{\mcl{A}_{2g}^o}\\
	\label{eq:T1u_T1g_direct_product}	&&\left[\mcl{T}_{1u}^o\right]^\Sigma \otimes \left[\mcl{T}_{1g}^e\right]^\zeta = \rim{\mcl{T}_{1u}^o} \oplus \rim{\mcl{T}_{2u}^o} \oplus \rim{\mcl{E}_u^o} \oplus \mcl{A}_{1u}^o\\
	\label{eq:T2u_T1g_direct_product}	&&\left[\mcl{T}_{2u}^o\right]^\Sigma \otimes \left[\mcl{T}_{1g}^e\right]^\zeta = \rim{\mcl{T}_{1u}^o} \oplus \rim{\mcl{T}_{2u}^o} \oplus \rim{\mcl{E}_u^o} \oplus \mcl{A}_{2u}^o.
	\end{eqnarray}
\end{subequations}

{The dimension of the representation depends non-trivially on the details of the spin-orbital locking, which, in turn, is reflected in the nature of different density wave phases that these masses lead to. These density wave phases mainly come in two varieties. Out of the total of 45 mixed masses, 18 (marked in black in Eq. \ref{eq:decomposition_4}) generically have at least four gapless fermionic modes protected by a subgroup of SU(8), often in conjunction with lattice symmetries. Thus they describe different kinds of {\it density wave Dirac semimetal} (summarised in Table \ref{tab_mixed_semimetals}). The rest 27 (marked in red in Eq. \ref{eq:decomposition_4}) generically consists of density wave-insulators (summarised in Table \ref{tab_mixed_insulators}). } Two of the insulators, both singlets -- $\mathcal{A}^e_{1g}$ and $\mathcal{A}^o_{2g}$, have edge modes whose signature is evident from appropriate Chern-Simons terms.

Before delving into the details of the resultant phases in this category in Section. \ref{subsec_group3}, we summarise the general structure of these masses and their classification here under various microscopic symmetries leading up to the decomposition in Eq. \ref{eq:decomposition_4}. To this end, starting with microscopic TR, $\mathbb{T}$ (Eq. \ref{eq:microtr}), the 45 bilinears are divided into two classes with 15 TR even given by
\begin{align}
    \{-i\langle \bar\chi\Sigma_{i}\zeta_1\chi\rangle, -i\langle \bar\chi\Sigma_{i}\zeta_2\chi\rangle, -i\langle \bar\chi\Sigma_{i}\zeta_3\chi\rangle\}
    \label{eq:mixed-treven}
\end{align}
{\rm where}~$i=1,2,3,4,5$ (same $\Sigma_i$ content as in Eq. \ref{eq_so5-vector}) and 30 TR odd given by
\begin{align}    
    \{-i\langle \bar\chi\Sigma_{i}\zeta_1\chi\rangle, -i\langle \bar\chi\Sigma_{i}\zeta_2\chi\rangle, -i\langle \bar\chi\Sigma_{i}\zeta_3\chi\rangle\}
    \label{eq:mixed-trodd}
\end{align}
with the $i$ indices being given in Eq.~\ref{eq_so5_generator_index}.

Each of these two sets, under lattice inversion, $\mathbf{I}$  (Table. \ref{tab:lat_sym}), break up into two subsets as odd and even under $\mathbf{I}$. Out of the set of 15 in Eq. \ref{eq:mixed-treven} the six odd ones are given by
\begin{align}
       \{-i\langle \bar\chi\Sigma_{4}\zeta_i\chi\rangle, -i\langle \bar\chi\Sigma_{5}\zeta_i\chi\rangle\}
       \label{eq:mixed-treven-iodd}
\end{align}
while nine even ones are given by
\begin{align}
       \{-i\langle \bar\chi\Sigma_{1}\zeta_i\chi\rangle, -i\langle \bar\chi\Sigma_{2}\zeta_i\chi\rangle, -i\langle \bar\chi\Sigma_{3}\zeta_i\chi\rangle\}
       \label{eq:mixed-treven-ieven}
\end{align}

where $i=1,2,3$.

On the other hand, the set of 30 masses in Eq.~\ref{eq:mixed-trodd} breaks up into two subsets. One of the subsets contains twelve masses that are even under $\mbf{I}$ and is given by
\begin{align}
    \{ -i\langle \bar\chi\Sigma_{45}\zeta_i\chi\rangle,
    -i\langle \bar\chi\Sigma_{12}\zeta_i\chi\rangle,
    -i\langle \bar\chi\Sigma_{23}\zeta_i\chi\rangle, -i\langle \bar\chi\Sigma_{13}\zeta_i\chi\rangle
    \}.
    \label{eq:mixed-trodd-ieven}
\end{align}
The other subset containing eighteen $\mbf{I}$ odd masses is 
\begin{align}
    &\{ -i\langle \bar\chi\Sigma_{14}\zeta_i\chi\rangle,
    -i\langle \bar\chi\Sigma_{15}\zeta_i\chi\rangle,
    -i\langle \bar\chi\Sigma_{24}\zeta_i\chi\rangle,\nonumber\\
    &\quad-i\langle \bar\chi\Sigma_{25}\zeta_i\chi\rangle,
     -i\langle \bar\chi\Sigma_{34}\zeta_i\chi\rangle,
    -i\langle \bar\chi\Sigma_{35}\zeta_i\chi\rangle
    \}.
    \label{eq:mixed-trodd-iodd}
\end{align}

Further application of lattice symmetries (Appendix \ref{appen_definition_of_space_group_representaiton}) break these up into singlets, doublets and triplets as follows. The TR even and inversion odd subset (Eq. \ref{eq:mixed-treven-iodd}) of six decomposes into two triplets given by Eq. \ref{eq:Eu_T1g_direct_product} which correspond to two different zig-zag spin-quadrupolar density wave insulators  given by Eq. \ref{eq_quadrupole_density_waves}.  Similarly, the nine TR and inversion even masses break up into two triplets, one doublet and one singlet as given by Eq. \ref{eq:T1g_T1g_direct_product}. They represent spin-quadrupole density waves. {While the singlet corresponds to an  insulator (Eq. \ref{eq_a1ge_mass_2}) with quantized spin-octupole filtered edge modes, the rest (Eqs. \ref{eq_massstripyquad}, \ref{eq_gapless_t1ge} and \ref{eq_ege_doublet_1}) are partially gapless semimetals.}

The subset of 12 TR odd and inversion even masses (Eq.~\ref{eq:mixed-trodd-ieven}) break up into three triplets, one doublet and one singlet under the action of the lattice symmetries (Eqs.~\ref{eq:A2g_T1g_direct_product}, \ref{eq:T2g_T1g_direct_product}). {The singlet (an anomalous Hall insulator) and one of the triplets represent insulating ferro (uniform) (Eq.~\ref{eq:ferro-spin-octupole-singlet}) and stripy (Eq.~\ref{eq_sigma45_density_waves}) density wave ordering of spin-octupoles respectively.}  The doublet (Eq.~\ref{eq_ego_doublet}) corresponds to ferro spin-octupole semimatal and the other two triplets (Eqs.~\ref{eq_t01gtriplet}, \ref{eq_gapless_t2go}) correspond to different stripy spin-octupole density wave semimetals.

Finally, the 18 TR and inversion odd masses, break up under lattice symmetries into four triplets, two doublets and two singlets given by Eqs. \ref{eq:T1u_T1g_direct_product} and \ref{eq:T2u_T1g_direct_product}. Out of them, the two singlets (Eqs. \ref{eq_neel_singlet1} and \ref{eq_neel_singlet2}) correspond to staggered (``N\'eel") spin-octupole density-wave semimetal. The two doublets (Eqs. \ref{eq_euo_mass_2} and \ref{eq_euo_mass_3}) on the other hand, both correspond to N\'eel spin-octupole density wave insulators. As discussed below Eq. \ref{eq_euo_mass_3}, they can be rotated into each other via a U(1) transformation generated by $\Sigma_{45}$ within the flavour space. Given this fact and they break the same symmetries, the two doublets correspond to the same phase and are not distinct from each other. Similar arguments hold for the two sets of triplets, each of which represents zig-zag spin-octupole density wave insulators. The two $\mathcal{T}^o_{1u}$ triplets (Eqs. \ref{eq_t1u0_masses_1} and \ref{eq_t1uo_masses_2}) can be continuously connected without change of symmetry and hence represent the same phase. Similarly the two $\mathcal{T}^o_{2u}$ triplets (Eqs. \ref{eq_t2uo_masses_1} and \ref{eq_t2uo_masses_2}) give the same phase.


\section{Group-1 : The chiral Masses}
\label{subsec_group1}
There are four chiral masses that are given in Eqs. \ref{eq_chiraldecomp_1} and \ref{eq_chiraldecomp_2}. Their transformations under microscopic symmetries are given in Appendix \ref{appen_definition_of_space_group_representaiton}. In accordance with Eqs. \ref{eq_chiraldecomp_1} and \ref{eq_chiraldecomp_2}, these are divided into a singlet (TR  and reflection odd) and a triplet (TR and inversion even) under the space-group symmetries while they are all singlets under the flavour SU(4). Here we discuss the physics of these masses. 

\subsection{The \texorpdfstring{SU(8)}{} symmetric Integer Chern Insulator}\label{sec:ICI-phase}

\begin{figure}
    \centering
    \includegraphics[scale=0.3]{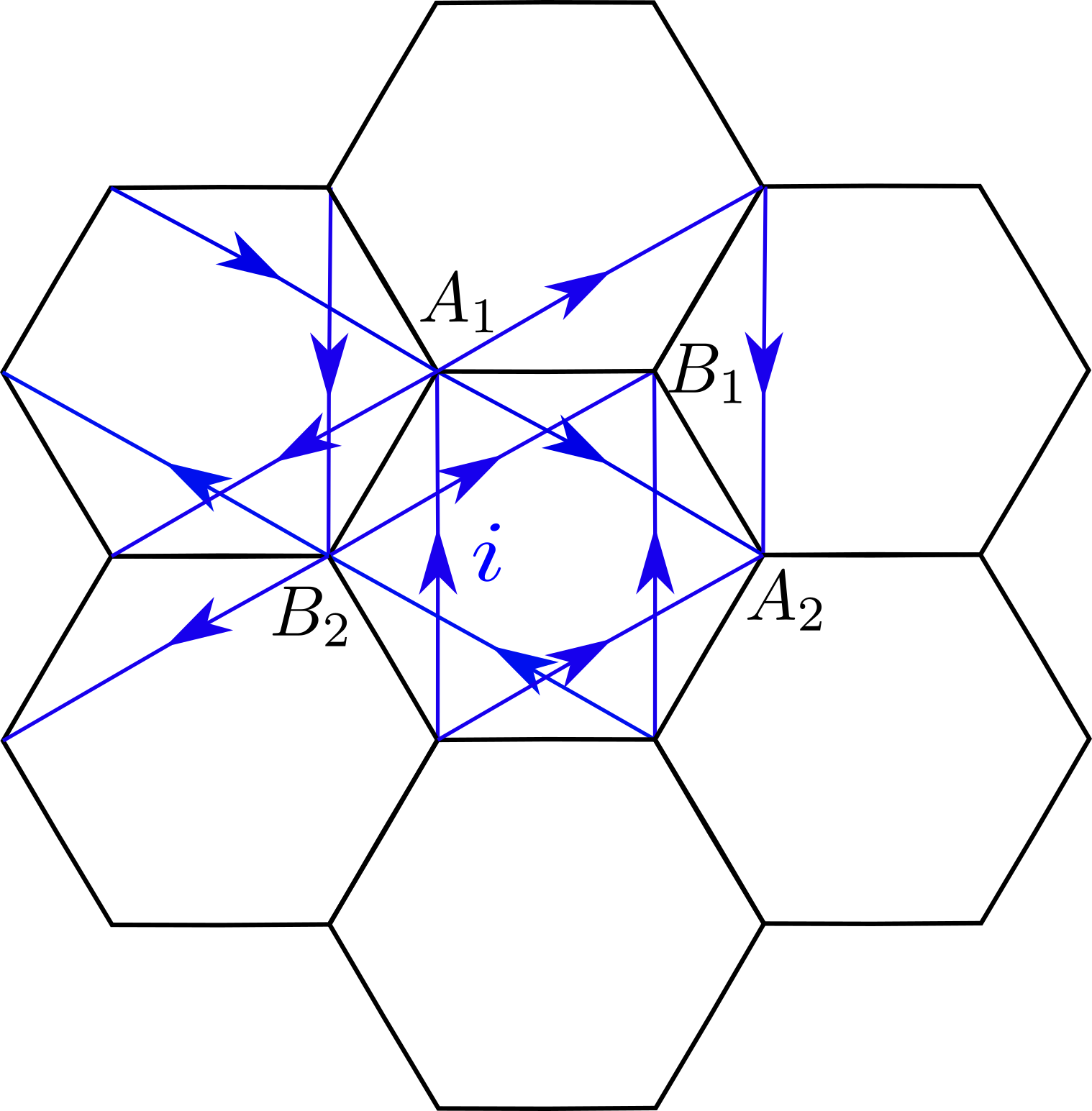}
    \caption{ The mean field hopping model for the Integer Chern insulator. The second neighbour hoppings (in blue) are generated via the spontaneous symmetry breaking  and \textcolor{violet}the hopping amplitudes along the directions of the arrows (in local basis) are $i=\sqrt{-1}$. The same hopping pattern also holds for the singlet spin-octupole Hall mass leading to the octupolar Hall effect (Eq. \ref{eq:octupolar_hall_action}) except in that case the hopping amplitude is given by $i\Sigma_{45}$.}
    \label{fig:chern_insulator_hopping}
\end{figure}

The SU(8) singlet mass, 
\begin{align}\label{eq_chern_insulator}
    \Delta_{ICI}=-i\langle\bar{\chi}\chi\rangle,
\end{align}
given by Eq. \ref{eq_chiraldecomp_1} is odd under the action of $\mbb{T}$, $\mbf{C_2'}$ and $\mbf{\sigma_d}$. This suggests that this mass is the analog of the Chern mass for graphene~\cite{PhysRevLett.61.2015,PhysRevB.74.085308} which shows the integer quantum Hall effect and hence represents the ICI phase. Indeed, minimally coupling the electrons in the massive Dirac action (Eq. \ref{eq_dirac_massiveaction}) for the above Chern mass, $\Delta_{ICI}$ to a  U(1) probe gauge field, $A_\mu$, that couples to the electronic charge and integrating out the gapped fermions, we get a U(1) Chern-Simons term with the (Euclidean) Lagrangian density :
\begin{align}
    \mathcal{L}_{\rm CS}=i\frac{N_F\rm{Sgn}[\Delta_{ICI}]}{4\pi} \epsilon^{\mu\nu\lambda} A_\mu\partial_\nu A_\lambda
    \label{eq_icilag}
\end{align}
where $N_F=4$ is the number of SU(4) flavors ({\it i.e.} number of four component Dirac fermion fields). Thus, all the flavors contribute the same amount to the charge Hall conductivity resulting in, $\sigma_{xy}=N_F\frac{e^2}{2\pi}$~\cite{PhysRevLett.49.405}.

The nature of symmetry breaking can be analysed by considering the low energy projection of the microscopic current operators on the lattice~\cite{PhysRevLett.61.2015,PhysRevLett.100.156401}. In particular, starting with the microscopic orbitals in the local basis, $\phi(\boldscriptr_\mcl{S})$, given by Eq. \ref{eq:definition of phi}, the hopping operator on the next nearest neighbour (NNN) bonds-- say the blue bond in Fig. \ref{fig:chern_insulator_hopping} from site $B_2$ to $B_1$-- is given by~\cite{PhysRevLett.61.2015} 
\begin{align}
    \mcl{B}_{B_2B_1}=\phi^\dagger(\boldscriptr_{B_1})\phi(\boldscriptr_{B_2})
    \label{eq:icilatticebond}
\end{align}
The form in the global basis (in terms of the $j=3/2$ orbitals) can be easily obtained via Eq. \ref{eq:definition of phi} and using the forms of $\mathcal{G}(\boldscriptr_\mathcal{S})$ given in Appendix \ref{sec:appen_g_matrices}. 

In the low energy limit, Eq. \ref{eq:icilatticebond} is equal to
\begin{align}
    \mathcal{B}_{B_2B_1}=\frac{1}{2\sqrt{3}}\bar\chi~\chi+\cdots
\end{align}
where $\cdots$ represent higher order terms. Therefore for $\Delta_{ICI}\neq 0$, we have an imaginary second neighbour hopping whose sign structure is given by Fig. \ref{fig:chern_insulator_hopping}. This leads to finite bond current such that the total gauge invariant loop current per hexagon is indeed zero (mod $2\pi$). The loop currents, therefore, form a Z$_2$ order parameter proportional to the mass, $\Delta_{ICI}$. Such Z$_2$ order parameters allow domain walls as one-dimensional topological defects in two-dimensional systems across which the sign of the mass changes. As is evident from Eq. \ref{eq_icilag}, the edge modes have opposite chirality in the two cases and hence the domain wall is associated with chiral gapless edge modes that are exponentially localised along the domain wall~\cite{callan1985anomalies}.


\subsection{The \texorpdfstring{SU(4)}{} symmetric stripy charge density wave Insulator}\label{sec:chiral-CDW-phase}

\begin{figure}
    \centering
    \includegraphics[scale=0.6]{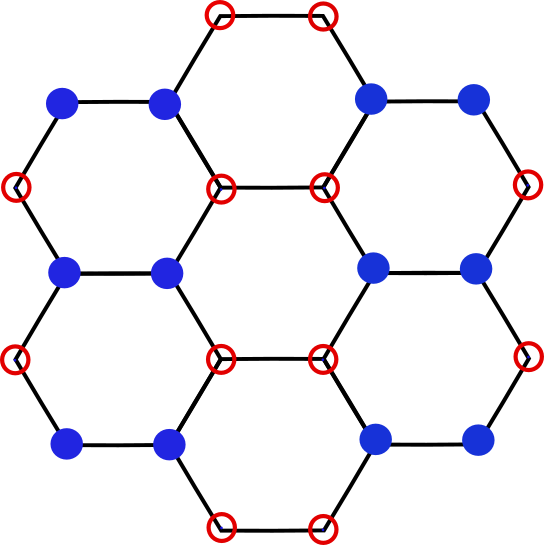}
    \caption{Stripy CDW for $-i\Braket{\bar{\chi}\zeta_1\chi}\neq 0$. Here red circles and blue dots represent opposite charge densities at the honeycomb sites.}
    \label{fig:stripy_CDW}
\end{figure}

The three other SU(4) flavour invariant masses given by Eq.~\ref{eq_chiraldecomp_2} form a triplet ($\mcl{T}^e_{1g}$) which is even under TR symmetry, $\mbb{T}$ as well as inversion, ${\bf I}$, about the plaquette centre. These are given by
\begin{align}
    \{-i \langle \bar{\chi} \zeta_1 \chi\rangle, -i \langle\bar{\chi} \zeta_2 \chi\rangle, -i\langle \bar{\chi} \zeta_3\chi\rangle\}.
    \label{eq_su4_cdw}
\end{align}

They transform into each other under various lattice rotations and reflections as a triplet as shown in Appendix \ref{appen_definition_of_space_group_representaiton}. $\zeta_i$ ($i=1,2,3$) of course generates an SU(2) (see Eq. \ref{eq_zetasu2}) which is broken down by the lattice symmetries to a triplet for the space group.

The three bilinears have the same symmetry as that of the three stripy CDW order as shown in Fig. \ref{fig:stripy_CDW} for $-i \langle \bar{\chi} \zeta_1 \chi\rangle$.  The other two can be obtained by $\mbf{C_3}$ rotations. Indeed the analysis of the above three fermion bilinears in the global basis (Eq. \ref{eq:4-component psi}) confirms the symmetry analysis in identifying the above stripy CDW masses. 
In particular, starting with the electron operators in the global basis \textit{i.e.,} $\psi$ (Eq.~\ref{eq:4-component psi}), the projected charge density operators on different sub-lattices have the following form~: 
\begin{eqnarray}
:\psi^\dagger(\boldscriptr_\mcl{S})\psi(\boldscriptr_\mcl{S}): ~=\begin{cases}
-i\bar{\chi}\xi_1\chi &\text{ For $\mcl{S} = B_2,A_2$ } \\
i\bar{\chi}\xi_1\chi &\text{ For $\mcl{S} = A_1,B_1$ } 
\end{cases}
\label{eq_scdw}
\end{eqnarray}
where $:\mathcal{O}:$ denotes normal ordering. Integrating out the gapped fermions in the presence of the mass does not lead to a finite charge Hall response. 

The three matrices $\gamma_0\zeta_i$ ($i=1,2,3$) pairwise anti-commute with each other such that the three stripy CDW masses are compatible in the sense that the fermion gap does not close as the three masses are rotated into each other under the chiral SU(2) (see Eq. \ref{eq_zetasu2}) transformations generated by $e^{i\theta\hat{\bf n}\cdot\mathbf{\zeta}}$. This would suggest that the order parameter manifold is a unit sphere, $\mcl{S}^2$ similar to collinear magnetic ordering (with one important difference that the present order parameter is even under TR unlike magnetic order). The above SU(2) is, however, broken down by the lattice symmetries which, in terms of the order parameter, selects out symmetry-allowed points on the sphere, $\mathcal{S}^2$. In particular, the leading order anisotropy of the form
\begin{align}
   \mathcal{L}_{aniso}\sim -\mathcal{w}_1\Delta_1\Delta_2\Delta_3+\mathcal{w}_2\left(\Delta_1^4+\Delta_2^4+\Delta_3^4\right) 
   \label{eq_cdwaniso}
\end{align}
is symmetry allowed (see Table \ref{tab_def_of_3d_irreps}) in the effective action with $\Delta_i$ being the amplitude for the three components of the CDW (Eq. \ref{eq_su4_cdw}). This reduces the order parameter manifold to discrete points on the sphere. The details of the ordering depend on the signs of the couplings $\mathcal{w}_1$ and $\mathcal{w}_2$. Due to the presence of the third-order invariant, the transition out of the semimetal is expected to be first-order.

We conclude the discussion of the chiral masses with two points. First, it is useful to compare the four chiral masses with the case of spinless fermions on the honeycomb lattice with $\pi$-flux at one-quarter filling presented in appendix.~\ref{appen:spinlessPiFluxHex}. The presence of the $\pi$-flux breaks up the chiral space as ${\bf 4} = {\bf 1} \oplus {\bf 3}$, as opposed to graphene where the chiral space is decomposed as ${\bf 4} = {\bf 1} \oplus {\bf 1} \oplus {\bf 2}$.  In the present case where the $\pi$-flux is a consequence of SOC, the above SU(4) singlet masses can be thought of as four copies of the spinless case in Appendix \ref{appen:spinlessPiFluxHex}. Second, while the three triplet stripy masses are compatible with each other, {\it i.e.}, the respective mass matrices mutually anticommute, all of them are incompatible with the singlet ICI mass since the corresponding matrices (Table \ref{tab:real_space_irreps}) pairwise commute. This ensures a phase transition~\cite{ryu2009masses} between the two phases which is accompanied by the change in the nature of broken symmetry as well as the Chern-Simons level (Eq.\ref{eq_icilag}) from $N_F=4$ (in the ICI) to $N_F=0$ (in the stripy CDW).


\section{Group-2 : The flavour masses}
\label{subsec_group2}

The 15 flavour masses of the form $-i\langle\bar\chi\Sigma_i\chi\rangle$ are divided into six irreducible representations by microscopic symmetries that are given by Eqs. \ref{eq:decomposition_21}, \ref{eq:decomposition_22} and \ref{eq:decomposition_3}. Each of these 15 masses individually breaks the flavor SU(4) down to U(1)~$\otimes$~SO(4). However linear combinations of them can reduce the symmetry further as we discuss below in the case of each subgroup. 
A notable feature of these residual sub-groups is that the generators depend on the particular {\it direction} of the mass matrix and hence are {\it locally} defined in the space of the order parameters. This is exactly like the case of a collinear ferromagnet/antiferromagnet where the particular generator of the residual U(1) depends on the direction of the ordering of the magnetic moments in the spin-space. We shall study the nature of these for each of the six phases separately, including the action of the lattice symmetries-- including the spontaneously broken ones-- as well as the nature of the residual symmetry group.

Out of the 15 masses, the ten TR even ones (Eq. \ref{eq_so5_generator_index}) correspond to four different types of spin-octupole Hall phases while the five TR odd ones (Eq. \ref{eq_so5-vector}) represent two spin-quadrupole Hall phases. We explain their features in turn.


\subsection{Quantum spin-octupole Hall insulators}
\label{subsec_qsop}

\subsubsection{{\texorpdfstring{$\mcl{A}^e_{1g}$}{} Singlet mass
} }\label{sec_45,33_mass}

For the singlet (Eq. \ref{eq:decomp_21 singlet}), the mass is given by the fermion bilinear 
\begin{align}
    \mcl{A}_{1g}^e~:&~-i\langle\bar\chi\Sigma_{45}\chi\rangle \label{eq_a1ge_mass}
\end{align} 
which fully gaps out all the fermions. This breaks the SU(4) flavour symmetry down to U(1)~$\otimes$~SO(4) with the U(1) being generated by $\Sigma_{45}$ and SO(4) by six other $\Sigma_i$'s that commute with $\Sigma_{45}$, {\it i.e.,} $\{\Sigma_1, \Sigma_2, \Sigma_3, \Sigma_{12}, \Sigma_{23}, \Sigma_{13}\}$.

In the microscopic $j=3/2$ basis (the transformation of the $\Sigma$ operators from the local to global basis is given in Table \ref{tab:matrix realtions of SU(4) and j=3/2 basis}), the $\Sigma_{45}$ operator is given by
\begin{align}
   \Sigma_{45} &= -\frac{4}{3\sqrt{3}}\left(J_xJ_yJ_z + J_yJ_zJ_x + J_zJ_xJ_y - \frac{15i}{8} \right)
\end{align}
where $(J_x,J_y, J_z)$  are the $j=3/2$ spin operators in the global basis (see Table \ref{tab:matrix realtions of SU(4) and j=3/2 basis}) such that $\Sigma_{45}$ is a spin-octupole. 

Starting with the Euclidean Dirac action in the presence of the mass term (Eq. \ref{eq_dirac_massiveaction}), we can diagonalise the spinors $\chi$ in terms of the eigenstates of $\Sigma_{45}$. The eigenstates consist of two pairs of Kramers doublets. The two members of each doublet have opposite eigenvalues of $\Sigma_{45}$, {\it i.e.} $\pm 1$. Each of the modes contributes to finite Hall edge current leading to {\it spin-octupole} filtered Hall edge modes similar to the quantum spin-Hall effect~\cite{PhysRevLett.95.226801}. The two TR partners carry current in the opposite direction and backscattering within each TR pair is disallowed by TR symmetry -- again just like quantum spin Hall effect. However, the scattering between the oppositely moving edge modes belonging to the two different Kramers doublets is not allowed because they necessarily have opposite eigenvalues of $\Sigma_{45}$.

A more formal derivation of the resultant symmetry-protected CS action is obtained by coupling probe charge and spin-octupolar gauge fields to Eq. \ref{eq_dirac_massiveaction}, {\it i.e.}, considering
\begin{align}\label{eq:octupolar_gauge_action}
    S[A_c,A_o]=-i\int d^2{\bf r}d\tau~\bar{\chi}({\bf r})\left[v_F \slashed{D}-\Delta~\Sigma_{45}\right]\chi({\bf r})
\end{align}
where
\begin{align}\label{eq_gauge_inv_derivative}
\slashed{D}=\gamma^\mu\left(i\partial_\mu-A_{c,\mu}-\Sigma_{45}A_{o,\mu}\right)
\end{align}
where ${\bf A}_c$ and ${\bf A}_o$ are charge and spin-octupole probe gauge fields respectively. Then integrating out the fermions leads to the mutual CS term given by
\begin{eqnarray}\label{eq:octupolar_hall_action}
S^{mutual}_{CS}&& = i\frac{N_F}{2\pi} ~sgn(\Delta)~  \int d^3x~\epsilon^{\mu\nu\lambda} A_{c,\mu}\partial_{\nu}A_{o,\lambda}
\end{eqnarray}
which characterises the {\it quantum spin-octupolar Hall response}. 

The lattice version of this mass can be analyzed in a similar way as done for the ICI mass. For that, we consider microscopic hopping operators on one of the NNN bonds and project that to the low-energy sector. We again take the blue bond in Fig.~\ref{fig:chern_insulator_hopping} from site $B_2$ to $B_1$ and write the following hopping operator
\begin{align}
\mcl{B}_{B_2B_1}^{(45)}=\phi^\dagger(\boldscriptr_{B_1})~\Sigma_{45}~\phi(\boldscriptr_{B_2}).
\label{eq:QOH_latticebond}
\end{align}

In term of the low-energy spinors, this has the following form
\begin{eqnarray}
  \mcl{B}_{B_2B_1}^{(45)}= \frac{1}{2\sqrt{3}} \bar{\chi}~\Sigma_{45}\chi + \cdots.
\end{eqnarray}
Thus the imaginary part of this hopping operator is proportional to the order parameter in this phase. This shows that there are non-zero bond currents in this phase. The hopping pattern on the other bonds are same as shown in Fig.~\ref{fig:chern_insulator_hopping} with the hopping amplitudes being $i\Sigma_{45}$ instead of $i$.

\begin{figure}
    \centering
        \subfigure[Zig-zag edge]{
        \includegraphics[scale=0.525]{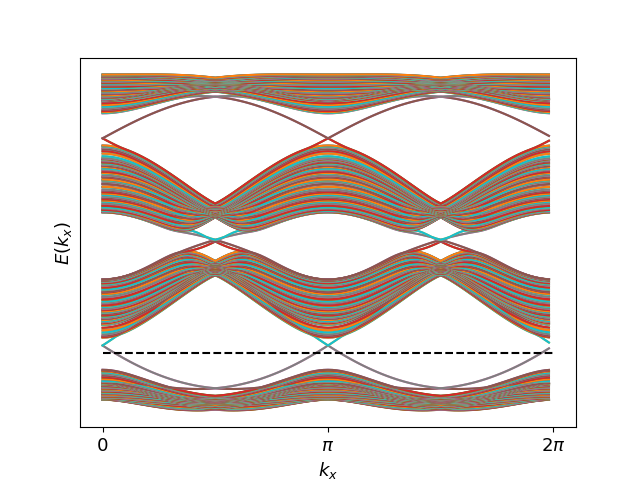}}
        \subfigure[Armchair edge]{
        \includegraphics[scale=0.525]{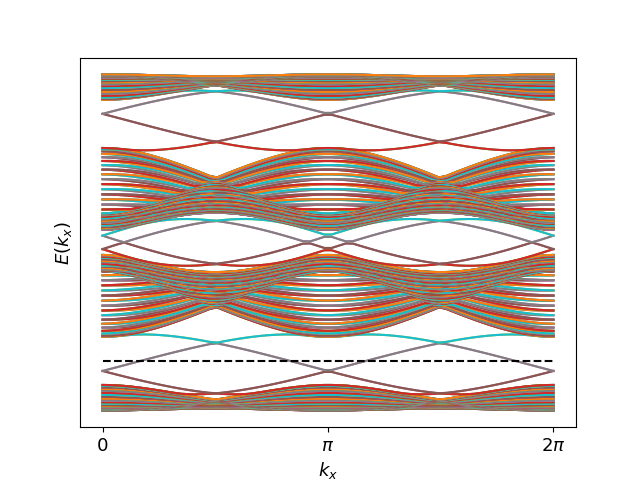}}
        \caption{Spectrum for $-i\bar{\chi}\Sigma_{45}\chi$ with zig-zag and armchair edges. Both of these plots show gapless states at $1/4th$ filling (shown by black dashed line). To get this spectrum, we consider a honeycomb lattice with cylindrical geometry where the edges of the cylinder have zigzag or armchair-like boundaries. Here $k_x$ is lattice momentum along the periodic direction. We take 32 magnetic unit cells along the length of the cylinder to perform this numerical calculation.
        }
        \label{fig_4533_zigzag}
\end{figure}

The lattice Hamiltonian can now be used to check for the edge modes by obtaining the spectrum with open boundary conditions. The spectrum on a cylinder for the zig-zag and armchair edges are shown in Fig. \ref{fig_4533_zigzag}.

Not surprisingly, in Fig. \ref{fig_4533_zigzag}, such edge modes are also observed for both 3/4th filling as well as 1/2 filling. While the case of 3/4th filling is expected to result from the microscopic particle-hole transformation that maps $1/4\leftrightarrow 3/4$, the physics of $1/2$ filling would be interesting to understand in future.

Similar to the ICI phase, the order parameter for the spin-octupole Hall phase is a $Z_2$ field and leads to gapless fermionic modes associated with the domain walls of the order parameter. Note, however, that such a field is TR even and is symmetric under all lattice transformations. Thus this mass is naturally allowed by the microscopics. This is analogous to the Kane-Mele mass~\cite{PhysRevLett.95.226801} for graphene which is symmetry allowed but is energetically suppressed due to the very small value of the SOC in that case.

\subsubsection{{The \texorpdfstring{$\mcl{T}_{1g}^e$}{} triplet masses} } 
\label{subsec_te1gsop}

The TR even triplet in Eq. \ref{eq:decomp_21 triplet} consists of three masses of the form
\begin{align} 
    \mcl{T}_{1g}^e~:&~\left\{\begin{array}{c}
         -i\braket{\bar{\chi}\Sigma_{12}\chi}  \\
         -i\braket{\bar{\chi}\Sigma_{13}\chi} \\ 
         -i\braket{\bar{\chi}\Sigma_{23}\chi}
    \end{array}\right.
    \label{eq_flavour-1}
\end{align} 
which fully gaps out the Dirac fermions. In terms of the $j=3/2$ operators, the three mass matrices are given by~:
\begin{align}
    \Sigma_{\alpha\beta}=\frac{7\epsilon^{\alpha\beta\gamma}}{3}\left( J_\gamma-\frac{4}{7}J_\gamma^3 \right)
    \label{eq_spin-oct-hall}
\end{align}
where $\alpha,\beta,\gamma=1,2,3$ with $\alpha\neq\beta$ such that they are a mixture of dipole and spin-octupole operators. Following Eq. \ref{eq_dirac_massiveaction}, the generic mass term is given by
\begin{align}
    -i\left(\Delta_1 \bar\chi\Sigma_{23}\chi+\Delta_2\bar\chi\Sigma_{13}\chi+\Delta_3\bar\chi\Sigma_{12}\chi\right)
    \label{eq_lin-spin-oct}
\end{align}
where $\Delta_i~(i=1,2,3)$ are the weights for each of the three components. Hence such masses lie on a 2-sphere with directional cosines given by $\cos\theta_i=\Delta_i/\sqrt{\Delta_1^2+\Delta_2^2+\Delta_3^2}$ where the different points can be rotated into each other via the SU(2) symmetry generated by Eq \ref{eq_so5-su2}. With reference to Eq. \ref{eq_flavour-1}, it is now clear that this SU(2) corresponds to the continuous rotation amongst the three spin-octupoles.

At each point on this {\it mass} sphere, the residual symmetry is U(1)~$\otimes$~SO(4). However, the particular generators of this residual symmetry group depend on the location of the point and are related to each other by the same SU(2) transformations (Eq. \ref{eq_so5-su2}). For example, the generators of the residual symmetry at the point $[001]$ are given by
\begin{eqnarray}
&& \{\Sigma_{12}, \Sigma_3,\Sigma_4,\Sigma_5,\Sigma_{34},\Sigma_{35},\Sigma_{45}\}
\end{eqnarray}
where the first generator corresponds to the U(1) (which is left-over of the SU(2) (Eq. \ref{eq_so5-su2})) and the rest generate the SO(4). The residual groups at other points on the mass sphere are obtained via SU(2) rotations generated by Eq. \ref{eq_so5-su2}.

The phase breaks the spin-octupole SU(2) symmetry (Eq. \ref{eq_so5-su2}) spontaneously and results in quantum spin-octupole Hall effect that is protected by U(1)$\rtimes$ Z$_{\rm 2}^{TR}$ and is similar to the quantum Spin Hall phase obtained via spontaneously broken spin-rotation symmetry discussed in Ref. \cite{PhysRevLett.100.156804} with interesting differences (see below). The presence of non-trivial spin-octupole filtered edge states is confirmed by calculating the mutual Hall response similar to Eq. \ref{eq:octupolar_gauge_action} for the singlet case above which leads to the mutual Chern-Simons action similar to Eq. \ref{eq:octupolar_hall_action}.

The presence of the gapless edge-modes can also be checked by going back to the mean-field Lattice Hamiltonian in presence of the lattice version of the mass (not shown). The lattice version of the Hamiltonian corresponding to the continuum bilinear $-i\langle\bar\chi\Sigma_{12}\chi\rangle$ is {the same as that for the quantum spin-octupolar Hall mass given in Eq.~\ref{eq:QOH_latticebond} with the hopping matrix(i.e., $\Sigma_{45}$) replaced by $\Sigma_{12}$.

An interesting fallout of the present implementation of the symmetry is the fact that the three component spin-octupolar order-parameter allows for Skyrmion configurations. Following the calculations of Refs. \cite{PhysRevLett.100.156804, abanov2000theta} (and references therein) it is rather straightforward to show that such Skyrmions carry $N_F(=4)$ units of electronic charge and are bosons. Hence, condensation of such Skyrmions within a framework discussed in Ref. \cite{PhysRevLett.100.156804} would lead to a novel $4e$ superconductor with  single electron excitations being gapped and the magnetic flux is quantized in units of $h c/4e$~\cite{berg2009charge}. This is tantamount to the {\it fractionalisation}~\cite{senthil2000z} of the elementary BCS $hc/2e$-vortex. The above mechanism to obtain a $4e$ superconductor is rather novel and differs from the usual mechanism of BCS superconductivity, where such a $4e$ superconductor is obtained by forming a 4-electron bound state and condensing them. The novel superconductor here seems to be a natural consequence of the SOC-mediated symmetry implementation in quarter-filled $j=3/2$ honeycomb lattices that allow binding of $4e$ charges to the topological texture of the spin-octupole order parameter. 

\subsubsection{{The \texorpdfstring{$\mcl{T}_{1u}^e$}{} and \texorpdfstring{$\mcl{T}_{2u}^e$}{} triplet masses}}
\label{subsec_flavour_special}

The two inversion odd TR even triplets (Eq.~\ref{eq:decomposition_22}) are given by 
\begin{align}
  \mcl{T}^e_{1u}~:&~\left\{\begin{array}{l}
  -i\braket{\bar{\chi}\Sigma_{34}\chi}\\ -i\braket{\bar{\chi}\frac{\Sigma_{14}+\sqrt{3}\Sigma_{15}}{2}\chi}\\ -i\braket{\bar{\chi}\frac{\Sigma_{24}-\sqrt{3}\Sigma_{25}}{2}\chi}\\
  \end{array}\right.\label{eq_flavour-21}\\
  \mcl{T}^e_{2u}~:&~ \left\{\begin{array}{l}
  -i\braket{\bar{\chi}\Sigma_{35}\chi}\\ -i\braket{\bar{\chi}\frac{\sqrt{3}\Sigma_{14}-\Sigma_{15}}{2}\chi}\\
  -i\braket{\bar{\chi}\frac{\sqrt{3}\Sigma_{24}+\Sigma_{25}}{2}\chi}\\
  \end{array}\right.
  \label{eq_flavour-22}
\end{align} 

In terms of the $j=3/2$ spin matrices, we have
\begin{subequations}
\begin{align}
\Sigma_{34} &= \frac{2}{3}\left( J_z^3-\frac{13}{4}J_z \right) \label{eq_te1u_m1} \\
-\frac{1}{2}\Sigma_{14}-\frac{\sqrt{3}}{2}\Sigma_{15} &= \frac{2}{3}\left( J_x^3-\frac{13}{4}J_x \right) \label{eq_te1u_m2} \\
-\frac{\sqrt{3}}{2}\Sigma_{25}+\frac{1}{2}\Sigma_{24} &= -\frac{2}{3}\left( J_y^3-\frac{13}{4}J_y \right)
\label{eq_te1u_m3}
\end{align}
\label{eq_te1u}
\end{subequations}
for the three $\mcl{T}^e_{1u}$ masses and
\begin{subequations}
\begin{align}
    &\Sigma_{35} =\frac{2}{3\sqrt{3}} \left[ (J_x^2J_z + J_xJ_zJ_x + J_zJ_x^2) \right. \nonumber\\
    & \left. \qquad \qquad \qquad \qquad - (J_y^2J_z + J_yJ_zJ_y + J_zJ_y^2) \right] \label{eq_te2u_m1}\\
     &\frac{\sqrt{3}}{2}\Sigma_{24} + \frac{1}{2}\Sigma_{25} =\frac{2}{3\sqrt{3}} \left[ (J_x^2J_y + J_xJ_yJ_x + J_yJ_x^2) \right.\nonumber\\
     & \left. \qquad \qquad \qquad \qquad- (J_z^2J_y + J_zJ_yJ_z + J_yJ_z^2) \right] \label{eq_te2u_m2}\\
     &\frac{\sqrt{3}}{2}\Sigma_{14} - \frac{1}{2}\Sigma_{15} =\frac{2}{3\sqrt{3}} \left[ (J_y^2J_x + J_yJ_xJ_y + J_xJ_y^2) \right.\nonumber\\
     & \left.\qquad \qquad \qquad \qquad- (J_z^2J_x + J_zJ_xJ_z + J_xJ_z^2) \right] .
     \label{eq_te2u_m3}
    \end{align}
    \label{eq_te2u}
\end{subequations}    
for the $\mcl{T}^e_{2u}$ triplet. Hence these masses represent two different sets of spin-octupole order. Note that while components of the two triplets can be rotated into each other by a U(1) rotation generated by $\Sigma_{45}$, the two triplets represent different phases since they have different transformations under lattice reflection, $\mathbf{C_2'}$.

The three masses in each of the triplets are incompatible, {\it i.e.} the matrices $(m_1, m_2, m_3)$ in Eq. \ref{eq_te1u_m1}, \ref{eq_te1u_m2}, \ref{eq_te1u_m3} or \ref{eq_te2u_m1}, \ref{eq_te2u_m2}, \ref{eq_te2u_m3} do not mutually pair-wise anticommute. This results in an interesting structure for the residual symmetry in the resultant massive phases. For a generic linear combination of the three masses, similar to Eq. \ref{eq_lin-spin-oct}  but now for the $\mcl{T}^e_{1u}$ and $\mcl{T}^e_{2u}$ triplets, {\it i.e.},
\begin{eqnarray}\label{eq_gen_mass_on_sphere}
-i\left(\Delta_1 \bar\chi m_1\chi + \Delta_2 \bar\chi m_2\chi + \Delta_3 \bar\chi m_3\chi\right),
\end{eqnarray}
where $m_1,m_2,m_3$ are the three matrices in Eq. \ref{eq_te1u} or \ref{eq_te2u}, the flavour SU(4) is broken down to U(1) $\otimes$ U(1) $\otimes$ U(1). However, to get more insights, it is useful to diagonalise the bilinear in Eq.~\ref{eq_gen_mass_on_sphere} for a generic point on the unit sphere described by the directional cosines $\Delta_i/\sqrt{\Delta_1^2+\Delta_2^2+\Delta_3^2}$ (middle panel of Fig. \ref{fig_spectrum_on_sphere}) to obtain
\begin{align}
    -\bar\chi'\mcl{D}\chi'
    \label{eq:chitilde}
\end{align}
where $\chi'$ are the fermions in the diagonalised basis and 
\begin{eqnarray}
\mcl{D}=\begin{pmatrix}
a_1\sigma_3 & 0  \\
0 & a_2\sigma_3
\end{pmatrix}\otimes \zeta_0
\label{eq_u1tu1tu1}
\end{eqnarray}
with $a_1,a_2$ are two real functions of $\Delta_i$s, $\sigma_3$ is the third Pauli matrix, and $\zeta_0$ is the identity matrix that acts in the valley-band space of the spinors, {\it i.e.} in the chiral SU(2) space (Eq. \ref{eq_zetasu2}). 

\begin{figure*}
    \centering
    \includegraphics[scale=0.5]{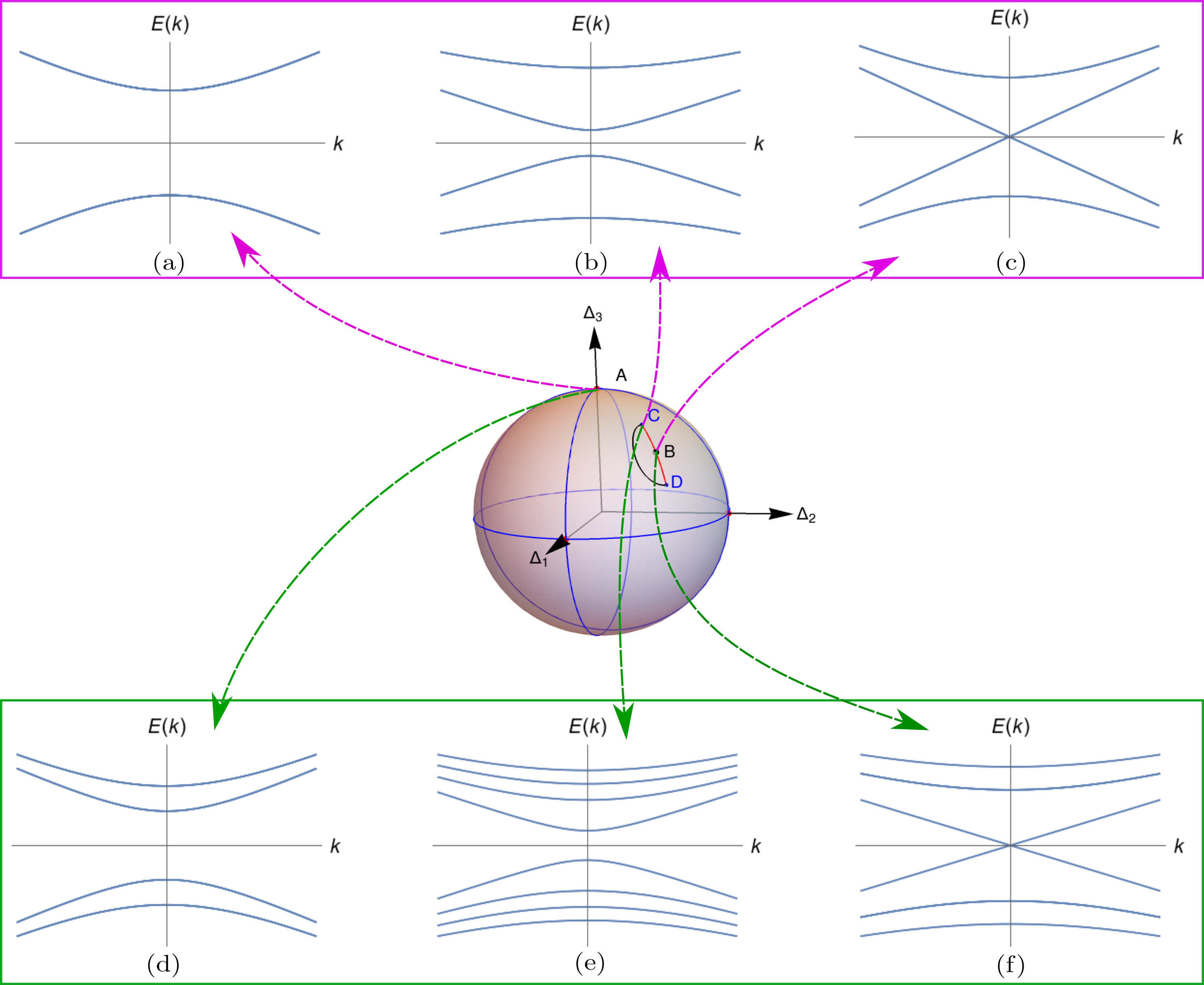}
    \caption{Figure shows the energy spectrum of low-energy fermions along the $k_x=k_y$ line for different combinations of the masses in a triplet. Any linear combination of the masses in a particular triplet can be represented on the surface of a unit sphere shown at the center. The energy spectrum shown in the upper panel of this figure corresponds to the $\mcl{T}_{1u}^e$ triplets in Eq.~\ref{eq_flavour-21},~\ref{eq_t1ue_masses} and the $\mcl{T}_{2u}^e$ triplets in Eq.~\ref{eq_flavour-22},~\ref{eq_t2ue_masses}. The spectrum in (a), (b) and (c) in the upper panel correspond to the spectrum at the points $A$, $C$ and $B$ on the sphere.
    Similarly, the spectrum in the bottom panel corresponds to the $\mcl{T}_{1u}^o$ and $\mcl{T}_{2u}^o$ triplets in Eq.~\ref{eq_t1u0_masses_1}, \ref{eq_t1uo_masses_2}, \ref{eq_t2uo_masses_1}, \ref{eq_t2uo_masses_2}. Here again, the spectrum in (d), (e) and (f) correspond to the spectrum at the points $A$, $C$, and $B$ on the sphere.}
    \label{fig_spectrum_on_sphere}
\end{figure*}

In this diagonalised basis, it is easy to see that there are three linearly independent matrices (other than the identity matrix) that commute with the $\mcl{D}$ matrix in Eq.~\ref{eq_u1tu1tu1}. These are this matrix, $\mcl{D}$, itself and
\begin{eqnarray}
\begin{pmatrix}
\sigma_3 & 0  \\
0  & 0
\end{pmatrix}\otimes \zeta_0,\quad
\begin{pmatrix}
0 & 0   \\
0 &\sigma_3
\end{pmatrix}\otimes \zeta_0.
\label{eq_u1tu1}
\end{eqnarray}

The above three matrices generate the residual U(1) $\otimes$ U(1) $\otimes$ U(1) symmetry on generic points on the sphere in the middle panel of Fig.~\ref{fig_spectrum_on_sphere} like $C$ and $D$.  The first U(1) results in conserved flavour (spin-octupole) currents along the NNN bonds with a flow pattern similar to that shown in Fig. \ref{fig:chern_insulator_hopping}. For such generic points, the fermions are fully gapped with each gapped band being 4-fold degenerate (spectrum (b) in the top panel of Fig.~\ref{fig_spectrum_on_sphere}). We can calculate the edge response, which is given by a mutual CS action similar to Eq. \ref{eq:octupolar_hall_action}. {These spin-octupole filtered edge modes are again protected by the U(1)$\rtimes$ Z$_{\rm 2}^{TR}$ as in the case of $\mcl{T}^e_{1g}$ mass discussed above.} 

Interestingly on putting two of the $\Delta_i$s to zero such as the point $A$ in Fig. \ref{fig_spectrum_on_sphere} (middle panel), while the above conclusions survive, the gapped bands have an enhanced 8-fold degeneracy (as shown in (a) of the top panel of Fig. \ref{fig_spectrum_on_sphere}) due to enhanced residual flavour symmetry of U(1) $\otimes$ SO(4). From the perspective of Eq. \ref{eq_u1tu1tu1}, the numbers $a_1$ and $a_2$ becomes equal at these points such that we can further basis transform $\mcl{D} \rightarrow (\sigma_3\otimes \sigma_0)\otimes \zeta_0$. Now there are six generators in addition to $\mcl{D}$, that commute with the mass which is given by
\begin{eqnarray}
\begin{pmatrix}
\sigma_i & 0  \\
0  & \sigma_i
\end{pmatrix}\otimes \zeta_0,\quad
\begin{pmatrix}
\sigma_i & 0   \\
0 &-\sigma_i
\end{pmatrix}\otimes \zeta_0.
\end{eqnarray}
with $i=1,2,3$. This generates SU(2)$\otimes$ SU(2) $\equiv$ SO(4) in addition to the  U(1) generated by $\mcl{D}$ itself.

A much more interesting situation arises when one moves from point $C$ (or $D$) to point $B$ (in the middle panel of Fig. \ref{fig_spectrum_on_sphere}) which is characterised by
\begin{eqnarray}\label{eq_gapless_point_condition}
|\Delta_1| = |\Delta_2| = |\Delta_3|,
\end{eqnarray}
and is one of the eight isolated special points. 

At these points, $a_2$ in Eq. \ref{eq_u1tu1tu1} becomes zero. We have assumed $a_1>a_2$ without any loss of generality. As a result four flavours of fermions belonging to the $a_2$ block become gapless while four others belonging to the $a_1$ block remain gapped. This leads to a partially gapped state. The resultant spectrum is shown in (c) of the top panel in Fig. \ref{fig_spectrum_on_sphere}. It is clear that at this special points when $a_2=0$, in addition to $\mcl{D}$ and the two matrices given by Eq. \ref{eq_u1tu1}, two additional matrices  
\begin{eqnarray}
\begin{pmatrix}
0  & 0  \\
0 & \sigma_1
\end{pmatrix}\otimes \zeta_0,\quad
\begin{pmatrix}
0  & 0  \\
0  & \sigma_2
\end{pmatrix}\otimes \zeta_0
\end{eqnarray}
also commute with the mass matrix at this special point. The above two matrices along with the last one of Eq.~\ref{eq_u1tu1} generate a SU(2) such that at these isolated points the symmetry is given by U(1) $\otimes$ U(1) $\otimes$ SU(2) and it is this last SU(2) which protects a subset of gapless Dirac fermions. On moving away from these special points infinitesimally, the SU(2) is broken down to U(1) as $a_2\neq 0$ and this gaps out the remaining fermions (spectrum (b) in the top panel of Fig. \ref{fig_spectrum_on_sphere}). 

The existence of such isolated gapless points is surprising and different from the usual incompatible masses such as the chiral masses~\cite{ryu2009masses}. In the case of chiral masses (Sec. \ref{subsec_group1}) in moving from the CDW masses to the ICI mass, one encounters an {\it unavoidable} line of bulk gap closing corresponding to a phase transition, across which the level of Chern-Simons term change. However, in the present case one can conceive two different classes of lines on the sphere joining the same two gapped end-points (C and D) as shown on the sphere in the middle panel of Fig.~\ref{fig_spectrum_on_sphere}, one not passing through the special point (the black path) and the other passing through the special point, B (the red path).

For the second path, one would naively conclude that the system goes through a {\it phase transition} via a critical point with higher symmetry. The situation can be understood by going back to the six inversion odd masses in Eq. \ref{eq_so5-6} and reminding ourselves that the six masses in $\mcl{T}_{1u}^e$ and $\mcl{T}_{2u}^e$ are mutually incompatible and together form a reducible representation  (Eqs. \ref{eq_so5-6-31} and \ref{eq_so5-6-32}) of a U(1)~$\otimes$~SU(2) subgroup of SO(5). A generic linear combination of the six masses in this case can be represented as points on the surface of a 5-dimensional sphere, $\mathcal{S}^5$, by extending Eq. \ref{eq_gen_mass_on_sphere} to all the six masses.  On this $\mathcal{S}^5$ due to the incompatibility, there are extended lower dimensional regions of parameter space where the fermions are partially gapless that separates the fully gapped regions as schematically depicted in Fig. \ref{fig_projection_from_S5_to_S2}. Projection of the gapless regions from $\mathcal{S}^5$ to $\mathcal{S}^2$ for the above two triplets results in the isolated {\it special} points. This is most easily seen by sitting at one of the special partially gapless points on the $\mathcal{S}^2$ for a particular triplet (say $\mcl{T}^e_{1u}$) and performing the U(1) $\otimes$ SU(2) transformation generated by Eq. \ref{eq_u145} and \ref{eq_so5-su2} as discussed above. The resultant mass necessarily involves the other triplet, $\mcl{T}^e_{2u}$ and hence does not lie on the $\mathcal{S}^2$ anymore but on the gapless manifold of $\mathcal{S}^5$. This is schematically shown in Fig. \ref{fig_projection_from_S5_to_S2}. Thus the special point B is the projected image of the gapless manifold on $\mathcal{S}^5$ to $\mathcal{S}^2$ and the two classes of paths between C and D mentioned above have a natural interpretation on $\mathcal{S}^5$ where the black (red) path avoiding (touching) the special isolated point corresponds to paths on $\mathcal{S}^5$ that lies within a single gapped phase but avoids (touches) the gapless manifold. Very importantly, the special point is mandated to exist under the microscopic symmetries  such that a system tuned to pass  through the special point B undergoes an {\it unnecessary phase transition}~\cite{PhysRevX.9.021034}. In this sense, the special points can be thought of as examples of {\it symmetry enforced  unnecessary multi-critical points}.

\begin{figure*}
    \centering
    \includegraphics[scale=0.75]{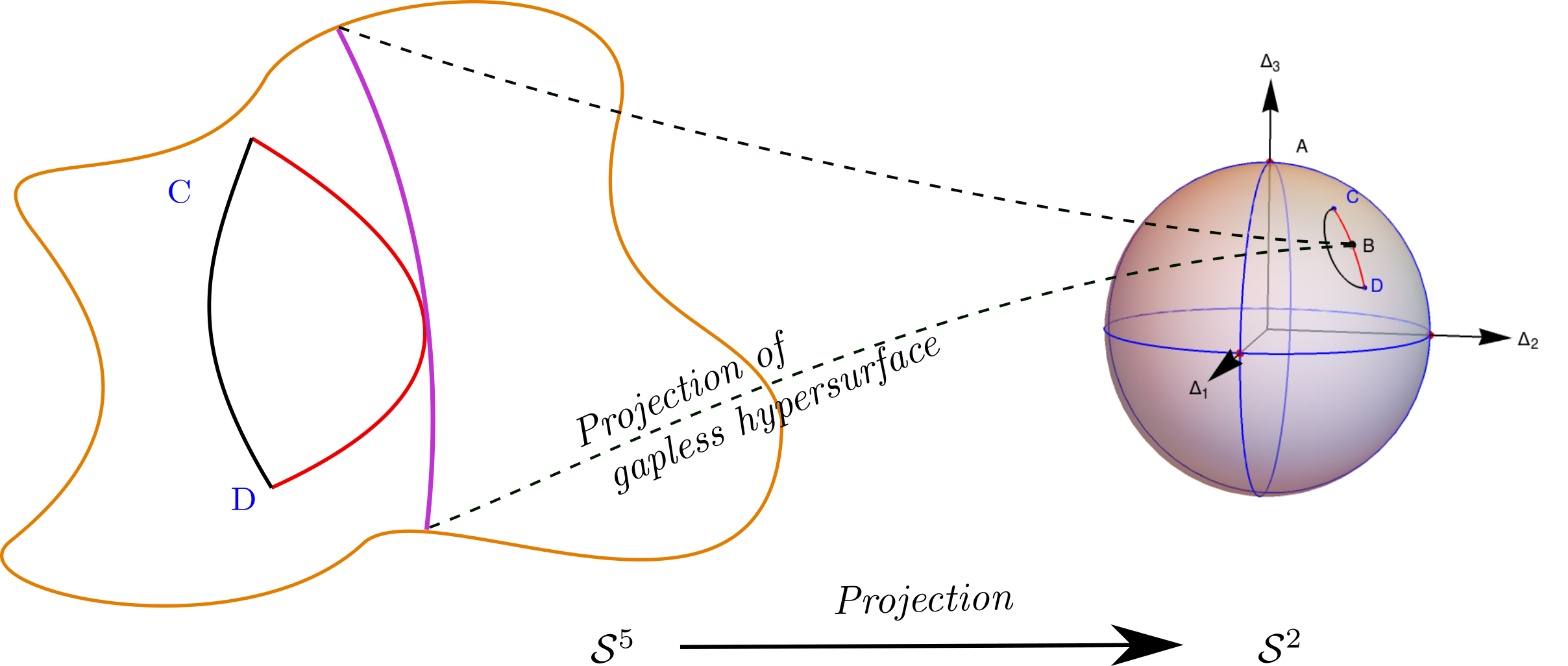}
    \caption{The six masses in Eq. \ref{eq_so5-6} are mutually incompatible. A generic linear combination of such masses (similar to Eq. \ref{eq_gen_mass_on_sphere} extended to the six masses) can be represented on a five-dimensional sphere, $\mathcal{S}^5$, as shown in the left-hand figure which has extended gapless critical hyper-lines (in magenta) separating two different phases described by the two triplets $\mcl{T}_{1u}^e$ and $\mcl{T}_{2u}^e$ (Eqs. \ref{eq_so5-6-31} and \ref{eq_so5-6-32}). The pink line on $\mcl{S}^5$ represents a hypersurface on which the fermionic spectrum is gapless (Fig. \ref{fig_spectrum_on_sphere}(c,f)). This hypersurface projects to the point $B$ on $\mcl{S}^2$ as shown by the dotted lines. The images of the two paths from C to D are also shown  in  $\mcl{S}^5$. }
    \label{fig_projection_from_S5_to_S2}
\end{figure*}

\subsection{Quantum spin-quadrupole Hall insulators}
\label{subsec_qsqp}

Turning to the five TR odd masses that form the $\mcl{E}^o_u$  doublet (Eq. \ref{eq_so5-2}) and $\mcl{T}^o_{2g}$ triplet (Eq. \ref{eq_so5-3}), the respective masses are given by

\begin{align}
    \mcl{E}^o_u~:~\left\{\begin{array}{c}
    -i\langle\bar{\chi}\Sigma_{4}\chi\rangle\\
    -i\langle\bar{\chi}\Sigma_{5}\chi\rangle\\
    \end{array}\right.
    \label{eq_flavour-3}
\end{align}
and
\begin{align}
    \mcl{T}^o_{2g}~:~\left\{\begin{array}{c}
    -i\Braket{\bar{\chi}\Sigma_{1}\chi}\\ -i\Braket{\bar{\chi}\Sigma_{2}\chi}\\ 
    -i\Braket{\bar{\chi}\Sigma_{3}\chi}\\
    \end{array}\right.
    \label{eq_flavour-4}
\end{align}

While  the doublet (Table \ref{tab_def_of_2d_irreps}) is odd under inversion symmetry and does not break lattice translation, the triplet (Table \ref{tab_def_of_3d_irreps}) is even under inversion, but breaks lattice translation. Transformation under other lattice symmetries is given in respective tables. Further in terms of the IR symmetries,  each of the two classes breaks SO(5) down to U(1) $\otimes$ SU(2) as mentioned above (Eq. \ref{eq_u145} and \ref{eq_so5-su2}) -- the doublet (triplet) is a U(1) (SU(2)) singlet. 

The two classes of masses fully gap out the fermionic spectrum and break the SU(4) flavour symmetry. Notably, in terms of the spin operators we have 
\begin{align}
\Sigma_4&=\frac{1}{\sqrt{3}}(J_x^2-J_y^2)~~{\rm and}~~~\Sigma_5=J_z^2-\frac{5}{4}.
\label{eq_quadrupole}
\end{align}
for the doublet and
\begin{align}
    \Sigma_1&=(J_yJ_z+J_zJ_y)/\sqrt{3};\nonumber\\
    \Sigma_2&=(J_zJ_x+J_xJ_z)/\sqrt{3}\nonumber\\
    \Sigma_3&=(J_yJ_x+J_xJ_y)/\sqrt{3}.
\end{align}
for the triplet, all of which correspond to spin-quadrupoles. In fact, as we show below, the two corresponds to different spin-quadrupole Hall phase protected by U(1) symmetry. Such a phase is an interesting generalisation of the QSH phase as the quadrupole Hall phase is TR odd, unlike the QSH phase. This can be traced to the fact that  unlike the spin-dipole and the spin-octupole currents, the spin-quadrupole currents are odd under TR.

The resultant non-zero Hall response can be obtained by performing a calculation similar to section~\ref{subsec_group2}. Sitting deep inside the gapped phase with $\braket{-i\bar{\chi}\Sigma_{4}\chi}\neq0$ (say), we can  introduce a spin-quadrupole probe gauge field $\mbf{A}_q$ in addition to a probe charge field, $\mbf{A}_c$,  and integrating out the fermions results in a mutual CS action 
\begin{eqnarray}\label{eq:quadrupolar_hall_action}
S^{mutual}_{CS}&& = i\frac{N_F}{2\pi} ~sgn(\Delta)~  \int d^3x~\epsilon^{\mu\nu\lambda} A_{c,\mu}\partial_{\nu}A_{q,\lambda}.
\end{eqnarray}
Such that there is a spin-quadrupolar edge current (corresponding to $\Sigma_4$). As in the case of the previously discussed spin-octupolar Hall phases, the presence of edge modes in the case of this mass can be confirmed by taking a mean-field lattice Hamiltonian and performing a band structure calculation on a finite-sized lattice. Similar results can be obtained for the triplet.

Focusing on the $\mathcal{E}^o_u$ doublet, we note that it breaks the SU(4) flavour symmetry to U(1) $\otimes$ SO(4). E.g., for $-i\braket{\bar{\chi}\Sigma_4 \chi} \neq 0$, the U(1) is generated by $\Sigma_4$ and the SO(4) is by $\{\Sigma_{12}, \Sigma_{13}, \Sigma_{15}, \Sigma_{23}, \Sigma_{25}, \Sigma_{35}\}$. Since the two-component order parameter (Eq. \ref{eq_flavour-3}) lives on a circle it supports point defects-- vortices characterized by the winding number. More precisely, consider the mass term
\begin{align}
    -i\left(\Delta_1\bar\chi\Sigma_4\chi+\Delta_2\bar\chi\Sigma_5\chi\right)
\end{align}
such that under the U(1) transformation generated by $\Sigma_{45}$ for an angle $\theta$,
\begin{align}
    \Delta=\Delta_1+i\Delta_2\rightarrow \Delta e^{i\theta}
\end{align}

At the core of such a vortex, the TRS is restored, and hence for a {\it fat} vortex with a sizeable core one expects  quadrupole-filtered zero modes around the vortex core. Further following each such unit vortex is expected to trap $N_F/2$ quanta of electronic charge~\cite{ryu2009masses,PhysRevLett.100.110405}. The transition mediated by the proliferation and condensation of such vortices are then expected to be novel~\cite{ran2008direct,ryu2009masses} and requires further understanding.


\section{Group-3 : The mixed masses}
\label{subsec_group3}

We now turn to the structure of the mixed masses which are obtained by nontrivial contributions from both the flavour and chiral sectors. The complex structure of the mass matrices and the intricate locking of the spin and real space symmetry transformations result in the rich properties of the resultant phases which we now discuss in detail. There are 45 masses divided into 19 irreducible representations summarised in Eq. \ref{eq:decomposition_4} and they give rise to 16 different phases. These are generic density wave phases which can be divided into two sub-sets depending on whether the fermions are generically fully gapped (insulators) or partially gapped (semimetals). Two of the insulators, have edge modes whose signature is evident from appropriate Chern-Simons terms. In most of the insulators and semimetals, the components of some of the multiplets are incompatible and hence they lead to gapless sub-manifold as the components of the masses are tuned (similar to the spin-octupole flavour triplet discussed in Subsection \ref{subsec_flavour_special}). 

\subsection{Density wave Insulators}

There are 27 such mass terms divided into two singlets ($\mcl{A}^e_{1g}, \mcl{A}^o_{2g}$), two doublets (2$\mcl{E}^o_u$) and seven triplets ($\mcl{T}_{2g}^o$,$\mcl{T}^e_{1u}, \mcl{T}^e_{2u}, 2\mcl{T}^o_{1u}, 2\mcl{T}^o_{2u}$). While the singlets and three triplets ($\mcl{T}_{2g}^o$,$\mcl{T}^e_{1u}, \mcl{T}^e_{2u}$) give rise to {\it five} distinct phases, the two doublets and the other two triplets with a multiplicity of two, {\it i.e.},  ($2\mcl{T}^o_{1u}, 2\mcl{T}^o_{2u}$) only give rise to {\it three} distinct phases since members of the same representation can be mixed without breaking any further symmetries. {Thus they give rise to a total of {\it eight} distinct flavour density wave insulating phases -- two of which have edge modes.}

\subsubsection{Ising ferro spin-quadrupolar insulator}\label{sec:ferro-spinQuadru-insulator}

The TR even mass for the $\mcl{A}_{1g}^e$ lattice singlet in Eq. \ref{eq:T1g_T1g_direct_product} is given by
\begin{align}\label{eq_a1ge_mass_2}
\Delta = -i\Braket{\bar{\chi} \left(\Sigma_{3}\zeta_{1} - \Sigma_{1}\zeta_{3} - \Sigma_{2}\zeta_{2} \right)\chi}/{\sqrt{3}}
\end{align}
While it is a lattice singlet, it breaks the flavour SU(4) down to U(1) (generated by $\Sigma_{45}$) and the chiral SU(2) down to $Z_2$.

This mass corresponds to a uniform {\it ferro} ordering in the spin-quadrupole density (in the global basis)
\begin{align}
\Sigma_1 + \Sigma_2 + \Sigma_3 = \frac{1}{\sqrt{3}} \left( \{J_x,J_y\} + \{J_y,J_z\} + \{J_z,J_x\}  \right),
\end{align}
as can be explicitly checked starting with the underlying lattice fermion bilinear similar to the case of CDW (Eq. \ref{eq_scdw}). { In addition, this singlet supports non-zero quantized spin-octupolar Hall response somewhat similar to that of the $\mcl{A}_{1g}^e$ mass in Eq.~\ref{eq_a1ge_mass}. To understand this, we write an action similar to that in  Eq.~\ref{eq:octupolar_gauge_action} and integrate out the fermions. This produces a mutual CS action of the form
\begin{eqnarray}
    S^{mutual}_{CS} = i \frac{N_F}{2} \frac{1}{2\pi} sgn(\Delta) \int d^3x \epsilon^{\mu \nu \lambda} A_{c,\mu} \partial_\nu A_{o,\lambda}.\nonumber \\
    \label{eq_cs_mutualising}
\end{eqnarray}
Here $N_F=4$ is the number of fermions flavors and $A_{c,\mu}$, $A_{o,\mu}$ are respectively electromagnetic and spin-octupole probe gauge fields as used in Eq.~\ref{eq_gauge_inv_derivative}. Thus, this mass too produces quantum spin-octupolar Hall response, but the CS level is half compared to that for the mass in Eq.~\ref{eq_a1ge_mass}.} The resultant counter-propagating edge modes (not shown) can be obtained for appropriate lattice models. These edge modes are protected by the microscopic time-reversal symmetry ($\mbb{T}$). 
Hence this corresponds to a gapped Ising ferro spin-quadrupolar phase with counter-propagating spin-octupole filtered edge modes. 

\subsubsection{Ising ferro spin-octupolar insulator}
\label{subsubsec:IsingFerroSpinOctopolar}

Similarly, the TR odd mass for the $\mcl{A}_{2g}^o$ lattice singlet in Eq.~\ref{eq:T2g_T1g_direct_product} given by 
\begin{align}\label{eq:ferro-spin-octupole-singlet}
\tilde{\Delta} = -i\Braket{\bar{\chi} \left(\Sigma_{12}\zeta_{1} - \Sigma_{23}\zeta_{3} + \Sigma_{13}\zeta_{2} \right)\chi}/{\sqrt{3}}
\end{align}
corresponds to uniform ordering for the spin-octupole density in 
\begin{align}
\Sigma_{12} - \Sigma_{13} + \Sigma_{23} = \frac{7}{3} \left(J_x+J_y + J_z\right) - \frac{4}{3} \left( J_x^3 + J_y^3 +J_z^3 \right)
\end{align}
However, unlike the above ferro spin-quadrupolar order, this breaks the flavour SU(4) down to SU(2) (generated by $\{\Sigma_4,\Sigma_5, \Sigma_{45}\}$) and is also odd under $\mbf{C_2'}$. The chiral SU(2) on the other hand is broken down to $Z_2$ similar to the ferro spin-quadrupolar case.

The above singlet leads to non-zero quantum Hall response in the presence of an external electromagnetic field. This can again be understood by writing an action of the form as in Eq.~\ref{eq_dirac_massiveaction} in the presence of an electromagnetic gauge field $A_{c,\mu}$ and integrating out the fermions.  This produces an effective action given by
\begin{eqnarray}
    S_{CS} = i \frac{N_F}{2} \frac{1}{4\pi} sgn(\tilde\Delta) \int d^3x \epsilon^{\mu \nu \lambda} A_{c,\mu} \partial_\nu A_{c,\lambda}.
    \label{eq_cs_anomaloushall}
    \end{eqnarray}
The CS level of this action is half compared to that for the ICI mass (Eq.~\ref{eq_icilag}) and hence represents a generalisation of an anomalous Hall insulator.

\subsubsection{Staggered (``N\'eel") spin-octupolar insulator}\label{sec:Neel-spinOctupole-insulator}

The four mixed masses that make up the two $\mcl{E}^o_u$ doublets in Eqs. \ref{eq:T1u_T1g_direct_product} and \ref{eq:T2u_T1g_direct_product} are respectively comprised of
\begin{align}\label{eq_euo_mass_2}
&-i\left<\bar{\chi} \frac{ 4\Sigma_{35}\zeta_1 - (\Sigma_{15}-\sqrt{3}\Sigma_{14})\zeta_3 - (\sqrt{3}\Sigma_{24}+\Sigma_{25})\zeta_2 }{2\sqrt{6}} \chi \right>, \nonumber\\
&-i\left<\bar{\chi} \frac{ (\Sigma_{15}-\sqrt{3}\Sigma_{14})\zeta_3 - (\Sigma_{25}+\sqrt{3}\Sigma_{24})\zeta_2 }{2\sqrt{2}} \chi \right> \nonumber\\
\end{align}

and

\begin{align}\label{eq_euo_mass_3}
    &-i\left<\bar{\chi} \frac{ 4\Sigma_{34}\zeta_1 - (\Sigma_{14}+\sqrt{3}\Sigma_{15})\zeta_3 + (\sqrt{3}\Sigma_{25}-\Sigma_{24})\zeta_2 }{2\sqrt{6}} \chi \right>,\nonumber\\
    &-i\left<\bar{\chi} \frac{ -(\Sigma_{14}+\sqrt{3}\Sigma_{15})\zeta_3 + (\Sigma_{24}-\sqrt{3}\Sigma_{25})\zeta_2 }{2\sqrt{2}} \chi \right>. \nonumber\\
\end{align}

The above four masses in the two doublets can be rotated into each other using a U(1) symmetry generated by $\Sigma_{45}$.  In particular, if $m_1 (m_1')$ and $m_2 (m_2')$ are the components of Eq. \ref{eq_euo_mass_2} (\ref{eq_euo_mass_3}), then the linear combinations $m_1^\pm= m_1 \pm im_1'$ and $m_2^\pm=m_2 \pm im_2'$ transform as one-dimensional representations of the above U(1). Hence they describe the same phase. 

These masses describe a fully gapped two-sublattice staggered ordering (as shown in Fig.~\ref{fig_Neel_pattern}) in the following spin-octupole operator respectively (whose representation in terms of spin operators are readily obtained using Appendix \ref{appen_gamma})
\begin{eqnarray}\label{eq:neel_density_1}
&&2\Sigma_{35} + \frac{\Sigma _{15}-\sqrt{3} \Sigma _{14}}{2}+\frac{\sqrt{3} \Sigma _{24}+\Sigma _{25}}{2},\nonumber\\
&&
\frac{\sqrt{3} \Sigma _{14}-\Sigma _{15}}{2} + \frac{\sqrt{3}\Sigma _{24}+\Sigma _{25}}{2}.
\end{eqnarray}
and 
\begin{eqnarray}\label{eq:neel_density_2}
&&2\Sigma_{34} + \frac{\Sigma _{14}+\sqrt{3} \Sigma _{15}}{2}+\frac{ \Sigma _{24}-\sqrt{3}\Sigma _{25}}{2},\nonumber\\
&&
\frac{ \Sigma _{14}+\sqrt{3}\Sigma _{15}}{2} - \frac{\Sigma _{24}-\sqrt{3}\Sigma _{25}}{2}.
\end{eqnarray}
This can be checked starting with the appropriate lattice bilinears similar to Eq.~\ref{eq_odw}.

\begin{figure}[h!]
    \includegraphics[scale=0.6]{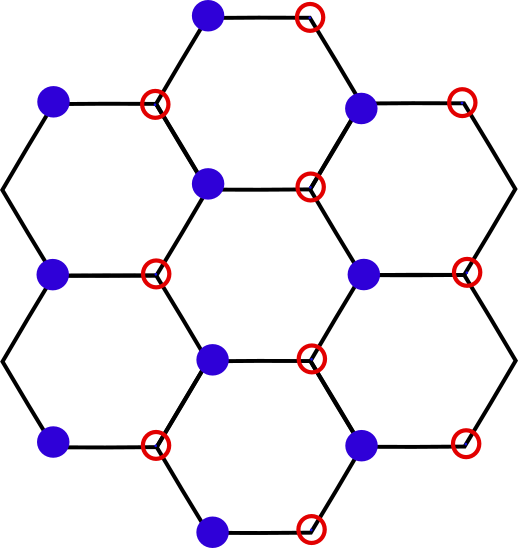}
    \caption{Density pattern for the staggered spin-octupole density waves corresponding to Eqs. \ref{eq_euo_mass_2} and \ref{eq_euo_mass_3}. The red circles and blue dots here represent opposite densities in the spin-octupole operators given by Eqs.~\ref{eq:neel_density_1}, \ref{eq:neel_density_2}.}
    \label{fig_Neel_pattern}
\end{figure}

A remarkable difference of the above sub-lattice staggered spin-octupolar orderings compared to {\it N\'eel} state in SU(2) spin-rotation invariant graphene is that the latter are given one-dimensional representations, $\mcl{A}_{1u}^o$,~\cite{ryu2009masses} under lattice transformations 
while transform as a O(3) vector under spin rotations. In the present case, due to SOC, we have doublets that transform non-trivially under both SU(4) and lattice symmetries. In fact, this allows for non-trivial quantum numbers for the vortices of the resultant doublet masses which forms an interesting avenue to explore in the future.

\subsubsection{Stripy spin-octupole density wave insulator}\label{sec:stripy-spinOctupole-insulator}

The three masses that form the $\mcl{T}_{2g}^o$ triplet (in Eq. \ref{eq:A2g_T1g_direct_product})  are given by

\begin{align}\label{eq_sigma45_density_waves}
   -i\langle\bar\chi\Sigma_{45}\zeta_i\chi\rangle\neq 0
\end{align}
with $i=1,2,3$.  Each mass in this triplet breaks down the flavor SU(4) to U(1) $\otimes$ SU(2) and the chiral SU(2) to U(1).

The transformation properties of the triplet components (see Table \ref{tab_def_of_3d_irreps}) under lattice symmetries is completely determined by $\zeta_i$ as $\Sigma_{45}$ is a lattice singlet (see Table \ref{tab:SU4_space_irreps} and \ref{tab:real_space_irreps}). Thus they are very similar to the triplet mass in group-1 (Eq. \ref{eq_su4_cdw}). However, unlike the CDW, here the density modulation occurs in the spin-octupole moment, {\it i.e.}, 
\begin{eqnarray}
\Sigma_{45} = -\frac{4}{3\sqrt{3}} \left(J_xJ_yJ_z + J_yJ_zJ_x + J_zJ_xJ_y - \frac{15i}{8} \right).\nonumber\\
\end{eqnarray}
Hence they are nothing but stripy spin-octupole density wave as shown in Fig. \ref{fig:stripy_CDW} with the modulation being in the spin-octupole density. This can be seen explicitly by looking at the low energy projection of the microscopic on-site spin-octupole density operator. Similar to the CDW case (Eq.~\ref{eq_scdw}), here we have
\begin{eqnarray}
:\psi^\dagger(\boldscriptr_\mcl{S})\Sigma_{45}\psi(\boldscriptr_\mcl{S}): ~=\begin{cases}
-i\bar{\chi}\Sigma_{45}\zeta_1\chi &\text{ For $\mcl{S} = A_2,B_2$ } \\
i\bar{\chi}\Sigma_{45}\zeta_1\chi &\text{ For $\mcl{S} = A_1,B_1$ } 
\end{cases}\nonumber\\
\label{eq_odw}
\end{eqnarray}

The two other members of the triplet describe stripy order along the other two directions rotated by $\pm 2\pi/3$ with respect to Fig. \ref{fig:stripy_CDW}.

\subsubsection{Zig-zag spin-quadrupole density wave insulator}\label{sec:zigzag-spinQuadru-insulator}
 
 Eq. \ref{eq:Eu_T1g_direct_product} consists of six masses of the form 
 \begin{eqnarray}\label{eq_quadrupole_density_waves}
\{-i\Braket{\bar{\chi}\Sigma_4\zeta_i \chi},~-i\Braket{\bar{\chi}\Sigma_5\zeta_i \chi}\},
\end{eqnarray}
for $i=1,2,3$. Under the action of the lattice symmetries, these six masses form two triplets with representations $\mcl{T}_{1u}^e$ and $\mcl{T}_{2u}^e$ which leads to two different types of spin-quadrupole density wave phases (Eq. \ref{eq_quadrupole}) which we discuss below. 

\paragraph{$\mcl{T}_{1u}^e$ masses~:} The $\mcl{T}_{1u}^e$ masses are given by
\begin{eqnarray}\label{eq_t1ue_masses}
&&-i\Braket{\bar{\chi} \Sigma_5 \zeta_1 \chi},\nonumber\\ &&-\frac{i}{2}\Braket{\bar{\chi} (-\sqrt{3}\Sigma_4+\Sigma_5)\zeta_3 \chi},\\ &&-\frac{i}{2}\Braket{\bar{\chi} (-\sqrt{3}\Sigma_4-\Sigma_5)\zeta_2\chi}.\nonumber
\end{eqnarray}
The relation between the first mass and the underlying $j=3/2$ orbitals is given by
\begin{eqnarray}\label{eq_sigma5_operator_lattice_to_continuum}
:\psi^\dagger(\boldscriptr_\mcl{S})\Sigma_{5}\psi(\boldscriptr_\mcl{S}): ~=\begin{cases}
-i\bar{\chi}\Sigma_{5}\zeta_1\chi &\text{ For $\mcl{S} = A_1,B_2$ } \\
i\bar{\chi}\Sigma_{5}\zeta_1\chi &\text{ For $\mcl{S} = B_1,A_2.$ } 
\end{cases}\nonumber\\
\end{eqnarray}
Notice the difference in the sign for the different sub-lattices compared to Eq. \ref{eq_scdw} and \ref{eq_odw}. Unlike in these earlier cases, Eq. \ref{eq_sigma5_operator_lattice_to_continuum} represents {\it zig-zag} pattern of spin-quadrupolar density wave as shown in Fig. \ref{fig:zig_zag_density_wave} which corresponds to spin-quadrupole order in $\Sigma_5$.  The other two masses are also zig-zag density waves of the $\frac{1}{2}\left(\sqrt{3}\Sigma_{4}-\Sigma_5 \right), ~\frac{1}{2}\left(\sqrt{3}\Sigma_{4}+\Sigma_5\right)$ operators whose patterns are rotated by $\pm 2\pi/3$ with respect to Fig. \ref{fig:zig_zag_density_wave}.  
\begin{figure}[h!]
	    \centering
	    \includegraphics[scale=0.6]{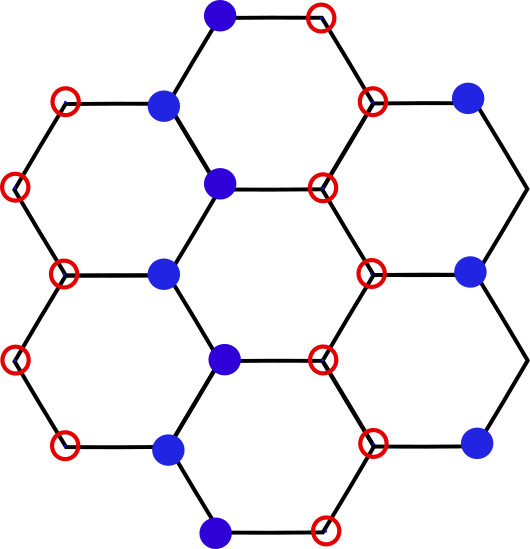}
	    \caption{Zig-zag Density wave pattern corresponding to the masses of Eqs. \ref{eq_t1ue_masses}, \ref{eq_t2ue_masses}, \ref{eq_t1u0_masses_1}, \ref{eq_t1uo_masses_2}, \ref{eq_t2uo_masses_1} and \ref{eq_t2uo_masses_2}. The blue dots and the red circles represent opposite densities of spin quadrupole operator $\Sigma_5$.}
	    \label{fig:zig_zag_density_wave}
	\end{figure}
	
\paragraph{$\mcl{T}_{2u}^e$ masses~:} 
The masses in the $\mcl{T}_{2u}^e$ triplet are given by
\begin{eqnarray}\label{eq_t2ue_masses}
&&-i\Braket{\bar{\chi} \Sigma_4 \zeta_1 \chi},\nonumber\\ 
&&-\frac{i}{2}\Braket{\bar{\chi} (\Sigma_4+\sqrt{3}\Sigma_5)\zeta_3 \chi},\\
&&-\frac{i}{2}\Braket{\bar{\chi} (\sqrt{3}\Sigma_5-\Sigma_4)\zeta_2\chi}\nonumber
\end{eqnarray}
Similar to $\mcl{T}_{1u}^e$ triplet, these are also zig-zag density waves but of different spin-quadrupole operators, namely
\begin{eqnarray}
\Sigma_4, ~\frac{1}{2}\left(\sqrt{3}\Sigma_{5}+\Sigma_4 \right), \frac{1}{2}\left(\sqrt{3}\Sigma_{5}-\Sigma_4\right),
\end{eqnarray}
respectively.

The members of both the above triplets are incompatible and hence generically one expects gapless points when tuning among different components of the masses similar to the $\mcl{T}_{1u}^e$ and $\mcl{T}_{2u}^e$ masses in Eq.~\ref{eq_flavour-21} and Eq.~\ref{eq_flavour-22}. In fact, similar to that case, we can consider labeling the linear combination of the six masses in Eq. \ref{eq_quadrupole_density_waves} as points on a 5-dimensional sphere, $\mcl{S}^5$. The points on this $\mcl{S}^5$ are closed under the action of the U(1) $\otimes$ SU(2) subgroup generated by $\{\Sigma_{45},\zeta_1,\zeta_2,\zeta_3\}$. Then arguments similar to those discussed for the $\mcl{T}_{1u}^e$ and $\mcl{T}_{2u}^e$ masses (in subsection \ref{subsec_flavour_special}) hold in the present case. 

The generic linear combination of the present two triplets (given by a form similar to Eq.~\ref{eq_gen_mass_on_sphere}) can then be parametrised on a 2-sphere, $\mathcal{S}^2$ which can be thought as two different projections of the points on  $\mathcal{S}^5$ and all the discussions of Fig. \ref{fig_spectrum_on_sphere} and  \ref{fig_projection_from_S5_to_S2} and the associated discussion also apply to the present case leading to unnecessary multi-critical points.

\subsubsection{ Zig-zag spin-octupole density wave insulator}\label{sec:zigzag-spinOctupole-insulator}

The four triplets (two $\mcl{T}_{1u}^o$ and two $\mcl{T}_{2u}^o$) in Eq.~\ref{eq:T1u_T1g_direct_product} and~\ref{eq:T2u_T1g_direct_product} correspond to two different types of spin-octupolar density wave patterns (of the type given by Fig. \ref{fig:zig_zag_density_wave}). This can be shown by an analysis similar to that in Eq.~\ref{eq_sigma5_operator_lattice_to_continuum}. Note that the two triplets in each of the representations break the same symmetries and hence they are not counted as distinct phases. 

Notably, each triplet is made up of non-compatible members. Thus while the generic linear combination of the masses (Eq. \ref{eq_gen_mass_on_sphere}) gap out all the fermions, there are special linear combinations, similar to  the zig-zag spin-quadrupole density waves (Eq. \ref{eq_quadrupole_density_waves}) where fermions become gapless giving rise to unnecessary multi-critical points.


\paragraph{The $\mcl{T}_{1u}^o$ triplets:} Two such triplets given by Eqs.~\ref{eq:T1u_T1g_direct_product} and Eq.~\ref{eq:T2u_T1g_direct_product} correspond to zig-zag ordering in 
\begin{align}
&\frac{\sqrt{3}\Sigma_{14}-\Sigma_{15}}{2} - \frac{\sqrt{3}\Sigma_{24} + \Sigma_{25}}{2},\nonumber\\ 
& \Sigma_{35} - \frac{\sqrt{3}\Sigma_{24} + \Sigma_{25}}{2},\nonumber\\ 
& \Sigma_{35} + \frac{\sqrt{3}\Sigma_{14}-\Sigma_{15}}{2}, 
\end{align}
and
\begin{align}
	 	&\frac{\sqrt{3}\Sigma_{15}+\Sigma_{14}}{2} + \frac{\sqrt{3}\Sigma_{25} - \Sigma_{24}}{2}, \nonumber\\
	 	& \Sigma_{34} + \frac{\Sigma_{24} - \sqrt{3}\Sigma_{25}}{2},  \nonumber\\
	 	& \Sigma_{34} - \frac{\sqrt{3}\Sigma_{15}+\Sigma_{14}}{2}
	 	\end{align}
respectively. The zig-zag patterns for the first mass of both the triplets are similar to the one shown in Fig.~\ref{fig:zig_zag_density_wave} while that for the other two are obtained by rotating this pattern by $\pm 2\pi/3$.

The fermion bilinear corresponding to the two triplets is given by
\begin{eqnarray}\label{eq_t1u0_masses_1}
&& -i\left<\bar{\chi}\left(\left(\Sigma_{15}-\sqrt{3} \Sigma_{14}\right)\zeta_2+\left(\sqrt{3} \Sigma_{24}+\Sigma_{25}\right)\zeta_3\right)\chi\right>/{2 \sqrt{2}},\nonumber\\
&&-i\left<\bar{\chi}\left(-\left(\sqrt{3} \Sigma_{24}+\Sigma_{25}\right)\zeta_1-2 \Sigma_{35}\zeta_2\right)\chi\right>/{2 \sqrt{2}}, \nonumber\\
&& -i\left<\bar{\chi}\left(\left(\Sigma_{15}-\sqrt{3} \Sigma_{14}\right)\zeta_1+2 \Sigma_{35}\zeta_3\right)\chi\right>/{2 \sqrt{2}}
\end{eqnarray}
for and
\begin{eqnarray}\label{eq_t1uo_masses_2}
&& -i\left< \bar{\chi} \left((-\Sigma_{24}+\sqrt{3}\Sigma_{25})\zeta_{3} + (\Sigma_{14} + \sqrt{3}\Sigma_{15})\zeta_{2}\right)\chi \right>/{2\sqrt{2}},\nonumber \\
&&-i\left<\bar{\chi} \left((-\Sigma_{24}+\sqrt{3}\Sigma_{25})\zeta_{1} + 2\Sigma_{34}\zeta_{2}\right) \chi \right>/{2\sqrt{2}}, \nonumber \\
&&-i\left<\bar{\chi} \left(-(\Sigma_{14}+\sqrt{3}\Sigma_{15})\zeta_{1} + 2\Sigma_{34}\zeta_{3}\right) \chi \right>/{2\sqrt{2}}.
\end{eqnarray} 

For a generic linear combination of the three masses (similar to Eq. \ref{eq_gen_mass_on_sphere}) for each of the two triplets, the degeneracy and the magnitude of the fermionic gap change for different points on the sphere, $\mathcal{S}^2$ (middle panel of Fig. \ref{fig_spectrum_on_sphere}), as members of each triplet are incompatible. The fermionic spectrum corresponding to the three points $A, B$, and  $C$ on the sphere in Fig.~\ref{fig_spectrum_on_sphere} are shown in the bottom panel of the same figure. This clearly shows the change in the degeneracy of each of the fermionic bands as well as the change in the fermionic gap. For a general point on $\mcl{S}^2$ (\textit{e.g.,} points C,D on the sphere in Fig.~\ref{fig_spectrum_on_sphere}), the fermionic spectrum has eight bands, each of which is 2-fold degenerate. 
However, for the points on the great circles obtained by setting one of the $\Delta_i$s to zero (blue circles on the sphere in Fig~\ref{fig_spectrum_on_sphere} which includes the point $A$), the spectrum has four bands and each of these is 4-fold degenerate. 
Finally, for the special eight isolated points given by Eq.~\ref{eq_gapless_point_condition} (such as point $B$ on the sphere in Fig.~\ref{fig_spectrum_on_sphere}), the fermion gap closes partially giving rise to {\it four} gapless fermionic modes while the rest of the bands remain gapped and two-fold degenerate.

The above pattern is best understood by performing a basis transformation (similar in spirit to Eq. \ref{eq:chitilde}) which allows useful insights into the breaking of the SU(8) symmetry by the above masses. We explicitly discuss this for the first triplet given by Eq. \ref{eq_t1u0_masses_1}. This basis transformation is defined by  
\begin{eqnarray}
\chi'' = U''\chi,
\label{eq_chidtbasis}
\end{eqnarray}
where
\begin{eqnarray}
U'' = \Sigma_0\otimes\begin{pmatrix}
\sigma_0\\
&i\sigma_2
\end{pmatrix}.
\end{eqnarray}
The form of the Dirac Matrices in this new basis is $\gamma_0'' =\Sigma_0\tau_0\sigma_3 ,\quad \gamma_1'' = \Sigma_0\tau_0\sigma_2,\quad \gamma_2'' = -\Sigma_0\tau_0\sigma_1$ such that the SU(8) generators ($\mcl{P}$ in Eq. \ref{eq_su8_generators}), that commute with the Dirac matrices, in the transformed basis, must have the form :
\begin{eqnarray}
\mcl{P}'' = \Sigma_a\tau_\beta\sigma_0. 
\end{eqnarray}
The six masses in Eqs. \ref{eq_t1u0_masses_1} and \ref{eq_t1uo_masses_2}, in this new basis have the following form
\begin{eqnarray}\label{eq_chi double prime basis}
 -i\bar{\chi}^{\prime\prime }~\gamma_0^{\prime\prime}(\mcl{R} \otimes\sigma_3 )~\chi'',
\end{eqnarray}
where $\mcl{R}$ are $8\times 8$ Hermitian matrices. 

The residual subgroup of the SU(8) in presence of these masses can be obtained from the set of $8\times 8$ linearly independent matrices that commute with $\mcl{R}$. As shown in Appendix~\ref{appen_symmetry_group_at_gapless_points}, this yields the following :
\begin{itemize}
    \item At generic points such as $C,D$ on the sphere in Fig.~\ref{fig_spectrum_on_sphere}, the SU(8) symmetry breaks down to U(1)$\otimes$ [U(1) $\otimes$ U(1) $\otimes$ U(1)]$^2$ and there are no zero modes (Fig. \ref{fig_spectrum_on_sphere}(e)).
    \item For the points on the blue great circles (e.g., point $A$), the SU(8) symmetry breaks down to U(1)$\otimes$ [U(1) $\otimes$ SO(4)] $\otimes$ [U(1) $\otimes$ SO(4)] and there are no zero modes but because of the larger residual symmetry, the gapped modes have a higher degeneracy (Fig. \ref{fig_spectrum_on_sphere}(d)).
    \item Finally, at the special points where all the $\Delta_i$ have equal magnitude (e.g., point $B$ in Fig.~\ref{fig_spectrum_on_sphere}), the SU(8) symmetry is broken to U(1) $\otimes$ U(1) $\otimes$ U(1) $\otimes$ U(2) $\otimes$ SO(4). Thus, the isolated gapless points have higher symmetry compared to it's nearby points. This high symmetry preserves a zero block in the $\mathcal{R}$ matrix (Eq.~\ref{eq_specialpta}) and this protects the four gapless fermion modes (Fig. \ref{fig_spectrum_on_sphere}(f)). 
\end{itemize}

\paragraph{The $\mcl{T}_{2u}^o$ triplets:} We now discuss the two $\mcl{T}_{2u}^o$ triplets. The geometric order and the SU(8) symmetry breaking of these masses are similar to the two $\mcl{T}_{1u}^o$ triplets leading respectively to zig-zag ordering of  
\begin{align}
	 	&\frac{\sqrt{3}\Sigma_{14}-\Sigma_{15}}{2} + \frac{\sqrt{3}\Sigma_{24} + \Sigma_{25}}{2}, \nonumber\\
	 	& \Sigma_{35} + \frac{\sqrt{3}\Sigma_{24} + \Sigma_{25}}{2},  \nonumber\\
	 	& \Sigma_{35} - \frac{\sqrt{3}\Sigma_{14}-\Sigma_{15}}{2},  \nonumber\\
	 	\end{align}
	 	and
	 	\begin{align}
	 	&\frac{\Sigma_{14}+\sqrt{3}\Sigma_{15}}{2} + \frac{\Sigma_{24} - \sqrt{3}\Sigma_{25}}{2}, \nonumber\\
	 	& \Sigma_{34} + \frac{\sqrt{3}\Sigma_{25} - \Sigma_{24}}{2},  \nonumber\\
	 	& \Sigma_{34} - \frac{\sqrt{3}\Sigma_{15}+\Sigma_{14}}{2},  \nonumber\\
	 	\end{align}
spin-octupole operators.

The corresponding masses are given by
\begin{eqnarray}\label{eq_t2uo_masses_1}
	&&\left\{  \left< -i\bar{\chi} \frac{-(\Sigma_{25}+\sqrt{3}\Sigma_{24})\zeta_{3} + (\Sigma_{15} - \sqrt{3}\Sigma_{14})\zeta_{2} }{2\sqrt{2}} \chi \right>, \right.\nonumber \\
	&& \left.\quad \left< -i\bar{\chi} \frac{-(\Sigma_{25}+\sqrt{3}\Sigma_{24})\zeta_{1} + 2\Sigma_{35}\zeta_{2} }{2\sqrt{2}} \chi \right>, \right. \nonumber \\
	&& \left.\quad \left< -i\bar{\chi} \frac{(\sqrt{3}\Sigma_{14}-\Sigma_{15})\zeta_{1} + 2\Sigma_{35}\zeta_{3} }{2\sqrt{2}} \chi \right> \right\} \neq 0,
\end{eqnarray}
for the $\mcl{T}_{2u}^o$ triplet in Eq.~\ref{eq:T1u_T1g_direct_product} and 
\begin{eqnarray}\label{eq_t2uo_masses_2}
&&\left\{  \left< -i\bar{\chi} \frac{(\Sigma_{24}-\sqrt{3}\Sigma_{25})\zeta_{3} + (\Sigma_{14} + \sqrt{3}\Sigma_{15})\zeta_{2} }{2\sqrt{2}} \chi \right>, \right.\nonumber \\
&& \left.\quad \left< -i\bar{\chi} \frac{(-\Sigma_{24}+\sqrt{3}\Sigma_{25})\zeta_{1} - 2\Sigma_{34}\zeta_{2} }{2\sqrt{2}} \chi \right>, \right. \nonumber \\
&& \left.\quad \left< -i\bar{\chi} \frac{(\Sigma_{14}+\sqrt{3}\Sigma_{15})\zeta_{1} + 2\Sigma_{34}\zeta_{3} }{2\sqrt{2}} \chi \right> \right\} \neq 0.
\end{eqnarray}
for the $\mcl{T}_{2u}^o$ triplet is given by Eq.~\ref{eq:T2u_T1g_direct_product}.

Similar to the $\mcl{T}^o_{1u}$ triplets discussed above, the SU(8) symmetry breaking for these two triplets depend on the position on the sphere described by the $\Delta_i$s in Eq.~\ref{eq_gen_mass_on_sphere}. In fact, the SU(8) symmetry breaking for these two triplets is the same as that of the $\mcl{T}_{1u}^o$ masses discussed before. 

The existence of the isolated gapless points in all the four above zig-zag triplets  can be understood as {\it unnecessary multi-critical points} as before (see discussion near Eq.~\ref{eq_gapless_point_condition}) within the context of microscopic symmetries and represent the non-trivial embedding of the underlying UV symmetries in the enlarged IR symmetry group. Similar to the case of $\mcl{T}_{1u}^e$ masses discussed above, the existence of such isolated points can be understood as a projection of higher dimensional critical surface on the triplet sphere. Here, however, here we have to consider a 17-dimensional sphere, $\mcl{S}^{17}$ arising from considering an 18-dimensional incompatible vector mass 
comprised of the masses given by Eqs. ~\ref{eq:T1u_T1g_direct_product} and ~\ref{eq:T2u_T1g_direct_product}. This is made up of the four zig-zag triplets along with two $\mcl{E}^o_u$ doublets  the $\mcl{A}^o_{1u}$ and $\mcl{A}^o_{2u}$ singlets. Then, on projecting back to the $\mathcal{S}^2$ spheres spanned by each zig-zag triplet, the isolated points are obtained as a projection of the gapless parts on $\mathcal{S}^{17}$. 

\subsection{Density wave semimetals}\label{subsec_semimetals}

 There are 18 density wave semimetals divided into four triplets ($\mcl{T}^e_{1g},\mcl{T}^o_{1g}, \mcl{T}^e_{2g}, \mcl{T}^o_{2g})$, two doublets ($\mcl{E}^e_g, \mcl{E}^o_g$) and two  singlets ($\mcl{A}^o_{1u}, \mcl{A}^o_{2u}$). The analogs of these semimetals are absent in graphene~\cite{ryu2009masses}. These 18 density wave semimetals can be divided up into two categories depending on the number of gapless fermionic modes which, for the first set is at least {\it four} and the second set is always {\it eight}. Insights into these two sets is best obtained by using the global basis (Eq. \ref{eq:4-component psi}) as discussed in Appendix \ref{appen:globalbasis}. As noted in Section \ref{sec_microscopics}, in the global basis, there are four doubly degenerate Dirac nodes at $\Gamma$, M$_1$, M$_2$, M$_3$ points in the Brillouin zone as shown in \figref{fig:globalBZBands}. To reiterate the crucial aspect, the IR space group does not mix the Dirac spinor at $\Gamma$ point with the other three at the M points, in other words, the former behaves as a ``singlet" and the latter behaves as a ``triplet" as mentioned before. As far as the irreducible masses go, this feature throws up the two categories mentioned above :  (1) Irreducible masses that vanish on for the spinor at the $\Gamma$-point and leave the Dirac cone at $\Gamma$ ungapped -- these are dubbed $\Gamma$-Dirac Semimetals ($\Gamma$-DSM) guaranteeing {\it at least four} gapless Dirac modes which do not depend on the mass parameters, and, (2) the masses that couple the Dirac spinors at each M$_i (i=1,\ldots,3)$ to that at the $\Gamma$-point but the Dirac spinors at the M-points do not directly couple to each other and this guarantees the existence of {\it eight} zero modes-- phases thus realized are dubbed M-Dirac Semimetals (M-DSM). Six masses that makeup two triplets ($\mathcal{T}^e_{2g}, \mathcal{T}^o_{1g}$) correspond to M-DSM that give rise to the stripy spin-quadrupole and spin-octupole density waves. The rest of the 12 masses are of $\Gamma$-DSM type. These consist of two singlets, two doublets and two triplets. The two singlets ($\mcl{A}_{1u}^o, \mcl{A}_{2u}^o$) give rise to staggered spin-octupole density waves, the two doublets ($\mcl{E}_{g}^e, \mcl{E}_{g}^o$) make up, respectively, ferro spin-quadrupole and spin-octupole density waves. The two triplets ($\mathcal{T}^o_{2g}, \mathcal{T}^e_{1g}$) form stripy spin-octupole and spin-quadrupole density waves respectively. Finally, the number of gapless Dirac nodes for the $\Gamma$-DSM can be greater than four for a specific linear combination of masses as discussed below. In Appendix \ref{appen_denwavsem}, we note an interesting structure of the above 18 masses with respect to their transformation under SU(8).

\subsubsection{Staggered (``N\'eel") spin-octupole density wave semimetal}\label{sec:Neel-spinOctupole-semimetal}
 
 The two TR odd masses that form the $\mcl{A}_{1u}^o$ and $\mcl{A}_{2u}^o$ singlet masses in Eqs. \ref{eq:T1u_T1g_direct_product} and \ref{eq:T2u_T1g_direct_product} respectively are given by
     \begin{align}\label{eq_neel_singlet1}
&-i\left<\bar{\chi} \left( \frac{\Sigma_{35}\zeta_{1} }{\sqrt{3}} - \frac{(\sqrt{3}\Sigma_{14}-\Sigma_{15})\zeta_3}{2\sqrt{3}} \right. \right. \nonumber \\
&\left.\left. \qquad \qquad+ \frac{(\sqrt{3}\Sigma_{24} + \Sigma_{25})\zeta_2}{2\sqrt{3}} \right)\chi\right>
\end{align}
and
\begin{align}\label{eq_neel_singlet2}
    &-i\left<\bar{\chi} \left(\frac{\Sigma_{34}\zeta_1}{\sqrt{3}} + \frac{(\Sigma_{14} + \sqrt{3}\Sigma_{15})\zeta_3}{2\sqrt{3}} \right. \right. \nonumber\\
    &\left.\left.\qquad \qquad+ \frac{(\Sigma_{24} - \sqrt{3}\Sigma_{25})\zeta_2}{2\sqrt{3}}\right) \chi \right>.
\end{align}
These represent spin-octupole ordering in 
\begin{align}
&\Sigma_{35} + 	\frac{\sqrt{3}\Sigma_{14} - \Sigma_{15} }{2} - \frac{\sqrt{3}\Sigma_{24} - \Sigma_{25} }{2} = \nonumber\\
&\frac{2}{3\sqrt{3}} \left[ (J_zJ_xJ_x + ~c.p) + (J_yJ_zJ_z +~c.p)+ (J_xJ_yJ_y +~c.p) \right.\nonumber\\
&\left.- (J_zJ_yJ_y + ~c.p) - (J_yJ_xJ_x +~c.p)- (J_xJ_zJ_z +~c.p) \right] 
\label{eq_cpdef}
\end{align}
and 
\begin{align}
& \Sigma_{34} - 	\frac{\Sigma_{14} + \sqrt{3}\Sigma_{15} }{2} - \frac{\Sigma_{24} - \sqrt{3}\Sigma_{25} }{2} = \nonumber \\
&\quad \frac{2}{3}(J_x^3 + J_y^3 + J_z^3) - \frac{13}{6}(J_x+J_y +J_z)
\end{align}
respectively where ``$c.p$" in Eq. \ref{eq_cpdef} refers to all possible cyclic permutations of the operators. The main difference between the two spin-octupolar orders is the fact that the former is odd under reflection, $\mbf{\sigma_d}$ (see Table \ref{tab_def_of_1d_irreps}) while the latter is even under it. Both, however, are odd under inversion. 

\begin{figure}[h!]
    \includegraphics[scale=0.5]{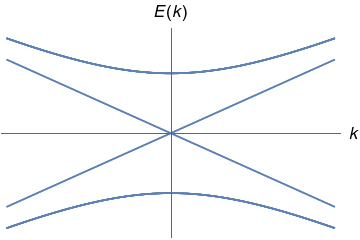}
    \caption{Energy spectrum for the fermions along the $k_x=k_y$ line in presence of either of the singlet masses written in Eq.~\ref{eq_neel_singlet1}, \ref{eq_neel_singlet2}. Each of the gapless bands is two-fold degenerate. So there are four gapless fermion.}
    \label{fig_Neel_singlet_1}
\end{figure}

In either case, the fermionic dispersion is given by Fig.~\ref{fig_Neel_singlet_1} with twelve of the fermionic modes are gapped while the other four are gapless which can be understood from Eq.~\ref{eq_massmatrixfour} discussed below. 
Hence they represent two-sublattice staggered spin-octupolar density wave semimetals of $\Gamma$-DSM type where the symmetry breaking pattern is given by Fig.~\ref{fig_Neel_pattern}.

The gapless fermionic modes are protected by $\mbb{T}\mbf{I}\times$ SU(2)$_{\rm IR}$, where $\mbb{T}({\bf I})$ is the microscopic time reversal (inversion) as given by Table \ref{tab:lat_sym} and SU(2)$_{\rm IR}$ is a subgroup of the emergent SU(8) which is best understood via a basis transformation for the spinors as
\begin{eqnarray}\label{eq_chi tilde basis}
    \tilde{\chi} = U\chi,
\end{eqnarray}
where $U$ is a $16\times 16$ unitary matrix given by Eq.~\ref{eq_U_matrix_for_new_spinor} of Appendix~\ref{appen_partially_gapless_masses}. This transformation relates the low energy Dirac fermions in the local basis (Eq. \ref{eq:localbasis}) with those in the global basis (Eq. \ref{eq:4-component psi} and Appendix \ref{appen:globalbasis}).

The free Dirac Hamiltonian (Eq. \ref{eq_freediracham}) in this new basis is given by
 \begin{eqnarray}
&H_D  &=
 v_F \int d^2x~ \tilde{\chi}^{\dagger}\left( i\mu_0\tilde\Sigma_{23}\partial_x  -i\mu_0\tilde\Sigma_{24}\partial_y \right)\tilde{\chi}.
 \label{eq_freediracchitilde}
\end{eqnarray}
where we have introduced two new set of $4\times 4$ matrices, $\mu_i$ and $\tilde\Sigma_i$ which mixes the flavour and chiral spaces non-trivially. While  the form of the  $\tilde\Sigma_i$ matrices is the same as the $\Sigma_i$ matrices defined in Appendix~\ref{appen_gamma}, unlike the latter, they do not exclusively act on the flavour space. The $\mu_i$ (for $ i=1,\cdots, 15$) matrices, on the other hand, are SU(4) Gell-Mann matrices which are defined in Ref.~\cite{sbaih2013lie} with $\mu_0=\mathbb{I}_4$. The combination of $\mu_i\tilde\Sigma_j$ then gives a {\it new} set of 256 linearly independent $16\times 16$ matrices. Such a combined basis is essential to capture the essence of the mixed masses that we are dealing with, which, in turn, stems from the underlying  SOC. Equivalently the free Dirac Lagrangian in this new basis is given by Eq. \ref{eq_dirac_l} where 
\begin{align}
    \tilde\gamma_0 = -\mu_0\tilde\Sigma_{34},\quad  \tilde\gamma_1 = -\mu_0\tilde\Sigma_{24},\quad \tilde\gamma_2 = \mu_0\tilde\Sigma_{23}.
\end{align}
are the new Dirac matrices.

In this new basis, the mass terms can be written as $-i\langle\bar{\tilde\chi}\tilde m\tilde\chi\rangle$, where $\tilde m$ is a $16\times 16$ Hermitian matrix. More explicitly, the $\mcl{A}^o_{1u}$ (Eq. \ref{eq_neel_singlet1}) and $\mcl{A}^o_{2u}$ (Eq. \ref{eq_neel_singlet2}) masses in this new basis are given by 
\begin{eqnarray}
    -i\Braket{\bar{\tilde{\chi}} \left( \frac{ \mu _8 - \sqrt{2} \mu _{15}}{\sqrt{3}}\tilde\Sigma _{5} -  \frac{\mu_0 + \sqrt{6} \mu _{15}- 2 \mu _3}{2\sqrt{3}}\tilde\Sigma_{15}\right) \tilde{\chi}}, \nonumber\\
    \label{eq_mass1}
\end{eqnarray}
and
\begin{eqnarray}
&&-i\left<\bar{\tilde{\chi}}\left( - \frac{\left(\sqrt{2} \mu _{15}-\sqrt{3} \mu _3+2 \mu _8\right)}{3}\tilde\Sigma_{5} \right. \right. \nonumber\\
&& \qquad \qquad  \left. \left. + ~\frac{\left(-\sqrt{3}\mu_0 + \sqrt{2}\mu _{15}+ 2 \mu
   _8\right)}{2\sqrt{3}}\tilde\Sigma_{15}\right)\tilde{\chi}\right>.\nonumber\\
   \label{eq_mass2}
\end{eqnarray}

The advantage of the new {\it mixed} basis is the fact that when we decompose the 16-component spinor, $\tilde\chi$, into a stack of four 4-component ones as
\begin{eqnarray}\label{eq_4_tilde_modes}
\tilde{\chi} = \left(\tilde{\chi}_1^T,~\tilde{\chi}_2^T,~,\tilde{\chi}_3^T,~\tilde{\chi}_4^T \right)^T, 
\end{eqnarray}
both the mass matrices in Eqs. \ref{eq_mass1} and \ref{eq_mass2} take the generic form
\begin{eqnarray}
 \tilde{m} = \left(
\begin{tabular}{c|c c}
$0_{4\times 4}$&\qquad\qquad&$0_{4\times 12}$ \qquad\qquad\\\hline
&&\\
$0_{12\times 4}$&\qquad&\#$_{12\times 12}$ \qquad\\
&&
\end{tabular}
\right),
\label{eq_massmatrixfour}
\end{eqnarray} 
where, $0_{m\times n}$ are $m\times n$ null matrices and $\#_{12\times 12}$ is some $12\times 12$ dimensional Hermitian matrix.

It is then clear that the mass matrix $\tilde{m}$ has a decoupled four-dimensional zero block belonging to $\tilde\chi_1$ which gives rise to the gapless modes.  In fact, the Dirac action for the $\tilde\chi_1$ sector is given by 
\begin{eqnarray}\label{eq_chi_tilde_1_hamiltonian}
&S_{\tilde{\chi}_1} &=v_F
\int d\tau d^2x~ \bar{\tilde{\chi}}_1\left( i\tilde\Sigma_{34}\partial_t + i\tilde\Sigma_{24}\partial_x - i\tilde\Sigma_{23}\partial_y \right)\tilde{\chi}_1,\nonumber\\
\end{eqnarray}
which is similar to that of spinless graphene~\cite{ryu2009masses} and hence there is an {\it emergent chiral SU(2)} which we call SU(2)$_{\rm IR}$.  This SU(2)$_{\rm IR}$ is generated by
\begin{eqnarray}
\{ \tilde\Sigma_1, \tilde\Sigma_5, \tilde\Sigma_{15} \}/2.
\end{eqnarray}
which is actually a projection of the SU(2) generated by $\{\Sigma_{4}/2,~\Sigma_{5}/2,~\Sigma_{45}/2\}$ into the $\tilde{\chi}_1$ sector. This SU(2)$_{\rm IR}$ along with $\mbb{T}$ and $\mbf{I}$ keeps the $\tilde{\chi}_1$ sector gapless.

It is interesting to consider the four fermion bilinear masses that can open up a gap in this sector. They are given by 
\begin{eqnarray}\label{eq_chi_tilde_1_masses}
-i\bar{\tilde{\chi}}_1 \tilde{\chi}_1,~ -i\bar{\tilde{\chi}}_1\tilde\Sigma_{15}\tilde{\chi}_1,~
-i\bar{\tilde{\chi}}_1\tilde\Sigma_{5}\tilde{\chi}_1,~
-i\bar{\tilde{\chi}}_1\tilde\Sigma_{1}\tilde{\chi}_1.\nonumber\\
\end{eqnarray}

The first one is actually a SU(2)$_{\rm IR}$ scalar, but, is odd under $\mbb{T}\mbf{I}$ and is actually a projection of a group-1, chiral mass, namely the ICI bilinear, $-i\bar{\chi}\chi$ (Eq.~\ref{eq_chern_insulator}) to the $\tilde\chi_1$ subspace and hence itself  transforms under a $\mcl{A}^o_{2g}$ singlet under the microscopic symmetries. Hence this mass breaks the $\mbb{T}\mbf{I}$ symmetry (or alternatively $\mbf{C_2'}$ symmetry for the $\mcl{A}^0_{1u}$ singlet) in the $\tilde\chi_1$ sector. The resultant massive bands for the $\tilde\chi_1$ fermions have a non-zero Chern number while the already gapped $\tilde\chi_2, \tilde\chi_3$ and $\tilde\chi_4$ remain topologically trivial. This is unlike the ICI phase where all the bands have a non-zero Chern number as is required in that case due to the fact that the ICI mass, (unlike in the present case) is a SU(4) singlet. Indeed, $-i\langle\bar{\tilde\chi}_1\tilde\chi_1\rangle\neq0$ leads to a $N_F=1$ CS term of the form in Eq. \ref{eq_icilag} leading to a single gapless edge mode carrying electronic charge instead of four as in the case of ICI and hence represents a different phase more akin to an anomalous Hall phase.

The last three masses in Eq.~\ref{eq_chi_tilde_1_masses} are $\mbb{T}\mbf{I}$ singlets but  transform as a triplet under SU(2)$_{\rm IR}$ and break it down to U(1) subgroup. These masses are best thought as projections of the group-2 and group-3 masses into the $\tilde\chi_1$ sector that are invariant under $\mathbb{T}{\bf I}$ that are simultaneously odd or even under both $\mathbb{T}$ and {\bf I}. In particular, both the $\mcl{A}_{1g}^e$ masses in Eq.~\ref{eq_a1ge_mass} and Eq.~\ref{eq_a1ge_mass_2} project to the fourth mass term in Eq.~\ref{eq_chi_tilde_1_masses}. Also, each of the three $\mcl{E}_u^o$ doublets in Eq.~\ref{eq_flavour-3}, \ref{eq_euo_mass_2}, \ref{eq_euo_mass_3} project to the second and the third masses of Eq.~\ref{eq_chi_tilde_1_masses}.  It is important to note that while it may appear that the resultant phases may have edge modes since they are obtained as a projection of a mass which in unprojected form lead to  symmetry-protected topological phase, this is not the case, because the respective symmetries are broken by Eq. \ref{eq_neel_singlet1} or \ref{eq_neel_singlet2}.


\subsubsection{Stripy spin-octupole density wave semimetal}\label{sec:stripy-spinOctupole-semimetal}

There are two stripy spin-octupole phases, both TR odd triplets with distinct lattice symmetries, which differ in the nature of the spin-octupolar densities. These are given by $\mcl{T}_{1g}^o$ and $\mcl{T}_{2g}^o$ irreps in  Eq.~\ref{eq:T2g_T1g_direct_product} which are respectively even and  odd under $\mbf{C_2'}$. They correspond to stripy pattern (similar to Fig. \ref{fig:stripy_CDW}) in the spin-octupole densities of 
\begin{align}
    \frac{1}{\sqrt{2}}(\Sigma_{13}+ \Sigma_{23}),\frac{1}{\sqrt{2}}(\Sigma_{12}+ \Sigma_{13})~~ {\rm and}~~\frac{1}{\sqrt{2}}(\Sigma_{12}- \Sigma_{23})
\end{align}
for the $\mcl{T}_{1g}^o$ and
\begin{align}
    \frac{1}{\sqrt{2}}(\Sigma_{13}- \Sigma_{23}),\frac{1}{\sqrt{2}}(\Sigma_{12}- \Sigma_{13})~~{\rm and}~~ \frac{1}{\sqrt{2}}(\Sigma_{12}+ \Sigma_{23})
\end{align}
for $\mcl{T}_{2g}^o$.

For all the masses of these two triplets, a certain number of fermionic modes are always gapless. However, due to the difference in the symmetry representation, the number, structure and stability of the remnant gapless fermions are different. While a generic linear combination like Eq. \ref{eq_gen_mass_on_sphere} for the $\mcl{T}^o_{1g}$ mass {\it always} lead to eight gapless fermions, the number of gapless modes for $\mcl{T}^o_{2g}$ triplet varies. In this latter case, generically there are four gapless modes. However, for special linear combinations, this number increases to eight. Thus, the $\mcl{T}_{1g}^o$ and the $\mcl{T}_{2g}^o$ masses represent M-DSM and $\Gamma$-DSM types of semimetals respectively. 
Since the structure of the remnant gapless fermions affects the fate of the low energy theory and the nature of possible phase transitions, we discuss it in some more detail for the two cases separately. 

In both cases, however, the structure and the symmetry protection of the fermions that remain gapless are best understood in the basis of $\tilde{\chi}$ spinors introduced in Eq.~\ref{eq_chi tilde basis}.

\paragraph{The $\mcl{T}_{1g}^o$ triplet :}

The three masses that form this triplet are
    \begin{align}
        &-i\Braket{\bar{\chi}\left(-\Sigma_{13}\zeta_3 - \Sigma_{23}\zeta_2 \right)\chi}/\sqrt{2},\nonumber\\
        &-i\Braket{\bar{\chi}\left(\Sigma_{12}\zeta_2 - \Sigma_{13}\zeta_1 \right)\chi}/\sqrt{2},\nonumber\\
        &-i\Braket{\bar{\chi}\left(\Sigma_{23}\zeta_1 + \Sigma_{12}\zeta_3 \right)\chi}/\sqrt{2}.
        \label{eq_t01gtriplet}
    \end{align}
which, in the $\tilde\chi$ basis (Eq.~\ref{eq_chi tilde basis}), become 
\begin{align}
&i\sqrt{2}\Braket{\bar{\tilde{\chi}} \mu_2\tilde\Sigma_5 \tilde{\chi} }, -i\sqrt{2}\Braket{\bar{\tilde{\chi}}\mu_5\tilde\Sigma_{0}\tilde{\chi} }, -i\sqrt{2}\Braket{\bar{\tilde{\chi}} \mu_{9}\tilde\Sigma_5\tilde{\chi} }.
\label{eq_massrotstripyoct}
\end{align}
such that the mass matrices have the generic form

\begin{eqnarray}
\tilde m = \left(
\begin{tabular}{c|c c}
$0_{4\times 4}$&\qquad\qquad&\#$_{4\times 12}$ \qquad\qquad\\\hline
&&\\
\#$^{\dagger}_{4\times 12}$&\qquad&$0_{12\times 12}$ \qquad\\
&&
\end{tabular}
\right)
\label{eq_massmatrixeight}
\end{eqnarray} 

This generic structure should be contrasted with Eq.~\ref{eq_massmatrixfour} which gave rise to four gapless modes from the $\tilde\chi_1$ sector. In the present case, Eq.~\ref{eq_massmatrixeight} however gives rise to eight gapless modes. This is because any matrix of the form given in Eq.~\ref{eq_massmatrixeight} always has eight zero eigenvalues. These gapless modes are protected by $\mbf{C_2'}\times$SO(4) symmetry, where the SO(4) is a subgroup of the SU(8) which acts non-trivially only on the gapless fermions.

One can now consider gapping out these fermions. This can be done by doing a similar analysis as done for the masses in Eq.~\ref{eq_neel_singlet1} and \ref{eq_neel_singlet2}.
As an example, for the first mass in this triplet, the $\tilde{\chi}_3$ and $\tilde{\chi}_4$ fermions are gapless. One can show that there are 16 independent fermion bi-linears that can gap out the $\tilde{\chi}_3$ and $\tilde{\chi}_4$ fermions in this case and the fate of the resultant phases can be analysed.

 
\paragraph{The $\mcl{T}_{2g}^o$ Triplet : }
The three masses in this triplet are
    \begin{eqnarray}\label{eq_gapless_t2go}
     &&   \left<-i\bar{\chi}\left(\Sigma_{13}\zeta_{3} - \Sigma_{23}\zeta_{2} \right)\chi\right>/\sqrt{2}, \nonumber \\ 
     && \left<-i\bar{\chi}\left(-\Sigma_{13}\zeta_{1} - \Sigma_{12}\zeta_{2} \right)\chi\right>/\sqrt{2}, \nonumber\\
  	&& \left<-i\bar{\chi}\left(-\Sigma_{23}\zeta_{1} + \Sigma_{12}\zeta_{3} \right)\chi\right>/\sqrt{2}.
    \end{eqnarray}
which, in term of the $\tilde{\chi}$ spinors (Eq. \ref{eq_chi tilde basis}),  are given by
\begin{eqnarray}
    -i\sqrt{2}\Braket{\bar{\tilde{\chi}}\mu_{14}\tilde\Sigma_{5} \tilde{\chi}}, i\sqrt{2}\Braket{\bar{\tilde{\chi}}\mu_{12}\tilde\Sigma_{0} \tilde{\chi}}, -i\sqrt{2}\Braket{\bar{\tilde{\chi}}\mu_{6}\tilde\Sigma_{5} \tilde{\chi}}.\nonumber\\
\end{eqnarray}
such that a generic linear combination of the form in Eq. \ref{eq_gen_mass_on_sphere}, but in $\tilde\chi$ basis, is given by $-i\bar{\tilde\chi}\tilde m\tilde\chi$ where the mass matrix has the generic form given by Eq. \ref{eq_massmatrixfour}. Hence the $\tilde\chi_1$ sector gives rise to four gapless fermion modes similar to the N\'eel spin-octupole density wave semimetal (see the discussion following Eq. \ref{eq_massmatrixfour}). The rest of the discussion proceeds similarly to that of N\'eel spin-octupole density wave semimetal. However here the $\mbf{C_2'}$ is already broken and the ICI mass term is generically allowed by symmetry.

In the present case, however, the $\#_{12\times12}$ block has a further rich structure that is immediately evident from writing the mass matrix, $\tilde m$, explicitly 
\begin{eqnarray}
\tilde{m} = \left(
\begin{array}{cccc}
 0 & 0 & 0 & 0 \\
 0 & 0 & -\Delta _3 \tilde\Sigma_{15} & -i \Delta _2  \\
 0 & -\Delta _3 \tilde\Sigma_{15} & 0 & -i \Delta _1 \tilde\Sigma_5 \\
 0 & i \Delta _2  & i \Delta _1 \tilde\Sigma_5 & 0 \\
\end{array}
\right)
\label{eq_12t12}
\end{eqnarray}

It is clear that if one or two of the $\Delta_i$ in Eq.~\ref{eq_12t12} are zero, then $\tilde{m}$ has extra four zero eigenvalues and hence total eight fermionic modes are gapless in this case. This is shown in Fig.~\ref{fig_line_degeneracy_triplet}, where we represent the linear combination of the masses on the sphere as before. For the three great circles (in Fig.~\ref{fig_line_degeneracy_triplet}) that lie in the three coordinate planes, there are eight gapless modes present. For any other point, the number of gapless modes is four.

\begin{figure}
    \includegraphics[scale=0.4]{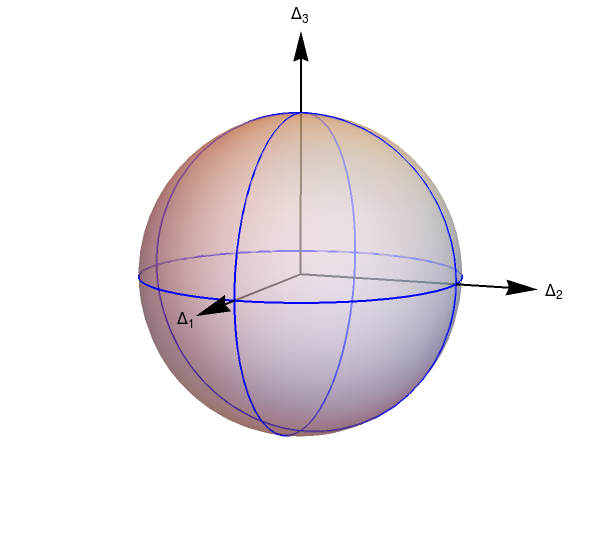}
     \caption{Gapless manifold for the triplet $\Gamma$-DSMs : Linear combinations of the form given in Eq.~\ref{eq_gen_mass_on_sphere} for the  triplet $\Gamma$-DSMs is represented on the surface of a unit sphere. The three great circles shown here are obtained by setting one of the $\Delta_i$s to zero in Eq.\ref{eq_gen_mass_on_sphere}. For the triplet $\Gamma$-DSMs in Eqs.~\ref{eq_gapless_t2go} and \ref{eq_gapless_t1ge}, on these great circles, the number of gapless modes is eight as opposed to only four at other points on the sphere.}
    \label{fig_line_degeneracy_triplet}
\end{figure}

\subsubsection{Stripy spin-quadrupole density wave semimetal}\label{sec:stripy-spinQuadru-semimetal}

There are two distinct stripy spin-quadrupolar density wave semimetal phases both of which are TR even but have distinct lattice symmetries. These are the two triplets given by  $\mcl{T}_{1g}^e$ and $\mcl{T}_{2g}^e$ irreps in Eq.~\ref{eq:T1g_T1g_direct_product}. While the first triplet is even under $\mbf{C_2'}$ and corresponds to the stripy pattern (similar to Fig. \ref{fig:stripy_CDW}) in the spin-quadrupoles
\begin{align}
    \frac{1}{\sqrt{2}}(\Sigma_1- \Sigma_2),\frac{1}{\sqrt{2}}(\Sigma_2- \Sigma_3)~{\rm and}~\frac{1}{\sqrt{2}}(\Sigma_1- \Sigma_3),
\end{align}
the second is inversion even and gives rise to stripy pattern in a different set of spin-quadrupoles given by
\begin{align}
    \frac{1}{\sqrt{2}}(\Sigma_1+ \Sigma_2),\frac{1}{\sqrt{2}}(\Sigma_2+ \Sigma_3)~{\rm and}~ \frac{1}{\sqrt{2}}(\Sigma_1+ \Sigma_3).
\end{align}

Each of these masses breaks the flavour SU(4) to SU(2). E.g., the residual SU(2) for the first mass in both the triplets is generated by $\{\Sigma_{34},\Sigma_{35},\Sigma_{45}\}$. 

In spite of the opposite behavior under time reversal, $\mbb{T}$, the structure of these masses is very similar to the two triplets in Eq.~\ref{eq_t01gtriplet}, \ref{eq_gapless_t2go} discussed above that represent stripy spin-octupoalr density wave semimetals. The analysis of the remnant gapless modes proceeds in the same way except for the fact that now the $\mcl{T}_{2g}^e$ triplet always has eight gapless fermionic modes and is a M-DSM type semimetal while the $\mcl{T}^e_{1g}$ triplet is a $\Gamma$-DSM type semimetal which generically has four gapless modes except at special combination of the mass as shown in Fig. \ref{fig_line_degeneracy_triplet}. Here we briefly summarise this structure for completeness. 

\paragraph{The $\mcl{T}_{2g}^e$ triplet :}

The three components of the $\mcl{T}_{2g}^e$ triplet are given by
\begin{align}
&-i\Braket{\bar{\chi}\left(-\Sigma_1\zeta_2 + \Sigma_2\zeta_3 \right)\chi}/\sqrt{2},\nonumber\\
&-i\Braket{\bar{\chi}\left(\Sigma_3\zeta_2 + \Sigma_2\zeta_1 \right)\chi}/\sqrt{2},\nonumber\\ 
&-i\Braket{\bar{\chi}\left(\Sigma_3\zeta_3 + \Sigma_1\zeta_1 \right)\chi}/\sqrt{2}.
\label{eq_massstripyquad}
\end{align}
which in the $\tilde{\chi}$ basis (Eq. \ref{eq_chi tilde basis})is given by
\begin{align}
-i\sqrt{2}\Braket{\bar{\tilde{\chi}} \mu_1\tilde\Sigma_{15}\tilde{\chi} }, i\sqrt{2}\Braket{\bar{\tilde{\chi}} \mu_5\tilde\Sigma_{1}\tilde{\chi} }, -i\sqrt{2}\Braket{\bar{\tilde{\chi}}\mu_{10}\tilde\Sigma_{15}\tilde{\chi} }.
\end{align}
These masses have the same form as in Eq. \ref{eq_massmatrixeight} and hence these give rise to eight gapless fermions. The gaplessness of these modes is similarly protected via lattice symmetries and various subgroups of SU(8).    

\paragraph{The $\mcl{T}_{1g}^e$ triplet : }

The three masses of the the $\mcl{T}_{1g}^e$ triplet are given by  
    \begin{eqnarray}\label{eq_gapless_t1ge}
     &&  \left<-i\bar{\chi}\left(-\Sigma_{1}\zeta_{2} - \Sigma_{2}\zeta_{3} \right)\chi\right>/\sqrt{2}, \nonumber \\ 
     && \left<-i\bar{\chi}\left(-\Sigma_{3}\zeta_{2} + \Sigma_{2}\zeta_{1} \right)\chi\right>/\sqrt{2}, \nonumber\\
  	&&  \left<-i\bar{\chi}\left(\Sigma_{3}\zeta_{3} - \Sigma_{1}\zeta_{1} \right)\chi\right>/\sqrt{2}.
    \end{eqnarray}

In $\tilde{\chi}$ basis, these masses have the form
    \begin{eqnarray}
    i\sqrt{2}\Braket{\bar{\tilde{\chi}}\mu_{13}\tilde\Sigma_{15} \tilde{\chi}}, i\sqrt{2}\Braket{\bar{\tilde{\chi}}\mu_{11}\tilde\Sigma_{1} \tilde{\chi}}, i\sqrt{2}\Braket{\bar{\tilde{\chi}}\mu_{7}\tilde\Sigma_{15} \tilde{\chi}}. \nonumber \\
    \end{eqnarray}
    Any linear combination (Eq. \ref{eq_gen_mass_on_sphere}) of these masses can be written as $-i\bar{\tilde{\chi}}\tilde{m}\tilde{\chi}$ where again $\tilde m$ has the structure given by Eq.~\ref{eq_massmatrixfour}, albeit with different entries, {\it i.e.},  
    \begin{eqnarray}
 \tilde{m} = \left(
\begin{array}{cccc}
 0 & 0 & 0 & 0 \\
 0 & 0 & -i \Delta _3 \tilde\Sigma_{15} & \Delta _2 \tilde\Sigma_1 \\
 0 & i \Delta _3 \tilde\Sigma_{15} & 0 & \Delta _1 \tilde\Sigma_{15} \\
 0 & \Delta _2 \tilde\Sigma_1 & \Delta _1 \tilde\Sigma_{15} & 0 \\
\end{array}
\right).
\end{eqnarray} 
Therefore it gives rise to four gapless Dirac fermions except for the three great circles where there are four additional gapless modes due to additional zeros in the $\#_{12\times12}$ sector similar to Eq. \ref{eq_12t12}. 

\subsubsection{Ferro spin-quadrupole semimetal}\label{sec:ferro-spinQuadru-semimetal}

{
The $\mcl{E}_g^e$ doublet in Eq.~\ref{eq:T1g_T1g_direct_product} corresponds to uniform (ferro) ordering of the spin-quadrupole densities in $(\Sigma_1-\Sigma_2)/\sqrt{2}$ and $(2\Sigma_3-\Sigma_1-\Sigma_2)/\sqrt{6}$. The corresponding two masses are given by~: 
\begin{eqnarray}\label{eq_ege_doublet_1}
  &&\Braket{-i\bar{\chi}(-\Sigma_1\zeta_3+\Sigma_2\zeta_2\chi})/\sqrt{2},  \nonumber \\
  &&\Braket{-i\bar{\chi}(\Sigma_1\zeta_3+\Sigma_2\zeta_2 + \Sigma_3\zeta_1)\chi}/\sqrt{6}.
\end{eqnarray}

These two masses do not fully gap out the fermions and hence represent ferro spin-quadrupolar density wave semimetals. Moreover, depending on the linear combination of these two masses, the number of gapless modes change due to the change in the residual symmetry-- similar to the case described above by Eq. \ref{eq_12t12}, but now on a circle, {\it i.e.} $\mathcal{S}^1$. This is a fallout of the fact that the two masses are non-compatible. Consider a generic linear combination of the two masses of the form akin to Eq. \ref{eq_gen_mass_on_sphere}, but now on a circle, {\it i.e.},    $-i\bar{\chi}m(\vartheta)\chi$, where
\begin{eqnarray}\label{eq_doublet_masses_linear_comb}
m(\vartheta) = \cos\vartheta ~m_1 + \sin \vartheta ~m_2,
\end{eqnarray}
and $\{m_1,m_2\}$ represent the two mass matrices in Eq.~\ref{eq_ege_doublet_1} and $\vartheta\in (0,2\pi]$. For a generic value of $\vartheta$, there are four gapless modes in the spectrum and thus, this doublet is a $\Gamma$-DSM type semimetal. The flavour SU(4) is broken down to U(1) at these points. However, for special isolated values of $\vartheta = \frac{n\pi}{3}$ (with $n=0,1,\cdots,5$), there are eight gapless modes since the flavour SU(4) is only broken down to SU(2). Thus, the residual symmetry is larger for the case where there are extra gapless modes.

The appearance of the gapless modes for these masses are better understood in the $\tilde{\chi}$ basis introduced in Eq.~\ref{eq_chi tilde basis}. In this basis, the masses in Eq. \ref{eq_ege_doublet_1} are given by 
\begin{align}
    &\Braket{-i\bar{\tilde{\chi}}\left(-\sqrt{3}\mu_0 + \mu _{15}+\sqrt{2} \mu _8\right)\tilde\Sigma_{1}\tilde\chi}/\sqrt{3},\nonumber \\ 
    &\Braket{-i\bar{\tilde{\chi}}( \mu_0 + \sqrt{3} \mu _{15}-\sqrt{2} \mu_3)\tilde\Sigma_{1} \tilde\chi}/\sqrt{3}. 
\end{align}
such that the generic mass matrix  in Eq.~\ref{eq_doublet_masses_linear_comb} is $-i\bar{\tilde{\chi}}\tilde{m}(\vartheta)\tilde{\chi}$ where 
 \begin{eqnarray}
&&\tilde{m}(\vartheta) =\nonumber\\
&&\quad \sqrt{2}\left(
\begin{array}{cccc}
 0 & 0 & 0 & 0 \\
 0 & \frac{2 \sin\vartheta}{\sqrt{3}} & 0 & 0 \\
 0 & 0 & \frac{\sin\vartheta}{\sqrt{3}}-\cos\vartheta & 0 \\
 0 & 0 & 0 & -\cos\vartheta-\frac{\sin\vartheta}{\sqrt{3}} \\
\end{array}
\right)\otimes \tilde\Sigma_1. \nonumber\\
\end{eqnarray}
Thus, the $\tilde{\chi}_1$ spinors defined in Eq.~\ref{eq_4_tilde_modes} are always gapless, accounting for the four gapless fermions for generic $\vartheta$. Also, for the special values of $\vartheta$ specified before, either the $\tilde{\chi_2}, \tilde{\chi}_3$ or the $\tilde{\chi}_4$ spinors also become gapless.   

The gaplessness of the $\tilde{\chi}_1$ modes for generic values of $\vartheta$ is protected by TR symmetry and the SU(2)$_{\rm IR}$ subgroup of the SU(8). For the special values of $\vartheta$ where there are extra gapless modes, the residual symmetry also becomes large compared to that for other values of $\vartheta$. This is evident because, as mentioned before, the flavour SU(4) breaks down to U(1) at generic values of $\vartheta$. But for the special values noted above, it only breaks down to SU(2). 
}

\subsubsection{Ferro spin-octupole semimetal}
\label{subsubsec:FSOPSWSM}

The $\mcl{E}_g^o$ doublet in Eq.~\ref{eq:T2g_T1g_direct_product} represent uniform (ferro) ordering in the spin-octupoles given by $\left(\Sigma _{13}+\Sigma _{23}\right)/{\sqrt{6}}$ and $\left(2\Sigma _{12}+\Sigma _{13}-\Sigma _{23}\right)/{3 \sqrt{2}}$ respectively. These masses are given by~:
\begin{eqnarray}\label{eq_ego_doublet}
  &&  \Braket{-i\bar{\chi}(\Sigma_{13}\zeta_2+\Sigma_{23}\zeta_3)\chi}/\sqrt{2},  \nonumber \\
  &&\Braket{-i\bar{\chi}(2\Sigma_{12}\zeta_1-\Sigma_{13}\zeta_2 + \Sigma_{23}\zeta_3)\chi}/\sqrt{6}.
\end{eqnarray}

Any linear combination of these masses of the form in Eq.~\ref{eq_doublet_masses_linear_comb} breaks down the flavour SU(4) to SU(2) generated by $\{\Sigma_{4},\Sigma_{5},\Sigma_{45}\}$. 

Similar to the ferro spin-quadrupole doublet in Eq.~\ref{eq_ege_doublet_1}, there are at least four gapless modes present for the masses in this doublet. Thus, this doublet is also a $\Gamma$-DSM type semimetal. The number of these gapless modes again depends on their linear combination because of the non-compatibility of these masses and are best analysed in the $\tilde\chi$ basis (Eq. \ref{eq_chi tilde basis}). In this basis, the two masses in Eq. \ref{eq_ego_doublet} are given by
\begin{eqnarray}
    &&\Braket{-i\bar{\tilde{\chi}}\left(2 \mu _{15}-\sqrt{2} \mu _8\right)\tilde\Sigma_{0}  \tilde\chi}/\sqrt{3},\nonumber \\ 
    && \Braket{-i\bar{\tilde{\chi}} \left(-2 \mu _{15}+\sqrt{6} \mu _3-2 \sqrt{2} \mu _8\right)\tilde\Sigma_{0}\tilde\chi}/3. 
\end{eqnarray}
such that the mass matrices have the generic form given by Eq. \ref{eq_massmatrixfour}. Hence the $\tilde{\chi}_1$ modes are always gapless. In addition, considering linear combinations as in Eq.~\ref{eq_doublet_masses_linear_comb}, for some special values of $\vartheta$ given by $\vartheta = (2n+1)\frac{\pi}{6}$ with $n=0,1,\cdots,5$, there are extra gapless modes whose existence can be understood via an analysis similar to the ferro spin-quadrupolar doublet in Eq.~\ref{eq_ege_doublet_1} discussed before. However, unlike this previous case, this doublet breaks TR symmetry. So, for generic values of $\vartheta$, the $\tilde{\chi}_1$ spinors can be gapped out by turning on the ICI mass which does not break any further symmetries. Hence the leftover gapless modes are less robust and can easily give way to a $N_F=1$ ICI phase.


\section{Discussion and Outlook}
\label{sec:Discussion}

In this paper, we have described a material context of realising SU(8) Dirac fermions in the presence of strong SOC in two-dimensional honeycomb lattices with quarter filled $j=3/2$ electronic orbitals. The resultant semimetal allows for non-trivial implementation of the microscopic symmetries at low energies that is reflected in the spontaneous symmetry broken and SPT phases proximate to the Dirac semimetal. The spontaneous symmetry-breaking phases reflect indelible signatures of the underlying strong SOC that intertwines the real and spin spaces via the nature of their symmetry-breaking while the SPTs represent interesting realisations of edge physics. Of particular interest are a class of density-wave semimetals where a subset of Dirac fermions remain gapless and are protected by subgroups of the emergent SU(8).

All these phases can be accessed via finite electron-electron interactions, which leads to a plethora of interaction-driven transitions of the Gross-Neveu-Yukawa type~\cite{PhysRevD.10.3235,zinn1991four,sachdev1999quantum,boyack2021quantum,PhysRevLett.97.146401}. Indeed, the implementation of symmetries allows for several unnecessary phase transitions\cite{PhysRevX.9.021034} within a single phase, in particular unnecessary multicritical points, arising due to non-compatibility of fermion bilinears. A second class of such transitions involves the spin-density wave semimetals where a subset of gapless fermions survive on the symmetry-broken side. Finally, the present classification of masses also predicts a newer class of transitions involving the topological defects of the order parameters. In particular, a triplet quantum spin-octupole Hall order parameter allows skyrmion defects that carry four units of electronic charge such that the condensation of such bosonic skyrmions naturally leads to $4e$ superconductivity. 

The SU(8) Dirac semimetal thus provides for new  phases engendered by the symmetry implementation owing to the SOC as well as offers opportunities to pose a more general set of questions. What aspects of these results hold in a more general setting? For example, in the case discussed here, the problem is reduced to studying a honeycomb system with $\pi$-flux and an SU(4) global symmetry. What aspects of the results will hold if a system realizes a $\pi$-flux state with SU(N) symmetry?

We address this question by constructing and studying a model which realizes a $\pi$-flux state with a global SU(2) symmetry in appendix~\ref{appen:jHalfModel}. We also find three groups of masses (chiral, flavor, and mixed) in the SU(2) case. Not unexpectedly, we find that the chiral masses (see appendix.~\ref{subsec:J2CM}) have exactly similar structure ${\bf 4} = {\bf 1} \oplus {\bf 3}$ with an integer Chern insulator and a triplet stripy density wave. The flavor masses (appendix.~\ref{subsec:J2CM}) also have a similar structure, with a single triplet corresponding to a quantum dipolar Hall mass. The SU(4) case discussed in the main text realizes other SU(4) symmetry broken phases as the flavour space is larger, offering more possibilities. In particular, some of the  SU(4) flavour masses offer the possibility of  unnecessary multicitical points, as mentioned above (owing their non-compatibility of the mass components); the SU(2) flavour space of  appendix.~\ref{subsec:J2CM} is not large enough to realize such possibilities. The findings of the main text and appendix~\ref{appen:jHalfModel} taken together suggest that the chiral masses have natural generalization to any SU(N) flavour system that realizes a $\pi$-flux through the honeycomb plaquette. Moreover, every flavor mass found for an SU($N$) will have a counterpart for SU($N'$) ($N' > N$), with $N'$ system offering a larger class of broken symmetry ordered phases. In this regard, it is useful to point out the importance of the underlying honeycomb lattice. It is easy to conceive models on a square lattice (in a fashion as done in appendix.~\ref{subsec:J2MM}) that uses SOC to produce a $\pi$-flux per plaquette. The chiral masses in that case will have a ${\bf 4} = {\bf 1} \oplus {\bf 1} \oplus {\bf 2}$ structure, which is distinct from the ${\bf 4} = {\bf 1} \oplus {\bf 3}$ structure realized in the honeycomb lattice. This will lead to different kinds of masses on a square lattice~\cite{PhysRevLett.95.036402,hermele2004stability,PhysRevB.104.245106,marston1989large,affleck1988large} than the honeycomb lattice. 

The true richness of the symmetry implementation arising SOC is found in the mixed masses. One finds semimetallic phases even in the SU(2) case (see appendix.~\ref{subsec:J2MM}), and they have corresponding counterparts in SU(4) case discussed in the main text. The SU(2) case also realizes both types of semimetals ({$\Gamma$-DSM and M-DSM}). {However, the SU(4) case produces new phases such as the zig-zag ordered phases, which are not found in the case SU(2).} Thus the larger flavor space offers more interesting possibilities for realizing representations of the IR space group that entangle the chiral and flavor spaces that obtain exotic phases.  

We end our discussion with two particular issues regarding the candidate $d^1$ materials -- (1) the sub-leading hopping pathways and anisotropies in addition to the indirect hopping model (Eq. \ref{eq:hamiltonian in j=3/2 orbitals}), and, (2) the strong coupling insulating limit. We discuss them in turn starting with the former. Eq. \ref{eq:hamiltonian in j=3/2 orbitals} disregards other hopping pathways like the direct overlap of $d$-orbitals and/or lattice distortions such as the trigonal distortion of the octahedral crystal field which are often relevant for candidate materials with stacked honeycomb structures. Such terms will generically explicitly break the SU(4) symmetry. The effect of such terms, if small, can be systematically captured by starting with the SU(4) limit and treating them in terms of the symmetry breaking they represent. In such cases, our classification of masses provides the list of instabilities of the Dirac semimetal, albeit with the explicit breaking of SU(8) symmetry, but not lattice symmetries or microscopic time reversal. Within our classification scheme, these, then, have to belong to $\mathcal{A}_{1g}^e$ irrep and there are precisely two such masses   corresponding to ferro spin-quadrupole (Eqs. \ref{eq_a1ge_mass_2}) and spin-octupole Hall (Eq. \ref{eq_a1ge_mass}) phases respectively. This should be observable in candidate materials which realise such perturbations. 
 In addition, there may be a loss of three fold rotation symmetry about the site due to unequal bond lengths as suggested in Ref. \cite{ushakov2020can} for ZrCl$_3$. In this case the indirect hopping model (Eq. \ref{eq:hamiltonian in j=3/2 orbitals}) retains SU(4) symmetry on changing the amplitude of one of the bonds (say $z$-bonds in Fig. \ref{fig:2-point honeycomb}) compared to the other two since the hopping matrix structure (Eq. \ref{eq_umatrices}) is preserved. Notably, the Dirac points do not gap out without breaking further symmetries in this case~\cite{yamada_2018_emergent_SU(4)} in accordance with the generalised Lieb-Schultz-Mattis-Affleck theorem~\cite{yamada_2018_emergent_SU(4), PhysRevB.104.224436,lajko2017generalization,lieb1961two,affleck1986proof,affleck1988spin,PhysRevLett.84.1535,hastings2005sufficient}. This robustness of the Dirac points to the breaking of three-fold rotation symmetry is different from graphene (and indeed the Dirac points at the half-filled SU(4) case~\cite{PhysRevB.90.075140}) where making one of the bond strengths different leads to the migration and eventual pairwise annihilation of the Dirac points~\cite{PhysRevB.74.033413}.

A second source of breaking of SU(8) symmetry of the low energy theory arises from generic four fermion interactions of the type given in Eq. \ref{eq_interactions} that are allowed by microscopic symmetries. They are given by the Lagrangian density $\sim \sum_{i,j} g_{ij}\left(\bar\chi\mathbf{\mathcal{M}}_i\chi\right)\cdot \left(\bar\chi\mathbf{\mathcal{M}}_j\chi\right)$ with $g_{ij}\neq g\delta_{ij}$ and $i,j$ are summed over different irreps. Depending on the values of coupling constants $g_{ij}$, such terms would preferentially enhance the possibility of one phase at the cost of another amongst those listed above leading to a complex phase diagram with interesting phase transitions between them. An exhaustive enumeration of microscopic-symmetry-allowed short-range four fermion terms requires a more careful analysis.

Finally, in this work, we have explored the intermediate coupling regime of the SU(8) Dirac semimetal. The strong coupling limit within a Hubbard model framework of the above model has been recently investigated~\cite{yamada_2018_emergent_SU(4)} and this leads to SU(4) Heisenberg spin model for $j=3/2$ spins. Such a model, as argued in Ref. \cite{yamada_2018_emergent_SU(4)}, necessarily has a non-trivial ground state unless SU(4) and/or translation symmetry is spontaneously broken. Indeed, a $\pi$-flux Dirac spin-orbital liquid state has been proposed~\cite{PhysRevX.2.041013,calvera2021theory} with strong numerical evidence. Very interestingly such a $\pi$-flux Dirac spin-orbital liquid can be obtained in a rather straightforward way from our approach via a parton decomposition of the electrons into charge-carrying bosonic rotors and fermionic spinons~\cite{PhysRevB.70.035114} and considering a state where the rotors are trivially gapped such that the spinons inherit the Dirac spectrum of the electrons. Several of the present classification of the masses then can be considered instabilities of such a spin-orbital liquid. This approach provides an interesting connection between the electronic phases and the magnetic phases that are worth exploring in the future.

\acknowledgements

We acknowledge fruitful discussions with F. Assaad, S. Chandrasekharan, S. Das Sarma, R. Gopakumar, M. Gupta, R. Moessner, A. Paramekanti, and T. Saha-Dasgupta. SB acknowledges adjunct fellow program at SNBNCBS, Kolkata for hospitality. We acknowledge the use of open-source software {\it Groups, Algorithms, Programming - a System for Computational Discrete Algebra} (GAP). BM and and SB acknowledge funding from Max Planck partner group grant at ICTS, SERB-DST (India) for funding through Swarna Jayanti grant SB/SJF/2021-22/12-G and the Department of Atomic Energy, Government of India, under Project No. RTI4001. VBS thanks DST for funding under MATRICS program.
\appendix
\section{The microscopic details}
\label{appen_microscopic}
\subsection{The \texorpdfstring{$j=3/2$}{} orbitals}
\label{appen_sub_j3/2}

The $t_{2g}$ orbitals behave effectively as $l=1$ states under rotations with
\begin{align}
|l_Z=0\rangle=|d_{XY}\rangle,~~~|l_Z=\pm 1\rangle=-\frac{1}{\sqrt{2}}(i|d_{XZ}\rangle\pm |d_{YZ}\rangle).
\end{align}

 So in the presence of SOC, these six degenerate (including spin degeneracy) states split into four with total angular momentum $j=3/2$ and the other two with $j=1/2$. The $j=3/2$ states in terms of the $t_{2g}$ states are given by 
\begin{align}
    |3/2\rangle&=\frac{1}{\sqrt{2}}(-|d_{YZ},\uparrow\rangle-i|d_{ZX},\uparrow\rangle)\nonumber\\
    |1/2\rangle&=\frac{1}{\sqrt{6}}(-|d_{YZ},\downarrow\rangle-i|d_{ZX},\downarrow\rangle+2|d_{XY},\uparrow\rangle)\nonumber\\
    |-1/2\rangle&=\frac{1}{\sqrt{6}}(|d_{YZ},\uparrow\rangle-i|d_{ZX},\uparrow\rangle+2|d_{XY},\downarrow\rangle)\nonumber\\
    |-3/2\rangle&=\frac{1}{\sqrt{2}}(|d_{YZ},\downarrow\rangle-i|d_{ZX},\downarrow\rangle)
\end{align}

In terms of the second quantized operators, the inverse relations, when projected to the $j=3/2$ orbitals are
\begin{subequations}\label{eq:projection from t2g to j=3/2}
	\begin{eqnarray}
		&&\Psi_{\sigma,x}^{\dagger}(\mathbf{r},\mcl{s}) = \frac{\sigma}{\sqrt{6}}\left(\psi_{\uparrow\bar{\sigma}}^{\dagger}(\mathbf{r},\mcl{s}) - \sqrt{3}\psi_{\downarrow \sigma}^{\dagger}(\mathbf{r},\mcl{s})\right)\nonumber\\ 
		&&\Psi_{\sigma,y}^{\dagger}(\mathbf{r},\mcl{s}) = \frac{i}{\sqrt{6}}\left(\psi_{\uparrow\bar{\sigma}}^{\dagger}(\mathbf{r},\mcl{s}) + \sqrt{3}\psi_{\downarrow \sigma}^{\dagger}(\mathbf{r},\mcl{s})\right)\nonumber\\ 
		&&\Psi_{\sigma,z}^{\dagger}(\mathbf{r},\mcl{s}) = \sqrt{\frac{2}{3}}\psi_{\uparrow \sigma}^{\dagger}(\mathbf{r},\mcl{s}), 
		\label{eqn_dtoj3/2}
	\end{eqnarray}
\end{subequations}
where on the LHS, $\Psi_{\sigma,x}^{\dagger}(\mathbf{r},\mcl{s}),\Psi_{\sigma,y}^{\dagger}(\mathbf{r},\mcl{s}),\Psi_{\sigma,z}^{\dagger}(\mathbf{r},\mcl{s})$ stand for the creation operators for the $|d_{YZ,\sigma}\rangle, |d_{ZX},\sigma\rangle, |d_{XY},\sigma\rangle$ orbitals respectively at the lattice sub-lattice site $\mcl{s}$ of the unit cell at $\mbf{r}$ and $\sigma = \uparrow, \downarrow$ are spin indices. The $\psi_{\uparrow \sigma}^{\dagger}$ and $\psi_{\downarrow \sigma}^{\dagger}$ in the RHS refer to the creations operators in the $j=3/2$ orbitals as~\cite{yamada_2018_emergent_SU(4)}
\begin{align}
    (\psi_{\uparrow \uparrow}^{\dagger},\psi_{\uparrow \downarrow}^{\dagger},\psi_{\downarrow \uparrow}^{\dagger},\psi_{\downarrow \downarrow}^{\dagger}) = (\psi^{\dagger}_{1/2},\psi^{\dagger}_{-1/2},\psi^{\dagger}_{3/2},\psi^{\dagger}_{-3/2}).
\end{align}
where $\bar\sigma=\downarrow(\uparrow)$ for $\sigma=\uparrow(\downarrow)$.

\subsection{The lattice}
\label{appen_sub_lattice}
Similar to Kitaev materials~\cite{PhysRevB.102.235124}, it is useful to consider the honeycomb lattice to lie in the plane perpendicular to the  $[111]$ direction of the global Cartesian coordinates $(X,Y,Z)$ (see Fig.~\ref{fig:3d crystal structure of ZrCl3}).  Therefore the two orthogonal directions in the honeycomb plane are spanned by
\begin{eqnarray}
	\hat{x} = \frac{1}{\sqrt{2}}(-\hat{X}+\hat{Y}),~~~~~\hat{y} = \frac{1}{\sqrt{6}}(2\hat{Z}-\hat{X} -\hat{Y})
\end{eqnarray}
while
\begin{eqnarray}
	&& \hat{z} = \frac{1}{\sqrt{3}}(\hat{X} + \hat{Y} + \hat{Z})
\end{eqnarray}
is normal to the honeycomb plane. In this new coordinate system, the $x,y$, and $z$ bonds in these new coordinates are shown in Fig. \ref{fig:2-point honeycomb}. 

The lattice vectors (with reference to Fig. \ref{fig:2-point honeycomb}) are 
\begin{align}
    \mathbf{b}_1 = \frac{3\mathcal{l}}{2}\hat{x} + \frac{\sqrt{3}\mathcal{l}}{2}\hat{y},~~~~~\mathbf{b}_2 = \frac{3\mathcal{l}}{2}\hat{x} - \frac{\sqrt{3}\mathcal{l}}{2}\hat{y}
    \label{eq_lattice_vectors}
\end{align}

\section{The Microscopic symmetries}
\label{appen_symm}

The transformation of the $t_{2g}$ orbitals under lattice translation and TR are straightforward and are given by
\begin{align}
\begin{array}{ll}
    {\bf T_{1(2)}} :& \Psi_{\sigma,\alpha}({\bf r},\mcl{s})\rightarrow \Psi_{\sigma,\alpha}({\bf r}',\mcl{s}')\\
    \mathbb{T} :& \Psi_{\sigma,\alpha}({\bf r},\mcl{s})\rightarrow \left(\iota\sigma^y_{\sigma\sigma'}\right)\Psi_{\sigma',\alpha}({\bf r},\mcl{s})\\
    \end{array}
\end{align}
$\forall~\alpha=x,y,z$ and $\sigma=\uparrow,\downarrow$, where ${\bf r}'={\bf T}_{1(2)}\left[{\bf r}\right]$ as discussed in the main text.

For the point group symmetries listed in Table \ref{tab:lat_sym}, the transformation of the $t_{2g}$ orbitals have a generic form of
\begin{align}
   \mathbb{S} : \Psi^\dagger_{\sigma,\alpha}(\mbf{r},\mcl{s}) \to \Psi^{\prime \dagger}_{\sigma,\alpha}(\mbf{r},\mcl{s}) = \left[\mathcal{R}_{\mathbb{S}}\right]_{\beta \alpha} \Psi^\dagger_{\sigma,\beta}(\mbf{r}',\mcl{s}').
   \label{eq:t2gtransform}
\end{align}
where $\mathbb{S}$ are the point group symmetry generators listed in Table \ref{tab:lat_sym} that takes $({\bf r'},\mathcal{s}')\rightarrow ({\bf r},\mathcal{s})$ on the honeycomb lattice. The form of the $3\times3$ matrices, $\mathcal{R}_\mathbb{S}$ for different symmetries are :
\begin{align}
    \mathcal{R}_{\bf C_3} = \begin{pmatrix}
0& 0& 1 \\
1& 0& 0 \\
0& 1& 0 \\
\end{pmatrix},
\mathcal{R}_{\bf S_6} = \begin{pmatrix}
0& 1& 0 \\
0& 0& 1 \\
1& 0& 0 \\
\end{pmatrix},
    \mathcal{R}_{\bf C'_2} = \begin{pmatrix}
		0& 1& 0 \\
		1& 0& 0 \\
		0& 0& 1 \\
\end{pmatrix},
\end{align}
while
\begin{align}
    \mathcal{R}_{\bf{I}} = \mathbb{I}_3,~~~~\mathcal{R}_{\mbf{\sigma_d}}= \mathcal{R}_{\bf C_2'}.
\end{align}

The transformation of the $j=3/2$ orbitals (Eq.~\ref{eq:4-component psi})  under the action of the microscopic symmetries can be obtained from the above relations. For TR, we have
\begin{eqnarray}\label{eq:transformation of psi under TR}
\psi(\mbf{r},\mcl{s}) \to i \Sigma_{13} ~K~ \psi(\mbf{r},\mcl{s}).
\end{eqnarray}

For the lattice symmetries (Table \ref{tab:lat_sym}), similar to Eq. \ref{eq:t2gtransform}, the transformation of  $\psi(\mbf{r},\mcl{s})$ have the following generic form
\begin{align}\label{eq:transformation of psi operators}
   \mathbb{S} : \psi^\dagger_{i}(\mbf{r},\mcl{s}) \to \psi^{\prime \dagger}_{i}(\mbf{r},\mcl{s}) = \left[\mathcal{U}_{\mathbb{S}}\right]_{ji} \psi^\dagger_{j}(\mbf{r'},\mcl{s}').
\end{align}

The $\mcl{U}_\mbb{S}$ are $4\times 4$ unitary matrices which for different symmetries are
\begin{eqnarray}
\label{eq:transformation of psi under translation}
&&\mcl{U}_{\mbf{T}_1} = \mcl{U}_{\mbf{T}_2} =\mathcal{U}_{\bf I}= \mathbb{I}_4,\\
&&\mcl{U}_{\mbf{C_3}} = \frac{1}{4}\left(
\begin{array}{cccc}
 -1+i & -1-i & \frac{-1-i}{\sqrt{3}} & \frac{-1+i}{\sqrt{3}} \\
 1-i & -1-i & \frac{-1-i}{\sqrt{3}} & \frac{1-i}{\sqrt{3}} \\
 \frac{-1+i}{\sqrt{3}} & \frac{1+i}{\sqrt{3}} & -1-i & 1-i \\
 \frac{1-i}{\sqrt{3}} & \frac{1+i}{\sqrt{3}} & -1-i & -1+i \\
\end{array}\right), \nonumber \\  \\
%
&&\mcl{U}_{\mbf{S_6}} = \frac{1}{4}\left(
\begin{array}{cccc}
 1+i & -1-i & \frac{1+i}{\sqrt{3}} & \frac{-1-i}{\sqrt{3}} \\
 1-i & 1-i & \frac{-1+i}{\sqrt{3}} & \frac{-1+i}{\sqrt{3}} \\
 \frac{1-i}{\sqrt{3}} & \frac{1-i}{\sqrt{3}} & 1-i & 1-i \\
 \frac{1+i}{\sqrt{3}} & \frac{-1-i}{\sqrt{3}} & -1-i & 1+i \\
\end{array}
\right), \nonumber \\  \\
\label{eq:transformation of psi under C2}
&& \mcl{U}_{\mbf{C_2'}} = \frac{1}{\sqrt{2}}\left(
\begin{array}{cccc}
 0 & -1+i & 0 & 0 \\
 1+i & 0 & 0 & 0 \\
 0 & 0 & 0 & -1-i \\
 0 & 0 & 1-i & 0 \\
\end{array}
\right), \\
%
%
&& \mcl{U}_{\mbf{\sigma_d}} = \mcl{U}_{\mbf{C_2'}}.
\end{eqnarray}

\section{Relation between the \texorpdfstring{$j=3/2$}{} matrices and the \texorpdfstring{$\Sigma$}{} matrices generating \texorpdfstring{SU(4)}{}}
\label{appen_gamma}

Following reference \onlinecite{PhysRevB.69.235206}, we define a basis for the set of 4-dimensional Hermitian matrices using the SU(2) generators for spin-3/2. The three spin-3/2 matrices written in the $J_z$ eigenbasis are the following: 
\begin{align}
J_x=\left(
\begin{array}{cccc}
 0 & 1 & \frac{\sqrt{3}}{2} & 0 \\
 1 & 0 & 0 & \frac{\sqrt{3}}{2} \\
 \frac{\sqrt{3}}{2} & 0 & 0 & 0 \\
 0 & \frac{\sqrt{3}}{2} & 0 & 0 \\
\end{array}\right),\\
J_y=\left(
\begin{array}{cccc}
 0 & -i & \frac{i \sqrt{3}}{2} & 0 \\
 i & 0 & 0 & -\frac{i \sqrt{3}}{2} \\
 -\frac{i \sqrt{3}}{2} & 0 & 0 & 0 \\
 0 & \frac{i \sqrt{3}}{2} & 0 & 0 \\
\end{array}\right),\\
J_z=\left(
\begin{array}{cccc}
 \frac{1}{2} & 0 & 0 & 0 \\
 0 & -\frac{1}{2} & 0 & 0 \\
 0 & 0 & \frac{3}{2} & 0 \\
 0 & 0 & 0 & -\frac{3}{2} \\
\end{array}\right).\label{eq:Sz_matrix}
\end{align}
Note that, instead of the standard practice, we have used a different ordering of the $J_z$ eigen-basis (see Eq. \ref{eq:4-component psi}) to write these matrices which is evident from the form of the $J_z$ matrix in Eq. \ref{eq:Sz_matrix}. In our choice of basis, the hopping matrices of the lattice Hamiltonian in Eq. \ref{eq:hamiltonian in j=3/2 orbitals} have a simpler form. 

With the above matrices, one can define the following five Hermitian matrices:
\begin{subequations}\label{eq:Sigma_i_matrices}
\begin{align}
\Sigma_1&=\frac{1}{\sqrt{3}}\{J_y,J_z\},\\
\Sigma_2&=\frac{1}{\sqrt{3}}\{J_z,J_x\},\\
\Sigma_3&=\frac{1}{\sqrt{3}}\{J_x,J_y\},\\
\Sigma_4&=\frac{1}{\sqrt{3}}(J_x^2-J_y^2),\\
\Sigma_5&=J_z^2-\frac{5}{4}\mathbb{I}_4,
\end{align}
\end{subequations}
with $-\Sigma_1\Sigma_2\Sigma_3\Sigma_4\Sigma_5=\mathbb{I}_4\equiv\Sigma_0$. The above five matrices satisfy
\begin{align}
  \{\Sigma_\alpha,\Sigma_\beta\}=2\delta_{\alpha\beta}  
\end{align}
and therefore generate a (Euclidean) Clifford algebra~\cite{PhysRevB.69.235206}. The following 10 operators
\begin{align}\label{eq:Sigma_ij_matrices}
\Sigma_{\alpha\beta}=\frac{1}{2\imath}[\Sigma_\alpha,\Sigma_\beta]
\end{align}
then generate SO(5) rotations. Eq. \ref{eq:Sigma_i_matrices} and Eq. \ref{eq:Sigma_ij_matrices} together define a basis for the  4-dimensional Hermitian matrices that generate SU(4).

The spin matrices can be written in terms of these $\Sigma_i$ as
\begin{align}
J_x&=\frac{\sqrt{3}}{2}\Sigma_{15}-\frac{1}{2}(\Sigma_{23}-\Sigma_{14});\\
J_y&=-\frac{\sqrt{3}}{2}\Sigma_{25}+\frac{1}{2}(\Sigma_{13}+\Sigma_{24});\\ J_z&=-\Sigma_{34}-\frac{1}{2}\Sigma_{12}
\end{align}
which generates an SU(2) subgroup of SU(4) with commutation relation
\begin{align}
    [J_i,J_j]=i\epsilon_{ijk} J_k.
\end{align}

\begin{table*}
    \centering
    \begin{eqnarray}
    \left(
\begin{array}{c|cccc|cccc}
    && \boldscriptr=\text{ even}&&&& \boldscriptr=\text{ odd}&&\\
     & \mcl{S} = B_2 & \mcl{S} = A_1 & \mcl{S} = B_1 & \mcl{S} = A_2 & \mcl{S} = B_2 & \mcl{S} = A_1 & \mcl{S} = B_1 & \mcl{S} = A_2 \\ \hline
     
 \rho_0(\boldscriptr_\mcl{S}) & +1 & +1 & +1& +1 &
   +1 & +1 & +1 & +1 \\
 \rho_1(\boldscriptr_\mcl{S}) & +1 & -1 & +1 & +1 &
   -1 & +1 & -1 & -1 \\
 \rho_2(\boldscriptr_\mcl{S}) & +1 & +1 & -1 & +1 &
   -1 & -1 & +1 & -1 \\
 \rho_3(\boldscriptr_\mcl{S}) & +1 & -1 & -1 & +1 &
   +1 & -1 & -1 & +1 \\
 \rho_4(\boldscriptr_\mcl{S}) & +1 & -1 & +1 & -1 &
   +1 & -1 & +1 & -1 \\
 \rho_5(\boldscriptr_\mcl{S}) & +1 & -1 & +1 & -1 &
   +1 & -1 & 1 & -1 \\
 \rho_{12}(\boldscriptr_\mcl{S}) & +1 & -1 & -1 & +1 & +1 & -1 & -1 & +1 \\
 \rho_{13}(\boldscriptr_\mcl{S}) & +1 & +1 & -1 & +1 & -1 & -1 & +1 & -1 \\
 \rho_{14}(\boldscriptr_\mcl{S}) & +1 & +1 & +1 & -1 & -1 & -1 & -1 & +1 \\
 \rho_{15}(\boldscriptr_\mcl{S}) & +1 & +1 & +1 & -1 & -1 & -1 & -1 & +1 \\
 \rho_{23}(\boldscriptr_\mcl{S}) & +1 & -1 & +1 & +1 & -1 & +1 & -1 & -1 \\
 \rho_{24}(\boldscriptr_\mcl{S}) & +1 & -1 & -1 & -1 & -1 & +1 & +1 & +1 \\
 \rho_{25}(\boldscriptr_\mcl{S}) & +1 & -1 & -1 & -1 & -1 & +1 & +1 & +1 \\
 \rho_{34}(\boldscriptr_\mcl{S}) & +1 & +1 & -1 & -1 & +1 & +1 & -1 & -1 \\
 \rho_{35}(\boldscriptr_\mcl{S}) & +1 & +1 & -1 & -1 & +1 & +1 & -1 & -1 \\
 \rho_{45}(\boldscriptr_\mcl{S}) & +1 & +1 & +1 & +1 & +1 & +1 & +1 & +1 \\
\end{array}
\right)\nonumber
\end{eqnarray}
    \caption{The values of $\rho_i(\boldscriptr_\mathcal{S})=\pm 1$ defined in Eq. \ref{eq:transformation of Sigma global to local} are written in this table.} 
    \label{tab:matrix realtions of SU(4) and j=3/2 basis}
\end{table*}

We now consider two kinds of lattice operators defined below
\begin{eqnarray}
\mcl{O}_i^{global}(\boldscriptr_\mcl{S}) = \psi^\dagger(\boldscriptr_\mcl{S})\Sigma_i \psi(\boldscriptr_\mcl{S}),
\end{eqnarray}
and 
\begin{eqnarray}
\mcl{O}_i^{local}(\boldscriptr_\mcl{S}) = \phi^\dagger(\boldscriptr_\mcl{S})\Sigma_i \phi(\boldscriptr_\mcl{S}),
\end{eqnarray}

Using the relation between the $\psi$ and the $\phi$ operators given in Eq.~\ref{eq:definition of phi}, we find that
\begin{eqnarray}\label{eq:transformation of Sigma global to local}
\mcl{O}_i^{local}(\boldscriptr_\mcl{S}) = \rho_i(\boldscriptr_\mcl{S}) \mcl{O}_{i}^{global}(\boldscriptr_\mcl{S})
\end{eqnarray}
where $\rho_i(\boldscriptr_\mcl{S}) = \pm 1$. In table~\ref{tab:matrix realtions of SU(4) and j=3/2 basis}, we write what $\rho_i(\boldscriptr_\mcl{S})$ are for different $\mcl{S}$ and $\boldscriptr$. In this table, we assume the form of $\boldscriptr$ as given in Eq.~\ref{eq:definition of r_mu} which is
\begin{eqnarray}\label{eq:def of boldscript_r}
\boldscriptr = n_x \mbf{R}_1 + n_y \mbf{R}_2.
\end{eqnarray}
We say $\boldscriptr =$ even (odd)  if $(n_x+n_y)$ is even (odd).

\section{The \texorpdfstring{$\mathcal{G}(\boldscriptr_\mathcal{S})$}{} matrices of transformation to manifestly SU(4) invariant local basis}\label{sec:appen_g_matrices}

The set of $\mathcal{G}(\boldscriptr_\mathcal{S})$ matrices that lead to the form of $\eta(\boldscriptr_\mcl{S},\boldscriptr'_{\mcl{S}'})$ is given in Eq. \ref{eq_etaexp} is written below. Here we again take $\boldscriptr$ to be of the form as in Eq.~\ref{eq:def of boldscript_r}. 

For $(n_x+n_y)=$even :
\begin{eqnarray}\label{eq:even g-matrices}
&\mathcal{G}(\boldscriptr_{B_2}) &=  (-1)^{\frac{n_x-n_y}{2}} \nonumber \\
&\mathcal{G}(\boldscriptr_{A_1}) &=  (-1)^{\frac{n_x-n_y}{2}}U_y \nonumber \\
&\mathcal{G}(\boldscriptr_{B_1}) &=  (-1)^{\frac{n_x-n_y}{2}}U_zU_y \nonumber \\
&\mathcal{G}(\boldscriptr_{A_2}) &=  (-1)^{\frac{n_x-n_y}{2}}U_xU_zU_y 
\end{eqnarray}

And for $(n_x+n_y)=$odd :
\begin{eqnarray}\label{eq:odd g-matrices}
&\mathcal{G}(\boldscriptr_{B_2}) &=  (-1)^{\frac{n_x-n_y+1}{2}}U_xU_y \nonumber \\
&\mathcal{G}(\boldscriptr_{A_1}) &=  (-1)^{\frac{n_x-n_y+1}{2}}(-U_x) \nonumber \\
&\mathcal{G}(\boldscriptr_{B_1}) &=  (-1)^{\frac{n_x-n_y+1}{2}}(-U_zU_x) \nonumber \\
&\mathcal{G}(\boldscriptr_{A_2}) &=  (-1)^{\frac{n_x-n_y+1}{2}}U_z 
\end{eqnarray}
Although we will be using this particular form, there are other choices for the $\mathcal{G}(\boldscriptr_\mcl{S})$ matrices which lead to same form for the $\eta(\boldscriptr_\mcl{S},\boldscriptr'_{\mcl{S}'})$.
    

\section{Low-energy Hamiltonian}
\label{appen_band}


\subsection{Band structure}

To diagonalize the Hamiltonian given by Eq. \ref{eq:Hamiltonian in SU(4) diagonal form}, we define Fourier space operators
\begin{align}
\phi_f(\mbf{k},\mathcal{S})=\frac{1}{\sqrt{N}}\sum_{\boldscriptr}e^{i{\bf k}\cdot\boldscriptr}\phi_f(\boldscriptr_\mathcal{S})
\end{align}
where $N$ is the total number of magnetic unit-cells, $f=1,2,3,4$ are the four SU(4) flavours and ${\bf k}$ runs over the magnetic Brillouin zone (Fig. \ref{fig:band structure}).

 In terms of these Fourier space operators, the Hamiltonian in Eq. \ref{eq:Hamiltonian in SU(4) diagonal form} can now be written as
	\begin{eqnarray}\label{eq:Hamiltonian in fourier space}
	H = -\frac{t}{\sqrt{3}} \sum_{f=1}^4 \sum_{\mbf{k}} \sum_{\mathcal{S},\mathcal{S}'} \phi_{f}(\mbf{k},\mathcal{S}) ~[h(\mbf{k})]_{\mathcal{S}\mathcal{S}'}~\phi_{f}(\mbf{k},\mathcal{S}')\nonumber\\
	\end{eqnarray}
	where
\begin{eqnarray}
&&h(\mathbf{k}) =\nonumber \\
&&  \begin{pmatrix}
						0&		1+e^{-i\mathbf{k\cdot R_2}}&		0&		e^{-i\mathbf{k \cdot R_1}}\\
						1+e^{i\mathbf{k\cdot R_2}}&		0&		1&		0\\
						0&		1&						0&		1-e^{i\mathbf{k\cdot R_2}}\\
						e^{i\mathbf{k\cdot R_1}}&			0&		1-e^{-i\mathbf{k\cdot R_2}}&		0		
						\end{pmatrix}\nonumber \\
\end{eqnarray}
	 Diagonalizing the $h(\mbf{k})$ matrix, we get the band structure shown in Fig.~\ref{fig:band structure}.

\subsection{The low energy Dirac Hamiltonian}

	At $1/4$th filling, the valence band touches the conduction band at two Dirac points in the BZ given by Eq.~\ref{eq:position of dirc points}. To get the low-energy Hamiltonian, we first write the $\phi(\boldscriptr_\mcl{S})$ operators in terms of the soft modes $\phi_{f\mcl{S}\tau}$ as
	\begin{align}\label{eqn_slow_mod_expansion_of_phi}
	\phi_{f}(\boldscriptr_\mcl{S}) = \sqrt{\mcl{A}}\sum_{\tau=\pm 1}e^{i\tau\mbf{Q} \cdot \boldscriptr} \phi_{f\mcl{S}\tau}(\boldscriptr).
	\end{align}
	Here, $\boldscriptr$ (defined in Eq.~\ref{eq:definition of r_mu}) denotes position of a particular magnetic unit cell and $\mcl{A}$ is the area of a single magnetic unit cell. The $\phi_{f\mcl{S}\tau}(\boldscriptr)$ operators are defined for each valley ($\tau=\pm 1$ labels the valleys) as 
	\begin{eqnarray}
	\phi_{f\mcl{S}\tau}(\boldscriptr) = \frac{1}{\sqrt{\mcl{A}}}\sum_{\mbf{q}} e^{i\mbf{q} \cdot \boldscriptr} \phi_{f}(\tau\mbf{Q}+\mbf{q},\mcl{S}). 
	\end{eqnarray}
	In  the above summation, $\mbf{q}$ runs over half of the magnetic Brillouin zone for each $\tau$ such that the Dirac point $\tau \mbf{Q}$ is contained in that half. {These $\phi_{f\mcl{S}\tau}$ fields vary slowly over the magnetic unit cells. }

Now to get the low-energy Hamiltonian, we use the form of $\phi(\boldscriptr_\mcl{S})$ as in Eq.~\ref{eqn_slow_mod_expansion_of_phi} to rewrite the Hamiltonian in Eq.~\ref{eq:Hamiltonian in SU(4) diagonal form} in terms of the $\phi_{f\mcl{S}\tau}(\boldscriptr)$ operators. We also use the following expansion for $\phi_{f\mcl{S}\tau}$~:
\begin{eqnarray}
\phi_{f\mcl{S}\tau} (\boldscriptr + \vec{\delta}) = \phi_{f\mcl{S}\tau}(\boldscriptr) + \vec{\delta}\cdot \mbf{\nabla} \phi_{f\mcl{S}\tau}(\boldscriptr) + \mcl{O}(\delta^2).
\end{eqnarray}
Here, $\vec{\delta}$ can be some magnetic translation vector ($\mbf{R}_1$ or $\mbf{R}_2$). 
The soft-mode continuum Hamiltonian is then obtained by rewriting the Hamiltonian in Eq.~\ref{eq:Hamiltonian in SU(4) diagonal form} using the above expression and keeping terms that are linear in the derivative. The final form of the Hamiltonian is given below in Eq.~\ref{eq_linear_hamltonian}.
    \begin{widetext}
	\begin{eqnarray}\label{eq_linear_hamltonian}
	H =  \sum_{f=1}^4\sum_{\tau=\pm1}~\sum_{\mcl{S},\mcl{S}'=A_1,A_2,B_1,B_2}~\int d^2\mbf{x}~ \phi_{f\mcl{S}\tau}^{\dagger}\left( \mbf{x}\right) \left[ h_0^{(\tau)} -i\tau h_x\partial_x -i \tau h_y\partial_y \right]_{\mcl{S}\mcl{S}'}\phi_{f\mcl{S}'\tau}\left( \mbf{x}\right).\nonumber\\
	\end{eqnarray}

	where, 
\begin{subequations}
\begin{eqnarray}
&h_0^{(\tau)} &= -\frac{t\mcl{l}}{\sqrt{3}}\begin{pmatrix}
				0&		1+i\tau&		0&		-i\tau\\
				1-i\tau&		0&		1&		0\\
				0&		1&		0&		1+i\tau\\
				i\tau&		0&		1-i\tau&		0
				\end{pmatrix},\nonumber
~~~h_x=\frac{t\mcl{l} }{\sqrt{3}}
				\begin{pmatrix}
				0&	0&	0&	3\\
				0&	0&	0&	0\\
				0&	0&	0&	0\\
				3 &	0&	0&	0\\
				\end{pmatrix},\nonumber
~~~h_y=\frac{t\mcl{l} }{\sqrt{3}}
				\begin{pmatrix}
				0&	-\sqrt{3}&	0&	0\\
				-\sqrt{3}&	0&	0&	0\\
				0&	0&	0&	\sqrt{3}\\
				0&	0&	\sqrt{3}&	0\\
				\end{pmatrix},	
\end{eqnarray}
\end{subequations}

\end{widetext}
where $\mcl{l}$ is the length of each side of a hexagon of the honeycomb lattice.

    As because the system is at 1/4th filling, we need to further project this Hamiltonian into the lowest two bands to get the low energy theory. For this, we take the eigenvectors corresponding to the lowest two eigenvalues of $h_0^{(\tau)}$ and project $h_x,h_y$ into the subspace of these two eigenvectors. This way of projecting the Hamiltonian is correct up to linear order in derivatives, which is sufficient in this case since the Hamiltonian~\ref{eq_linear_hamltonian} is also linear in derivatives.

	With this, the final form of the low-energy Dirac Hamiltonian is the following
	\begin{widetext}
	\begin{eqnarray}\label{eq:continuum hamiltonian1}
	H_D = v_F \sum_{f=1}^4\sum_{\tau=\pm1}\int d^2\mbf{x}~ \sum_{\alpha,\beta=1}^2\chi_{f\alpha\tau}^{\dagger}\left( \mbf{x}\right) \left[ -i \tau  \sigma_x\partial_x -i\sigma_y\partial_y \right]_{\alpha\beta}\chi_{f\beta\tau}\left( \mbf{x}\right)\nonumber\\
	\end{eqnarray}
	\end{widetext}
	Here, $v_F = \frac{t\mcl{l}}{\sqrt{2}}$ and  $\sigma_x,\sigma_y,\sigma_z$ are the three Pauli matrices. Also the operators $\chi_{f\alpha\tau}(\mbf{x})$ are defined as
	\begin{eqnarray}\label{eq:final definition of chi}
	\chi_{f\alpha\tau}(\mbf{x}) = \sum_{\mcl{S}=A_1,A_2,B_1,B_2}~\mcl{W}^{(\tau)\dagger}_{\mcl{S\alpha}}~ \phi_{f\mcl{S}\tau}(\mbf{x}).
	\end{eqnarray}
	which is the inverse of Eq. \ref{eq:expansion of phi in soft modes} of the main text.

	In the above equation,
	\begin{eqnarray}
	\mcl{W}^{(\tau)\dagger}_{\mcl{S}\alpha} = \sum_{\beta=1}^2~ [W_{\tau}]_{\alpha\beta} [T_{\tau}]_{\beta \mcl{S}}.
	\end{eqnarray}
	Here,
\begin{eqnarray}
	T_{\tau} = \begin{pmatrix}
		\frac{i\tau}{\sqrt{2}}&	-\frac{1-i\tau}{\sqrt{6}}&	0&	\frac{1}{\sqrt{6}}\\
		0&	\frac{1}{\sqrt{6}}&		\frac{1}{\sqrt{2}}&		\frac{1+i\tau}{\sqrt{6}} 
	\end{pmatrix}
\end{eqnarray}
	projects the annihilation operators to the lowest two bands. The matrices
	\begin{align}
	W_{\tau} = \left[exp\left({-i \tau \theta \frac{\sigma_x}{2}}\right)~exp\left({-i\tau\frac{\pi}{4}\frac{\sigma_z}{2}}\right) \right]
	\end{align} 
	with $\theta = cos^{-1}(\frac{1}{\sqrt{3}})$ are used to perform some extra unitary rotations on the spinors to bring the Dirac Hamiltonian in its canonical form.  


\section{Symmetry transformation of the soft modes}\label{sec:Symmetry_transformation_calculations}

\subsection{Action of lattice symmetry transformations on $\phi(\boldscriptr_\mcl{S})$ operators}
The transformation properties of the soft modes $\chi(\mbf{x})$ under the action of lattice symmetries can be determined from that of the original $j=3/2$ operators $\psi(\mbf{r},\mcl{s})$ as given in Eq.~\ref{eq:transformation of psi under TR}-\ref{eq:transformation of psi under C2}. Using these, the transformations of the $\phi(\boldscriptr_\mcl{S})$ operators are obtained as follows

\begin{align}\label{eq:transformation for phi}
   \mbb{S} : \phi^\dagger_{f}(\boldscriptr_\mcl{S}) \to \phi^{\prime \dagger}_{f}(\boldscriptr_\mcl{S}) = \left[ \mcl{G}(\boldscriptr_\mcl{S})^{\dagger}~ \mcl{U}_{\mbb{S}}~ \mcl{G}(\boldscriptr'_{\mcl{S}'}) \right]_{f'f} \phi^\dagger_{f'}(\boldscriptr'_{\mcl{S}'}).
\end{align}

Here, the site at $\boldscriptr'_{\mcl{S}'}$ goes to $\boldscriptr_\mcl{S}$ under the action of the lattice symmetry $\mbb{S}$.

Using this, one can now derive the transformations of the soft modes $\chi(\mathbf{x})$ which are defined in terms of the Fourier transforms of the $\phi(\boldscriptr_\mcl{S})$ operators(Eq.~\ref{eq:final definition of chi}). Below we provide some details of the transformations of both the $\phi(\boldscriptr_\mcl{S})$ and the $\chi(\mbf{x})$ operators under the action of various lattice symmetries. For $\phi(\boldscriptr_\mcl{S})$, we write down how the operators in a particular magnetic unit cell at $\boldscriptr$ transform where  
\begin{eqnarray}
\boldscriptr = n_x \mbf{R_1} + n_y\mbf{R_2}.
\end{eqnarray}
with $n_x,n_y$ being integers.

\textit{Transformation under translations $\mathbf{T}_1$ and $\mbf{T_2}$:} 
Under the action of $\mbf{T_1}$,
\begin{eqnarray}\label{eq:transformation under translation}
&&\phi(\boldscriptr_{B_2}) \to (-1)^{(n_x+n_y)}\left[\Omega^f_{\mathbf{T}_1}\right] \phi(~(\boldscriptr-\mbf{R_1-R_2})_{B_1}~), \nonumber \\
&&\phi(\boldscriptr_{A_1}) \to (-1)^{(n_x+n_y+1)}\left[\Omega^f_{\mathbf{T}_1}\right] \phi((\boldscriptr-\mbf{R_1})_{A_2}), \nonumber \\
&&\phi(\boldscriptr_{B_1}) \to (-1)^{(n_x+n_y+1)}\left[\Omega^f_{\mathbf{T}_1}\right] \phi(\boldscriptr_{B_2}), \nonumber \\
&&\phi(\boldscriptr_{A_2}) \to (-1)^{(n_x+n_y+1)}\left[\Omega^f_{\mathbf{T}_1}\right] \phi((\boldscriptr-\mbf{R_2})_{A_1}),   \nonumber \\
\end{eqnarray}
where, \begin{eqnarray}
\Omega^f_{\mathbf{T}_1} = -i \Sigma_{23}.
\end{eqnarray}

Similarly, under the action of $\mbf{T_2}$, 
\begin{eqnarray}
&&\phi(\boldscriptr_{B_2}) \to (-1)^{(n_x+n_y)}\left[\Omega^f_{\mathbf{T}_2}\right] \phi((\boldscriptr-\mbf{R_1})_{B_1}) \nonumber \\
&&\phi(\boldscriptr_{A_1}) \to (-1)^{(n_x+n_y+1)}\left[\Omega^f_{\mathbf{T}_2}\right] \phi((\boldscriptr-\mbf{R_1+R_2})_{A_2}) \nonumber \\
&&\phi(\boldscriptr_{B_1}) \to (-1)^{(n_x+n_y+1)}\left[\Omega^f_{\mathbf{T}_2}\right] \phi((\boldscriptr+\mbf{R_2})_{B_2}) \nonumber \\
&&\phi(\boldscriptr_{A_2}) \to (-1)^{(n_x+n_y+1)}\left[\Omega^f_{\mathbf{T}_2}\right] \phi(\boldscriptr_{A_1}).   
\end{eqnarray}
where
\begin{eqnarray}
\Omega^f_{\mathbf{T}_2} = i\Sigma_{13}.
\end{eqnarray}


\textit{Transformation under $\mbf{C'_2}$: } Under a $\mbf{C_2'}$ rotation, 
\begin{eqnarray}
&&\phi(\boldscriptr_{B_2}) \to (-1)^{n_x}\left[\Omega^f_{\mbf{C_2'}}\right] \phi(\boldscriptr'_{B_2}) \nonumber \\
&&\phi(\boldscriptr_{A_1}) \to (-1)^{n_x}\left[\Omega^f_{\mbf{C_2'}}\right] \phi((\boldscriptr'-\mbf{R_2})_{A_1}) \nonumber \\
&&\phi(\boldscriptr_{B_1}) \to (-1)^{n_x}\left[\Omega^f_{\mbf{C_2'}}\right] \phi((\boldscriptr'-\mbf{R_2})_{B_1}) \nonumber \\
&&\phi(\boldscriptr_{A_2}) \to (-1)^{n_x+1}\left[\Omega^f_{\mbf{C_2'}}\right] \phi(\boldscriptr'_{A_2}) 
\end{eqnarray}
Here,
\begin{eqnarray}
\boldscriptr' = n_x \mbf{R_1} - n_y\mbf{R_2},
\end{eqnarray}
and
\begin{eqnarray}
\Omega^f_{\mbf{C_2'}} = \frac{i}{\sqrt{2}} \left(\Sigma _{14}-\Sigma _{24}\right).
\end{eqnarray}


\begin{widetext}
	\textit{Transformation under $\mbf{C_3}$: }Under the action of $\mbf{C_3}$ rotation, transformation of the $\phi(\boldscriptr)$ operators are given below.
	\begin{center}
		\begin{tabular}{c | c | c}
			Sublattice&   $(n_x+n_y)=$ even& $(n_x+n_y)=$ odd \\
			\hline
			$\phi(\boldscriptr_{B_2}) \to$ & $(-1)^{n_x}[\Omega^f_{\mbf{C_3}}]~ \phi\left(~(\mbf{C_3}[\boldscriptr])_{B_1}~\right)$ & $(-1)^{n_x}[\Omega^f_{\mbf{C_3}}]~\phi\left(~(\mbf{C_3}[\boldscriptr])_{B_2}~\right)$ 	  \\
			$\phi(\boldscriptr_{A_1}) \to$& $(-1)^{n_x}[\Omega^f_{\mbf{C_3}}]~ \phi\left(~(\mbf{C_3}[\boldscriptr])_{A_2}~\right)$ & $(-1)^{n_x}[\Omega^f_{\mbf{C_3}}]~ \phi\left(~(\mbf{C_3}[\boldscriptr]-\mbf{R_2})_{A_1}~\right)$  \\
			$\phi(\boldscriptr_{B_1}) \to $& $(-1)^{n_x+1}[\Omega^f_{\mbf{C_3}}]~ \phi\left(~(\mbf{C_3}[\boldscriptr]-\mbf{R_2})_{B_1}~\right)$ & $(-1)^{n_x}[\Omega^f_{\mbf{C_3}}]~ \phi\left(~(\mbf{C_3}[\boldscriptr]-\mbf{R_2})_{B_2}~\right)$  \\
			$\phi(\boldscriptr_{A_2}) \to $& $(-1)^{n_x+1}[\Omega^f_{\mbf{C_3}}]~ \phi\left(~(\mbf{C_3}[\boldscriptr]-\mbf{R_2})_{A_1}~\right)$ & $(-1)^{n_x}[\Omega^f_{\mbf{C_3}}]~ \phi\left(~(\mbf{C_3}[\boldscriptr]-\mbf{R_1-R_2})_{A_2}~\right)$ 
		\end{tabular}
	\end{center}
	Here 
	\begin{eqnarray}
	\mbf{C_3}[\boldscriptr] = \begin{cases}
	                    \frac{-n_x+ n_y}{2}\mbf{R_1} -\frac{3n_x+n_y}{2}\mbf{R_2} & \text{if $n_x+n_y=$ even}\\
	                    \frac{-n_x+ n_y+1}{2}\mbf{R_1} -\frac{3n_x+n_y-1}{2}\mbf{R_2} & \text{if $n_x+n_y=$ odd}
	                    \end{cases}
	\end{eqnarray}
	and
	\begin{eqnarray}
	\Omega^f_{\mbf{C_3}} = \frac{1}{4} \left( \Sigma_0 -\sqrt{3}\Sigma_1 -\sqrt{3} \Sigma_2 +\sqrt{3} \Sigma_3 -i\Sigma_{12} -i\Sigma_{13} +i\Sigma_{23} +i\sqrt{3}\Sigma_{45}\right).
	\end{eqnarray}

	\textit{Transformation under $\mbf{S_6}$: }Transformation of $\phi(\boldscriptr_\mcl{S})$ operators under the action of $\mbf{S_6}$ are as follows
	\begin{center}
		\begin{tabular}{c | c | c}
			&   $(n_x+n_y)=$ even& $(n_x+n_y)=$ odd \\
			\hline
			$ \phi(\boldscriptr_{B_2}(n_x,n_y))$ $\to$&  $(-1)^{\frac{n_x+n_y}{2}}[\Omega^f_{\mbf{S_6}}]~ \phi\left(~(\mbf{S_6}[\boldscriptr])_{A_1}~\right)$&	 $(-1)^{\frac{n_x+n_y+1}{2}}[\Omega^f_{\mbf{S_6}}]~ \phi\left(~(\mbf{S_6}[\boldscriptr])_{A_2}~\right)$ \\
			$ \phi(\boldscriptr_{A_1}(n_x,n_y))$$\to$& $(-1)^{\frac{n_x+n_y}{2}}[\Omega^f_{\mbf{S_6}}]~ \phi\left(~(\mbf{S_6}[\boldscriptr])_{B_1}~\right)$& $(-1)^{\frac{n_x+n_y+1}{2}}[\Omega^f_{\mbf{S_6}}]~ \phi\left(~(\mbf{S_6}[\boldscriptr]+\mbf{R_1})_{B_2}~\right)$ \\
			$\phi(\boldscriptr_{B_1}(n_x,n_y))$$\to$& $(-1)^{\frac{n_x+n_y}{2}}[\Omega^f_{\mbf{S_6}}]~ \phi\left(~(\mbf{S_6}[\boldscriptr])_{A_2}~\right)$& $(-1)^{\frac{n_x+n_y+1}{2}}[\Omega^f_{\mbf{S_6}}]~ \phi\left(~(\mbf{S_6}[\boldscriptr]+\mbf{R_1-R_2})_{A_1}~\right)$\\
			$\phi(\boldscriptr_{A_2}(n_x,n_y))$$\to$& $(-1)^{\frac{n_x+n_y}{2}+1}[\Omega^f_{\mbf{S_6}}]~ \phi\left(~(\mbf{S_6}[\boldscriptr]-\mbf{R_2})_{B_1}~\right)$& $(-1)^{\frac{n_x+n_y+1}{2}}[\Omega^f_{\mbf{S_6}}]~ \phi\left(~(\mbf{S_6}[\boldscriptr]+\mbf{R_1-R_2})_{B_2}~\right)$
		\end{tabular}
	\end{center}		
	\begin{eqnarray}
	\end{eqnarray}
\end{widetext}
	Here, 
	\begin{eqnarray}
	\mbf{S_6}[\boldscriptr] = \begin{cases}
	                    \frac{n_x+n_y}{2} \mbf{R_1} -\frac{3n_x-n_y}{2}\mbf{R_2} & \text{if $n_x+n_y=$ even}\\
	                    \frac{n_x+n_y-1}{2} \mbf{R_1} -\frac{3n_x-n_y-1}{2} \mbf{R_2} & \text{if $n_x+n_y=$ odd}
	                    \end{cases} \nonumber\\
	\end{eqnarray}
	and 
	\begin{eqnarray}
	&&\Omega^f_{\mbf{S_6}} = -\frac{1}{4} \left( \sqrt{3}\Sigma_0 +\Sigma_1 +\Sigma_2 -\Sigma_3 +i \sqrt{3} \Sigma_{12} \right. \nonumber \\
	&&\left.\qquad \qquad \qquad +i \sqrt{3} \Sigma_{13} -i \sqrt{3} \Sigma_{23} +i \Sigma_{45}\right)
	\end{eqnarray}

\textit{Transformation under $\mbf{I}$~: }Under inversion, 
\begin{eqnarray}
&&\phi(\boldscriptr_{B_2}) \to (-1)^{n_x+n_y+1}\left[\Omega^f_{\mbf{I}}\right]~ \phi\left(~(\mbf{I}[\boldscriptr])_{A_2}~\right) \nonumber \\
&&\phi(\boldscriptr_{A_1}) \to (-1)^{n_x+n_y}\left[\Omega^f_{\mbf{I}}\right] \phi\left(~(\mbf{I}[\boldscriptr]-\mbf{R_2})_{B_1}~\right) \nonumber \\
&&\phi(\boldscriptr_{B_1}) \to (-1)^{n_x+n_y}\left[\Omega^f_{\mbf{I}}\right] \phi\left(~(\mbf{I}[\boldscriptr]-\mbf{R_2})_{A_1}~\right) \nonumber \\
&&\phi(\boldscriptr_{A_2}) \to (-1)^{n_x+n_y}\left[\Omega^f_{\mbf{I}}\right] \phi\left(~(\mbf{I}[\boldscriptr])_{B_2}~\right) 
\end{eqnarray}
where
\begin{eqnarray}
\mbf{I}[\boldscriptr] = -n_x\mbf{R_1} -n_y\mbf{R_2},
\end{eqnarray}
and 
\begin{eqnarray}
\Omega^f_{\mbf{I}} = i \Sigma_{45}.
\end{eqnarray}

\textit{Transformation under $\mbf{\sigma_d}$: }Under the action of reflection,
\begin{eqnarray}
&&\phi(\boldscriptr_{B_2}) \to (-1)^{n_y}\left[\Omega^f_{\mbf{\sigma_d}}\right] \phi\left(~(\mbf{\sigma_d}[\boldscriptr])_{A_2}~\right) \nonumber \\
&&\phi(\boldscriptr_{A_1}) \to (-1)^{n_y}\left[\Omega^f_{\mbf{\sigma_d}}\right] \phi\left(~(\mbf{\sigma_d}[\boldscriptr])_{B_1}~\right) \nonumber \\
&&\phi(\boldscriptr_{B_1}) \to (-1)^{n_y}\left[\Omega^f_{\mbf{\sigma_d}}\right] \phi\left(~(\mbf{\sigma_d}[\boldscriptr])_{A_1}~\right) \nonumber \\
&&\phi(\boldscriptr_{A_2}) \to (-1)^{n_y}\left[\Omega^f_{\mbf{\sigma_d}}\right] \phi\left(~(\mbf{\sigma_d}[\boldscriptr])_{B_2}~\right) 
\end{eqnarray}
where
\begin{eqnarray}
\mbf{\sigma_d}[\boldscriptr] = -n_x \mbf{R_1} + n_y \mbf{R_2},
\end{eqnarray}
and 
\begin{eqnarray}
\Omega^f_{\mbf{\sigma_d}} = \frac{i}{\sqrt{2}} \left(\Sigma _{15}-\Sigma _{25}\right).
\end{eqnarray}

\subsection{Transformation of the \texorpdfstring{$\chi(\mbf{x})$}{} operators under lattice symmetries}
\label{appen_softmodetrans}
Under the microscopic time-reversal, we have
\begin{eqnarray}
\mbb{T}~:~\chi(\mbf{x}) \to \left(\Omega_{\mbb{T}}^f \otimes \Omega_{\mbb{T}}^c\right)~K~\chi(\mbf{x}),
\label{eq:chitr}
\end{eqnarray} 	
where
\begin{eqnarray}
&&\Omega_{\mbb{T}}^f = i \Sigma_{13}, \\
&&\Omega_{\mbb{T}}^c = \gamma_1\zeta_2.
\end{eqnarray}	
and $K$ represents complex conjugation. 

Under the action of $\mbf{T_{1(2)}}$,
\begin{eqnarray}
\mbf{T_{1(2)}}~:~\chi(\mathbf{x}) \to \left( \Omega^f_{\mathbf{T}_{1(2)}}\otimes~ \Omega^c_{\mathbf{T}_{1(2)}} \right)\chi(\mathbf{x}).
\label{eq:transchi12}
\end{eqnarray}
where, \begin{eqnarray}
&&\Omega^f_{\mathbf{T}_{1}} = -i \Sigma_{23}.\\
&&\Omega^c_{\mathbf{T}_1} = -i\zeta_3.
\end{eqnarray}
and 
\begin{eqnarray}
&&\Omega^f_{\mathbf{T}_2} = i\Sigma_{13},\\
&&\Omega^c_{\mathbf{T}_2} = -i\zeta_2
\end{eqnarray}	
The superscripts $f$ and $c$ in $\Omega^f_{\mathbf{T}_1},\Omega^c_{\mathbf{T}_1} $ stand for ``flavor" and ``chiral" respectively since these matrices act on the SU(4) flavor space and the chiral space.

Under the action of $\mbf{C_2'}$,
\begin{eqnarray}
\mbf{C_2'}~:~\chi(\mathbf{x})  \to \left( \Omega^f_{\mbf{C_2'}} \otimes~\Omega^c_{\mbf{C_2'}} \right) \chi(\mbf{C_2}^{\prime -1}\mathbf{x}).\nonumber \\
\end{eqnarray}

where, 
\begin{eqnarray}
&&\Omega^f_{\mbf{C_2'}} = \frac{i}{\sqrt{2}} \left(\Sigma _{14}-\Sigma _{24}\right)\\
&&\Omega^c_{\mbf{C_2'}} = \frac{1}{\sqrt{2}}\gamma_2(\zeta_3 - \zeta_2).
\end{eqnarray}

Under the action of $\mbf{C_3}$, the soft modes transform in the following way:
	\begin{eqnarray}
	\mbf{C_3}~:~\chi(\mathbf{x}) \to \left(\Omega^f_{\mbf{C_3}} \otimes~ \Omega^c_{\mbf{C_3}} \right) \chi(\mbf{C_3}^{-1}\mathbf{x}). 
	\end{eqnarray}
	Here,
	\begin{eqnarray}
	&&\Omega^f_{\mbf{C_3}} = \frac{1}{4} \left( \Sigma_0 -\sqrt{3}\Sigma_1 -\sqrt{3} \Sigma_2 +\sqrt{3} \Sigma_3 -i\Sigma_{12} \right. \nonumber \\
	&&\left.\qquad \qquad \qquad -i\Sigma_{13} +i\Sigma_{23} +i\sqrt{3}\Sigma_{45}\right),
	\end{eqnarray}
	and
	\begin{eqnarray}
	 \Omega^c_{\mbf{C_3}} = 
 \frac{1}{4} \left(-\mathbb{I}_4+i
\sqrt{3} \gamma_0\right)\left(\mathbb{I}_4+i\zeta_1+i\zeta_2+i\zeta_3 \right)
	\end{eqnarray}
	
Under $\mbf{S_6}$, the soft modes transform in the following way:
	\begin{eqnarray}
	\mbf{S_6}~:~\chi(\mathbf{x}) \to \left(\Omega^f_{\mbf{S_6}} \otimes~ \Omega^c_{\mbf{S_6}}\right) \chi(\mbf{S_6}^{-1}\mathbf{x}).
	\end{eqnarray}
	
	Here, 
	\begin{eqnarray}
	&&\Omega^f_{\mbf{S_6}} = -\frac{1}{4} \left( \sqrt{3}\Sigma_0 +\Sigma_1 +\Sigma_2 -\Sigma_3 +i \sqrt{3} \Sigma_{12} \right. \nonumber \\
	&&\left.\qquad \qquad \qquad +i \sqrt{3} \Sigma_{13} -i \sqrt{3} \Sigma_{23} +i \Sigma_{45}\right)
	\end{eqnarray}
	and
	\begin{eqnarray}
    \Omega^c_{\mbf{S_6}} = \frac{1}{4} \left(\sqrt{3}\mathbb{I}_4-i\gamma_0\right) \left(\mathbb{I}_4-i\zeta_1-i\zeta_2-i\zeta_3\right)
	\end{eqnarray}	
	
Under the action of inversion,
\begin{eqnarray}
\mbf{I}~:~\chi(\mathbf{x})  \to \left(\Omega^f_{\mbf{I}}\otimes~\Omega^c_{\mbf{I}}\right) \chi(-\mathbf{x}).
\end{eqnarray}
where, 
\begin{eqnarray}
&&\Omega^f_{\mbf{I}} = i \Sigma_{45},\\
&&\Omega^c_{\mbf{I}} = i\gamma_0
\end{eqnarray}

Under the action of $\mbf{\sigma_d}$,
\begin{eqnarray}
\mbf{\sigma_d}~:~\chi(\mathbf{x})  \to \left(\Omega^f_{\mbf{\sigma_d}} \otimes~  \Omega^c_{\mbf{\sigma_d}} \right) \chi(\mbf{\sigma_d}^{-1}\mathbf{x}).\nonumber \\
\label{eq:chisigd}
\end{eqnarray}
where, 
\begin{eqnarray}
&&\Omega^f_{\mbf{\sigma_d}} = \frac{i}{\sqrt{2}} \left(\Sigma _{15}-\Sigma _{25}\right)\\
&&\Omega^c_{\mbf{\sigma_d}} = \frac{1}{\sqrt{2}}\gamma_1(\zeta_2 - \zeta_3).
\end{eqnarray}


\section{Irreducible representations of the IR space group}\label{appen_definition_of_space_group_representaiton}
As mentioned in the main text, the IR space group has total 96 elements and these can be divided into 20 conjugacy classes. So, there are 20 irreducible representations of the IR space group. Among these, 10 have $+ve$ trace for $2\pi$ rotations. In the tables~\ref{tab_def_of_1d_irreps}, \ref{tab_def_of_2d_irreps}, \ref{tab_def_of_3d_irreps}, we write down these irreducible representations by showing how fermions bilinears in Eq. \ref{eq_massexp} (which we symbolically denote as $X_1, X_2,X_3$ etc) transform.

\begin{table}[h!]
	\begin{tabular}{c|c|c|c|c|c|c|c|c}
		Irrep& mass& ${\bf T_1}$& ${\bf T_2}$& ${\bf I}$& ${\bf C_3}$& $\boldsymbol{S_6}$& ${\bf C_2'}$& $\boldsymbol{\sigma_d}$\\\hline
		$\mcl{A}_{1g}$& $X\to$& $X$& $X$& $X$& $X$& $X$& $X$& $X$\\ \hline
		$\mcl{A}_{2g}$& $X\to$& $X$& $X$& $X$& $X$& $X$& $-X$& $-X$\\ \hline
		$\mcl{A}_{1u}$& $X\to$& $X$& $X$& $-X$& $X$& $-X$& $X$& $-X$\\ \hline
		$\mcl{A}_{2u}$& $X\to$& $X$& $X$& $-X$& $X$& $-X$& $-X$& $X$\\ \hline						
	\end{tabular}
\caption{List of 1-dimensional irreps of the IR space group}
\label{tab_def_of_1d_irreps}
\end{table}

\begin{table}[h!]
	\begin{tabular}{c|c|c|c|c|c|c|c|c}
		Irrep& mass& ${\bf T_1}$& ${\bf T_2}$& ${\bf I}$& ${\bf C_3}$& $\boldsymbol{S_6}$& ${\bf C_2'}$& $\boldsymbol{\sigma_d}$\\\hline
		$\mcl{E}_{g}$& $X_1\to$& $X_1$& $X_1$& $X_1$& $-\frac{X_1}{2} + \frac{\sqrt{3}X_2}{2}$& $-\frac{X_1}{2} - \frac{\sqrt{3}X_2}{2}$& $-X_1$& $-X_1$\\
		& $X_2\to$& $X_2$& $X_2$& $X_2$& $-\frac{\sqrt{3}X_1}{2}-\frac{X_2}{2}$& $\frac{\sqrt{3}X_1}{2}-\frac{X_2}{2}$& $X_2$& $X_2$\\ \hline		
		$\mcl{E}_{u}$& $X_1\to$& $X_1$& $X_1$& $-X_1$& $-\frac{X_1}{2} + \frac{\sqrt{3}X_2}{2}$& $\frac{X_1}{2} + \frac{\sqrt{3}X_2}{2}$& $-X_1$& $X_1$\\
		& $X_2\to$& $X_2$& $X_2$& $-X_2$& $-\frac{\sqrt{3}X_1}{2}-\frac{X_2}{2}$& $-\frac{\sqrt{3}X_1}{2}+\frac{X_2}{2}$& $X_2$& $-X_2$\\ \hline		
	\end{tabular}
	\caption{List of 2-dimensional irreps of the IR space group}
	\label{tab_def_of_2d_irreps}
\end{table}

\begin{table}[h!]
	\begin{tabular}{c|c|c|c|c|c|c|c|c}
		Irrep& mass& ${\bf T_1}$& ${\bf T_2}$& ${\bf I}$& ${\bf C_3}$& ${\bf S_6}$& ${\bf C_2'}$& $\boldsymbol{\sigma_d}$\\\hline
		
		$\mcl{T}_{1g}$& $X_1\to$& $-X_1$& $-X_1$& $X_1$& $X_3$& $-X_2$& $X_1$& $X_1$\\
		& $X_2\to$& $X_2$& $-X_2$& $X_2$& $-X_1$& $-X_3$& $-X_3$& $-X_3$\\ 
		& $X_3\to$& $-X_3$& $X_3$& $X_3$& $-X_2$& $X_1$& $-X_2$& $-X_2$\\ \hline	

		$\mcl{T}_{2g}$& $X_1\to$& $-X_1$& $-X_1$& $X_1$& $X_3$& $-X_2$& $-X_1$& $-X_1$\\
& $X_2\to$& $X_2$& $-X_2$& $X_2$& $-X_1$& $-X_3$& $X_3$& $X_3$\\ 
& $X_3\to$& $-X_3$& $X_3$& $X_3$& $-X_2$& $X_1$& $X_2$& $X_2$\\ \hline	

		$\mcl{T}_{1u}$& $X_1\to$& $-X_1$& $-X_1$& $-X_1$& $X_3$& $X_2$& $X_1$& $-X_1$\\
& $X_2\to$& $X_2$& $-X_2$& $-X_2$& $-X_1$& $X_3$& $-X_3$& $X_3$\\ 
& $X_3\to$& $-X_3$& $X_3$& $-X_3$& $-X_2$& $-X_1$& $-X_2$& $X_2$\\ \hline	

		$\mcl{T}_{2u}$& $X_1\to$& $-X_1$& $-X_1$& $-X_1$& $X_3$& $X_2$& $-X_1$& $X_1$\\
& $X_2\to$& $X_2$& $-X_2$& $-X_2$& $-X_1$& $X_3$& $X_3$& $-X_3$\\ 
& $X_3\to$& $-X_3$& $X_3$& $-X_3$& $-X_2$& $-X_1$& $X_2$& $-X_2$\\ \hline	
	\end{tabular}
	\caption{List of 3-dimensional irreps of the IR space group}
	\label{tab_def_of_3d_irreps}
\end{table}

\section{Determination of broken symmetry group}\label{appen_symmetry_group_at_gapless_points}

For the first $\mcl{T}_{1u}^o$ triplet given by  Eq.~\ref{eq_t1u0_masses_1} in the main text, at a generic point on the sphere in Fig.~\ref{fig_spectrum_on_sphere} such as C or D, the corresponding $\mcl{R}$ matrix (introduced in Eq.~\ref{eq_chi double prime basis}), when diagonalised, has the form
\begin{eqnarray}
\mcl{R} = \begin{pmatrix}
a_1\sigma_3\\
&a_2\sigma_3\\
&&a_3\sigma_3\\
&&&a_4\sigma_3
\end{pmatrix}.
\end{eqnarray}
Here $a_1,\cdots,a_4$ are four real numbers which are not equal to each other. There are seven linearly  independent matrices that commute with $\mcl{R}$ in the above equation. These are
\begin{eqnarray}
&&\begin{pmatrix}
\sigma_0\\
&-\sigma_0\\
&&\sigma_0\\
&&&-\sigma_0
\end{pmatrix},
\begin{pmatrix}
\sigma_0\\
&-\sigma_0\\
&&0\\
&&&0
\end{pmatrix},
\begin{pmatrix}
0\\
&0\\
&&\sigma_0\\
&&&-\sigma_0
\end{pmatrix}, \nonumber
\end{eqnarray}
\begin{eqnarray}
&&\begin{pmatrix}
0\\
&\sigma_3\\
&&0\\
&&&0
\end{pmatrix},
\begin{pmatrix}
\sigma_3\\
&0\\
&&0\\
&&&0
\end{pmatrix},
\begin{pmatrix}
0\\
&0\\
&&\sigma_3\\
&&&0
\end{pmatrix},
\begin{pmatrix}
0\\
&0\\
&&0\\
&&&\sigma_3
\end{pmatrix}\nonumber
\end{eqnarray}
All these matrices commute with each other and hence these generate a subgroup of SU(8) which contains seven mutually commuting U(1) groups. Thus, the SU(8) at a general point on the sphere is broken down to $[U(1)]^7$.

For the points on the three great circles obtained by setting one of the $\Delta_i$ to zero, the corresponding $\mcl{R}$ matrix for the masses look like 
\begin{eqnarray}
\mcl{R} = \begin{pmatrix}
a_1\sigma_3\\
&a_1\sigma_3\\
&&a_3\sigma_3\\
&&&a_3\sigma_3
\end{pmatrix}.
\end{eqnarray}
There are 15 linearly independent matrices that commute with this $\mcl{R}$ which form a U(1) $\otimes$ [U(1) $\otimes$ SO(4)]$^2$ of the SU(8).

Finally, at the special eight points where all the $\Delta_i$s have equal magnitude such as point B on the sphere in Fig.~\ref{fig_spectrum_on_sphere}, the $\mcl{R}$ matrix has the following form
\begin{eqnarray}
\mcl{R} = \begin{pmatrix}
a_1\sigma_3\\
&a_1\sigma_3\\
&&a_3\sigma_3\\
&&&0
\end{pmatrix}.
\label{eq_specialpta}
\end{eqnarray}
The SU(8) symmetry group at these points breaks down to U(1) $\otimes$ SO(4) $\otimes$ U(1) $\otimes$ U(1) $\otimes$ U(2).

\section{Basis transformation for the density wave semimetals}\label{appen_partially_gapless_masses}

In Eq. \ref{eq_chi tilde basis}, the $16\times 16$ matrix $U$ is given by Eq. \ref{eq_U_matrix_for_new_spinor}.

\begin{figure*}[!htp]
\begin{eqnarray}\label{eq_U_matrix_for_new_spinor}
U = \frac{1}{\sqrt{2}} \left(
\begin{array}{cccccccccccccccc}
0 & 0 & 0 & 0 & 0 & 0 & 0 & 0 & 0 & 0 & 0 & -i & i & 0 & 0 & 0 \\
0 & 0 & 0 & 0 & 0 & 0 & 0 & 0 & 0 & 0 & i & 0 & 0 & i & 0 & 0 \\
-i & 0 & 0 & 0 & 0 & 0 & 0 & i & 0 & 0 & 0 & 0 & 0 & 0 & 0 & 0 \\
0 & -i & 0 & 0 & 0 & 0 & -i & 0 & 0 & 0 & 0 & 0 & 0 & 0 & 0 & 0 \\
0 & 0 & 0 & 0 & 0 & 0 & 0 & 0 & 0 & 0 & 0 & i & i & 0 & 0 & 0 \\
0 & 0 & 0 & 0 & 0 & 0 & 0 & 0 & 0 & 0 & -i & 0 & 0 & i & 0 & 0 \\
i & 0 & 0 & 0 & 0 & 0 & 0 & i & 0 & 0 & 0 & 0 & 0 & 0 & 0 & 0 \\
0 & i & 0 & 0 & 0 & 0 & -i & 0 & 0 & 0 & 0 & 0 & 0 & 0 & 0 & 0 \\
0 & 0 & 0 & 0 & 0 & 0 & 0 & 0 & -i & 0 & 0 & 0 & 0 & 0 & 0 & i \\
0 & 0 & 0 & 0 & 0 & 0 & 0 & 0 & 0 & -i & 0 & 0 & 0 & 0 & -i & 0 \\
0 & 0 & 0 & -i & i & 0 & 0 & 0 & 0 & 0 & 0 & 0 & 0 & 0 & 0 & 0 \\
0 & 0 & i & 0 & 0 & i & 0 & 0 & 0 & 0 & 0 & 0 & 0 & 0 & 0 & 0 \\
0 & 0 & 0 & 0 & 0 & 0 & 0 & 0 & i & 0 & 0 & 0 & 0 & 0 & 0 & i \\
0 & 0 & 0 & 0 & 0 & 0 & 0 & 0 & 0 & i & 0 & 0 & 0 & 0 & -i & 0 \\
0 & 0 & 0 & i & i & 0 & 0 & 0 & 0 & 0 & 0 & 0 & 0 & 0 & 0 & 0 \\
0 & 0 & -i & 0 & 0 & i & 0 & 0 & 0 & 0 & 0 & 0 & 0 & 0 & 0 & 0 \\
\end{array}
\right).
\end{eqnarray}
\end{figure*}

\subsection{The structure of the density wave semimetal masses}
\label{appen_denwavsem}

In terms of internal symmetry transformations, the 18 density wave semimetals can be divided up into two categories depending on whether the number of gapless fermionic modes changes depending on the particular linear combination of the mass term 
\begin{align}
   \sum_i \Delta_i\bar\chi m_i\chi,
\end{align}
where $i$ sums over the appropriate number of components depending on the dimension of the irreducible representation. The first class where the number of gapless modes remains unchanged for all values of $\Delta_i$ consists of eight masses belonging to the two singlets (of $\Gamma$-DSM type) and  two triplets (of M-DSM type). They respectively make up :
\begin{itemize}
    \item $\mathcal{A}^0_{1u}$ and $\mathcal{A}^o_{2u}$  (Eqs. \ref{eq_neel_singlet1} and \ref{eq_neel_singlet2}) : Staggered spin-octupole density wave semimetal.
    \item $\mathcal{T}_{1g}^o$ (Eq. \ref{eq_t01gtriplet}) : Stripy spin-octupole density wave semimetal
    \item $\mathcal{T}_{2g}^e$ (Eq. \ref{eq_massstripyquad}) : Stripy spin-quadrupole density wave semimetal.
\end{itemize}

The second class involves the rest of the 10 masses (of $\Gamma$-DSM type) whose number of gapless modes changes as one tunes $\Delta_i$. These consist of two doublets and two triplets given by
\begin{itemize}
    \item $\mathcal{T}^o_{2g}$ (Eq. \ref{eq_gapless_t2go}) : Stripy spin-octupolar density wave semimetal.
    \item $\mathcal{T}^e_{1g}$ (Eq. \ref{eq_gapless_t1ge}) : Stripy spin-quadrupolar density wave semimetal.
    \item $\mathcal{E}_g^e$ (Eq. \ref{eq_ege_doublet_1}) : Ferro spin-quadrupole density wave semimetal.
    \item $\mathcal{E}_g^o$ (Eq. \ref{eq_ego_doublet}) : Ferro spin-octupole density wave semimetal.
\end{itemize}

For the first class, leaving out the two singlets the six masses in the $\mathcal{T}_{1g}^o$ and $\mathcal{T}_{2g}^e$ representations form a reducible representation of a SO(4) subgroup of the low-energy SU(8). This SO(4)($\equiv$ SU(2)$\otimes$ SU(2)) subgroup is generated by the following six generators:
\begin{subequations}
\begin{eqnarray}\label{eq_so4_subgroup_gapless}
&&\mathcal{g}^+_1 = \frac{1}{2} \left(\mu _6 \tilde\Sigma_{15}-\mu _7 \tilde\Sigma_5\right)\\
&&\mathcal{g}^+_2 = \frac{1}{2} \left(\mu _{14} \tilde\Sigma_{15}-\mu _{13} \tilde\Sigma_5\right)\\
&& \mathcal{g}^+_3 = \frac{1}{2} \mu _{11} \left(\tilde\Sigma_1-\tilde\Sigma_0\right)\\ 
&&\mathcal{g}^-_1 = \frac{1}{2} \left(-\mu _6 \tilde\Sigma_{15}-\mu _7 \tilde\Sigma_5\right)\\
&&\mathcal{g}^-_2 = \frac{1}{2} \left(-\mu _{14} \tilde\Sigma_{15}-\mu _{13} \tilde\Sigma_5\right)\\
&&\mathcal{g}^-_3 = \frac{1}{2} \mu _{11} \left(\tilde\Sigma_1+\tilde\Sigma _0\right)
\end{eqnarray}
\end{subequations}

The $\mathcal{g}^+_i$ and $\mathcal{g}^-_i$ separately satisfy su(2) algebra which we call as su(2)$_+$ and su(2)$_-$ respectively. Also, these two su(2)s commute with each other, i.e., $[\mathcal{g}^+_i,\mathcal{g}^-_j] = 0$~~$\forall i,j$. Now we form the following linear combinations of the masses in the $\mathcal{T}_{1g}^o$ and $\mathcal{T}_{2g}^e$ triplets : 
\begin{subequations}
\begin{eqnarray}
    &&m^+_i = (\mathcal{T}_{2g}^e)_i - (\mathcal{T}_{1g}^o)_i \\
    &&m^-_i = (\mathcal{T}_{2g}^e)_i + (\mathcal{T}_{1g}^o)_i, 
\end{eqnarray}
\end{subequations}
for $i=1,2,3.$ Here $(\mathcal{T}_{1g}^o)_i,(\mathcal{T}_{2g}^e)_i$ are the masses in the triplets $\mathcal{T}_{1g}^o$ and $\mathcal{T}_{2g}^e$. The $m^+_i$ ($m^-_i$) masses transform in spin-1 (spin-0) representation under the action of su(2)$_+$ and in spin-0 (spin-1) representation under su(2)$_-$. Thus, the six masses in the two triplets transform in $(1,1)$ representation under the action of the SO(4). 

It is interesting to note that the $m^+_i$ masses go to $m^-_i$ under the action of the microscopic time-reversal (TR). Thus, TR symmetry enforces the two representations of SO(4) to mix resulting in the two triplets resulting in the TR even and odd spin-quadrupole and spin-octupole phases respectively. 

For the second group, the ten masses making up the two doublets ($\mathcal{E}_g^e,\mathcal{E}_g^o$) and two triplets ($\mathcal{T}_{1g}^e,\mathcal{T}_{2g}^o$) mix among themselves and actually form a $(2,2)$ representation of the SO(4) in Eq.~\ref{eq_so4_subgroup_gapless}. This is easy to see by writing the 10 masses in two sub-groups each consisting of five masses as
\begin{subequations}
\begin{eqnarray}
 &&\tilde{m}^+_i = (\mathcal{T}_{1g}^e)_i - (\mathcal{T}_{2g}^o)_i \quad \forall~ i=1,2,3, \\
 &&\tilde{m}^+_4 = (\mathcal{E}_{g}^e)_1 - (\mathcal{E}_{g}^o)_1 \\
 &&\tilde{m}^+_5 = (\mathcal{E}_{g}^e)_2 - (\mathcal{E}_{g}^o)_2
 \end{eqnarray}
 \end{subequations}
 and
 \begin{subequations}
 \begin{eqnarray}
 &&\tilde{m}^-_i = (\mathcal{T}_{1g}^e)_i + (\mathcal{T}_{2g}^o)_i \quad \forall~ i=1,2,3, \\
 &&\tilde{m}^-_4 = (\mathcal{E}_{g}^e)_1 + (\mathcal{E}_{g}^o)_1 \\
 &&\tilde{m}^-_5 = (\mathcal{E}_{g}^e)_2 + (\mathcal{E}_{g}^o)_2 
\end{eqnarray}
\end{subequations}

The first (second) sub-group of masses, $\tilde m_i^+ (\tilde m_i^-)$ transforms as a spin-$2 (0)$ representation under  su(2)$_+$ and in spin-0 (spin-2) representation under su(2)$_-$.


\section{Analysis in the global basis}
\label{appen:globalbasis}

As mentioned in the main text, Bloch diagonalizing the Hamiltonian in global basis (Eq. \ref{eq:hamiltonian in j=3/2 orbitals}) obtains four bands arising from the four $j=3/2$ orbitals (Eq. \ref{eq:4-component psi}) and each two-fold degenerate due to inversion symmetry. The first set of bands touches the second set of bands at {\em four}  distinct points with a Dirac cone structure, see \figref{fig:globalBZBands}. With $1/4$th filling of the bands, the chemical potential is tuned to the Dirac points at the four $\mathbf{Q}_g$ vectors, $\Gamma$, M$_1$, M$_2$, M$_3$ termed as valleys, in the original honeycomb lattice Brillouin zone. 

Following an approach similar to that adopted in the main text, the IR physics can be obtained by expanding in terms of the four Dirac modes at quarter filling; one obtains four flavours of {\it two component} Dirac fermions $\chi_{g;\nu}({{\bf x}})$ where $\nu(=\Gamma,{\textup{M}_1},{\textup{M}_2},{\textup{M}_3})$ refers to the four valleys (Fig. \ref{fig:globalBZBands}). Combining them together, we get the $16$-component Dirac spinor
\beq\label{eqn:globalChi}
\chi_g(\boldsymbol(x)) = (\chi^T_{g\Gamma}({{\bf x}}), \;\; \chi^T_{g M_1}({{\bf x}}), \;\;  \chi^T_{g M_2}({\bf x}), \;\; \chi^T_{g M_3}({{\bf x}}))^T
\eeq
in the global basis. This should be contrasted with the 16-component spinor in the local basis obtained by stacking the four $4$-component spinors in Eq. \ref{eq_16spinor}. The low-energy action in the global basis reads
\beq\label{eqn:DiracGlobal}
{\cal L}_g = v_F\bar{\chi}_g  (-i \slashed{\dou}) \chi_g
\eeq
repeated  are summed over the spatial directions.
The gamma matrices in the global basis are
\beq
\gamma^0_g = M_{0003}, \gamma^1_g = M_{0002}, \gamma^2_g = - M_{0001} 
\eeq
with
\beq\label{eqn:Mdef}
M_{\mu \nu \rho \tau} = \sigma_\mu \sigma_\nu \sigma_\rho \sigma_\tau, \;\;\; \mu,\nu,\rho,\tau \in \{0,1,2,3\}
\eeq
where $\sigma_\mu$ are the Pauli matrices $\sigma_0 = 1_{2 \times 2}, \sigma_1 = \sigma_x, \sigma_2 = \sigma_y$ and $\sigma_3 = \sigma_z$. The Dirac action obtained from \eqnref{eqn:DiracGlobal} has an emergent global SU(8) symmetry, much like  in the local basis. However, the crucial point is that the SU(4) symmetry of the transformed microscopic Hamiltonian in the local basis does not directly manifest in the Dirac lagrangian \eqnref{eqn:DiracGlobal}. This is the reason why we choose to represent the relevant matrices $M_{\mu \nu \sigma \tau}$ using the products of Pauli matrices as in \eqnref{eqn:Mdef}, as there is no natural choice of flavor and chiral spaces in the global basis. Despite the burden of this additional notation, we will  see that the global formulation provides key insights, particularly the semimetallic phases obtained in the main text.

The Dirac action obtained from  \eqnref{eqn:DiracGlobal} is invariant under the space group symmetry operations $\mathbb{S}$, the global basis spinors transform as $\chi_g({\bf x}) \to \Omega_\mathbb{S} \chi_g(\mathbb{S}^{-1} {\bf x})$. The matrices matrices $\Omega_{\mathbb S}$ are obtained as (analogous to that discussed in appendix~\ref{appen_softmodetrans} )

\begin{widetext}
\beq\label{eqn:SymUGlob}
\begin{split}
	\Omega_{\mbox{${\bf C}_3$}}  = &  \frac{1}{16}  \left(-M_{0000}-i \sqrt{3} M_{0003}+i \sqrt{3} M_{0010}-3 M_{0013}+\sqrt{3} M_{0120}+3 i M_{0123}+M_{0130}+i \sqrt{3} M_{0133}+i \sqrt{3} M_{0220} \right. \\
		 & -3 M_{0223}+i M_{0230}-\sqrt{3} M_{0233}-M_{0300}-i \sqrt{3} M_{0303}+i \sqrt{3} M_{0310}-3 M_{0313}+\sqrt{3} M_{1020} \\
		 &   +3 i M_{1023}-M_{1030}-i \sqrt{3} M_{1033}+M_{1100}+i \sqrt{3} M_{1103}+i \sqrt{3} M_{1110}-3 M_{1113}-i M_{1200} \\
		 & +\sqrt{3} M_{1203}+\sqrt{3} M_{1210}+3 i M_{1213}-\sqrt{3} M_{1320}-3 i M_{1323}+M_{1330}+i \sqrt{3} M_{1333}-i \sqrt{3} M_{2020} \\
		 & +3 M_{2023}+i M_{2030}-\sqrt{3} M_{2033}+i M_{2100}-\sqrt{3} M_{2103}-\sqrt{3} M_{2110}-3 i M_{2113}+M_{2200} \\
		 & +i \sqrt{3} M_{2203}+i \sqrt{3} M_{2210}-3 M_{2213}+i \sqrt{3} M_{2320}-3 M_{2323}-i M_{2330}+\sqrt{3} M_{2333}-M_{3000} \\
		 & -i \sqrt{3} M_{3003}+i \sqrt{3} M_{3010}-3 M_{3013}-\sqrt{3} M_{3120}-3 i M_{3123}-M_{3130}-i \sqrt{3} M_{3133}-i \sqrt{3} M_{3220} \\
		 &\left. +3 M_{3223}-i M_{3230}+\sqrt{3} M_{3233}-M_{3300}-i \sqrt{3} M_{3303}+i \sqrt{3} M_{3310}-3 M_{3313}\right) \\
\Omega_{\mbox{$\boldsymbol{\sigma}_d$}}= & \displaystyle{-\frac{1}{2} i \left(M_{0322}-M_{1122}-M_{2222}+M_{3022}\right)},\;\;\; \Omega_{\mbox{${\bf C}_2'$}}=  \displaystyle{\frac{1}{2} i \left(M_{0331}+M_{1131}+M_{2231}+M_{3031}\right)}, \;\;\;
\Omega_{{\bf I}}= \displaystyle{-M_{3313}} \\
\Omega_{\mbox{${\bf S}_6$}}= & \frac{1}{16} \left(3 M_{0000}-i \sqrt{3} M_{0003}+i \sqrt{3} M_{0010}+M_{0013}+\sqrt{3} M_{0120}-i M_{0123}-3 M_{0130}+i \sqrt{3} M_{0133}  \right. \\
	& +i \sqrt{3} M_{0220}+M_{0223}-3 i M_{0230}-\sqrt{3} M_{0233}+3 M_{0300}-i \sqrt{3} M_{0303}+i \sqrt{3} M_{0310}+M_{0313}+\sqrt{3} M_{1020} \\
	& -i M_{1023}+3 M_{1030}-i \sqrt{3} M_{1033}-3 M_{1100}+i \sqrt{3} M_{1103}+i \sqrt{3} M_{1110}+M_{1113}+3 i M_{1200} \\
	& +\sqrt{3} M_{1203}+\sqrt{3} M_{1210}-i M_{1213}-\sqrt{3} M_{1320}+i M_{1323}-3 M_{1330}+i \sqrt{3} M_{1333}-i \sqrt{3} M_{2020}\\
	& -M_{2023}-3 i M_{2030}-\sqrt{3} M_{2033}-3 i M_{2100}-\sqrt{3} M_{2103}-\sqrt{3} M_{2110}+i M_{2113}-3 M_{2200} \\
	& +i \sqrt{3} M_{2203}+i \sqrt{3} M_{2210}+M_{2213}+i \sqrt{3} M_{2320}+M_{2323}+3 i M_{2330}+\sqrt{3} M_{2333}+3 M_{3000} \\
	& -i \sqrt{3} M_{3003}+i \sqrt{3} M_{3010}+M_{3013}-\sqrt{3} M_{3120}+i M_{3123}+3 M_{3130}-i \sqrt{3} M_{3133}-i \sqrt{3} M_{3220} \\
	&\left. -M_{3223}+3 i M_{3230}+\sqrt{3} M_{3233}+3 M_{3300}-i \sqrt{3} M_{3303}+i \sqrt{3} M_{3310}+M_{3313}\right) \\
	\Omega_{\mbox{${\bf T}_1$}} &=  \displaystyle{M_{0300}}, \;\;\; \Omega_{\mbox{${\bf T}_2$}}=  \displaystyle{M_{3000}} \;\;\;	\Omega_{\mbox{$\mathbb{T}$}}= \displaystyle{i M_{0032}} 
\end{split}
\eeq
\end{widetext}
An important feature here is that  $\chi_g$ transforms under the action of ${\mathbb{S}}$ such that the space of spinors $\chi_{g\Gamma}$ is an invariant subspace, i.~e., symmetry transformations do not mix $\chi_{g \Gamma}$ with any other $\chi_{g \nu}, \nu\in \{\textup{M}_1,\textup{M}_2,\textup{M}_3\}$.

We now see that any fermion bilinear of the form
\beq\label{eqn:MassGlobal}
-i \bar{\chi}_g \gamma^0_g M_{\mu\nu\rho3} \chi_g
\eeq
described by a mass matrix $M_{\mu\nu\rho3}, \mu,\nu,\rho \in \{0,1,2,3\}$ which anticommutes with $\gamma^1_g$ and $\gamma^2_g$, gaps out the Dirac fermions. The $64$ masses can be classified by the irreducible representations of the space group using the fact that the masses $M_{\mu\nu\rho3}$ transform adjointly under the action of the operations \eqnref{eqn:SymUGlob}. On carrying out the classification of mass terms according to the irreducible representations of the space group, the analysis based on the global basis produces identical results as those in tables.~\ref{tab_chiral_masses} to \ref{tab_mixed_semimetals}.

\begin{figure*}
	\centerline{\includegraphics[height=6.0truecm]{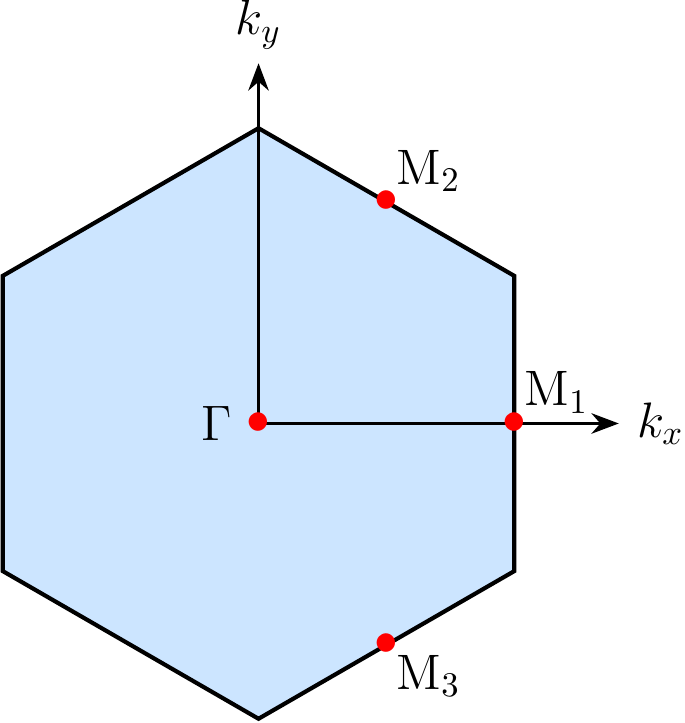} ~~~~~\includegraphics[height=8.0truecm]{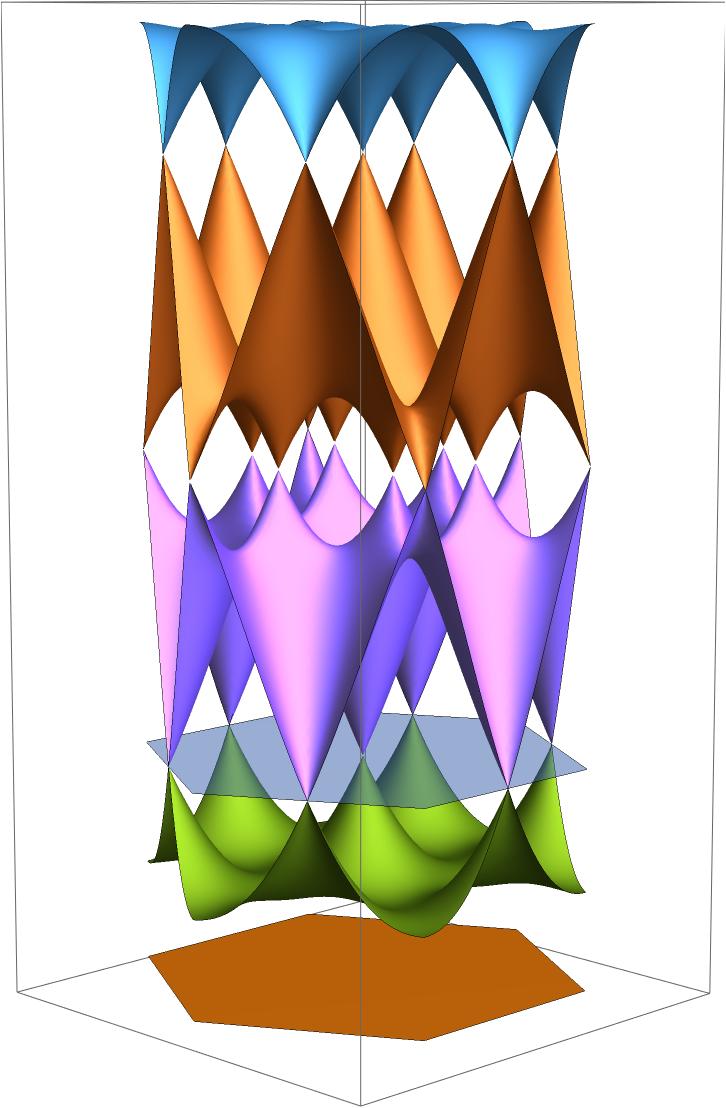}}
	\caption{Brillouin zone and band structure in the global basis. Each of the four bands indicated in a different color are twofold degenerate. At quarter filling, distinct Dirac cones appear at four points (valleys) $\Gamma$, M$_1$, M$_2$, M$_3$. The light blue plane indicates the chemical potential at quarter-filling.}
	\label{fig:globalBZBands}
\end{figure*}

The global basis offers illuminating insights  into understanding the phases, particularly the semimetallic ones. Central to this is the fact that symmetry operations $\mathbb{S}$ do not mix $\chi_{g \Gamma}$ with any other $\chi_{g \nu}, \nu\in \{\textup{M}_1,\textup{M}_2,\textup{M}_3\}$. Thus, a spinor can be decomposed into 
\beq\label{eqn:GammaProj}
\chi_g =  \underbrace{P_\Gamma \chi_g}_{\chi_{g \Gamma}} + \underbrace{(1-P_\Gamma) \chi_{g}}_{\chi_{gM}}
\eeq
where the operator $P_\Gamma$ projects a general spinor to the valley $\Gamma$.  The space of spinors $\chi_{g \Gamma}$ carries some irreps $D^a$ (in the fundamental representation) of the space group labeled by index $a$. Similarly, the space of spinors $\chi_{gM}$ (spinors belonging to valleys $\textup{M}_1,\textup{M}_1,\textup{M}_1$) may be decomposed into space group irreps $D^b$ labelled by $b$. Also, the adjoint representation on the space of masses is decomposed into irreps $D^c$ labeled by $c$. We can now study the structure of the masses in one of the representations $D^c$, by exploring which product representations $D^{P*} \otimes D^c \otimes D^{Q}$  where $P,Q \in \{a,b\}$ contain a singlet representation.
 Several interesting possibilities arise, of which two are crucially important: 

 \begin{enumerate}
 	\item An  irreducible mass matrix $M^c$ ($c$ labels the irrep $D^c$ in the adjoint representation) is such that there is no identity representation in the decomposable tensor product representation $D^{a'*}\otimes D^c \otimes D^a$ for all $a,a'$ representations carried by  the $\chi_{g \Gamma}$-space. Further, there is atleast one identity representation in $D^{b'*}\otimes D^c \otimes D^b$, where $b,b'$ are irreps of the $\chi_{gM}$-space. In such a scenario, the mass $M^c$ acts like a ``zero matrix'' on spinors $\chi_{g \Gamma}$, and has the following structure, 
 	\beq
 	M^c_{\textup{$\Gamma$-DSM}} = \bordermatrix{ & \Gamma &  \textup{M}_1 &  \textup{M}_2 &  \textup{M}_2 \cr
 		\Gamma & 0_{4\times4} &  0_{4\times4} &  0_{4\times4} &  0_{4\times4} \cr
 		\textup{M}_1 & 0_{4\times4} &  \blacksquare &  \blacksquare  &  \blacksquare \cr
 		\textup{M}_2 & 0_{4\times4} &  \blacksquare &  \blacksquare  &  \blacksquare \cr
 		\textup{M}_3 & 0_{4\times4} &  \blacksquare &  \blacksquare  &  \blacksquare \cr
 	}, \;\;\; 
 	\eeq
  where $\blacksquare \equiv \textup{non-zero entry}$. This guarantees that the Dirac cones at $\Gamma$ are ungapped, leading to a semimetallic phase, and we dub such a phase as ``Gamma-Dirac semimetal ($\Gamma$-DSM)''. Examples of such semimetals are entries No. 17, 18, 20, 22, 23 and 24 in Table~\ref{tab_mixed_semimetals}. 
  As discussed in the main text, additional gapless modes may be possible if such masses arise in a doublet or triplet representation when the coefficients of the mass matrices satisfy special criteria (see, for example, \ref{eq_12t12}). 

	\item The second interesting possibility for the mass $M^c$ is such that while there is no identity representation in $D^{a'*}\otimes D^c \otimes D^a$ ($a,a'$ are representations in $\chi_{g\Gamma}$-space) or  $D^{b'*}\otimes D^c \otimes D^b$ ($b,b'$ are representations in $\chi_{gM}$-space), but there is at least one identity representation in  $D^{a'*}\otimes D^c \otimes D^b$. This implies that the mass $M^c$ mixes $\chi_{g \Gamma}$ with $\chi_{gM}$, but since mixing between spinors at $\Gamma$ is forbidden as is the mixing between spinors between the $M$ valleys, the remaining possibility is that of mixing between spinors at $\Gamma$ with those of $M$ leading  to the mass matrix structured as
	\beq
	M^c_{\textup{M-DSM}} = \bordermatrix{ & \Gamma &  \textup{M}_1 &  \textup{M}_2 &  \textup{M}_2 \cr
		\Gamma & 0_{4\times4} & \blacksquare & \blacksquare & \blacksquare \cr
		\textup{M}_1 & \blacksquare & 0_{4\times4} &  0_{4\times4} &  0_{4\times4} \cr
		\textup{M}_2 & \blacksquare & 0_{4\times4} &  0_{4\times4} &  0_{4\times4} \cr
		\textup{M}_3 & \blacksquare & 0_{4\times4} &  0_{4\times4} &  0_{4\times4} \cr
	}, \;\;\; 
	\eeq
 This type of mass matrix has an emergent sub-lattice symmetry where
	 \beq
	 U_{SL}^\dagger M^c_{\textup{M-DSM}} U_{\textup{SL}} = -M^c_{\textup{M-DSM}}
	 \eeq
	 with
	 \beq
	 U_{\textup{SL}} = \bordermatrix{ & \Gamma &  \textup{M}_1 &  \textup{M}_2 &  \textup{M}_2 \cr
	 	\Gamma & 1_{4\times4} &  0_{4\times4} &  0_{4\times4} &  0_{4\times4} \cr
	 	\textup{M}_1 & 0_{4\times4} & -1_{4\times4} &  0_{4\times4} &  0_{4\times4} \cr
	 	\textup{M}_2 &0_{4\times4}  & 0_{4\times4} &  -1_{4\times4} &  0_{4\times4} \cr
	 	\textup{M}_3 & 0_{4\times4} & 0_{4\times4} &  0_{4\times4} &  -1_{4\times4} \cr
	 }
	 \eeq
	 which guarantees that there are at least 8 zero eigenvalues leading to (at least) 8 gapless modes for any such mass. Note that the gapless modes will be a linear combination of spinors from all the valleys and, in particular will depend on the non-zero entries denoted by $\blacksquare$. Such semimetals are dubbed "M-Dirac semimetals (M-DSM)". In Table \ref{tab_mixed_semimetals}, the entries 19 and 21 are of this type.
\end{enumerate}

{\section{Spinless fermions on a honeycomb lattice with \texorpdfstring{$\pi$}{}-flux at \texorpdfstring{$1/4$}{}-th filling}\label{appen:spinlessPiFluxHex}

\begin{figure*}
\centerline{\includegraphics[height=6.0truecm]{small_piflux_unit_cell.pdf}~\includegraphics[height=8.0truecm]{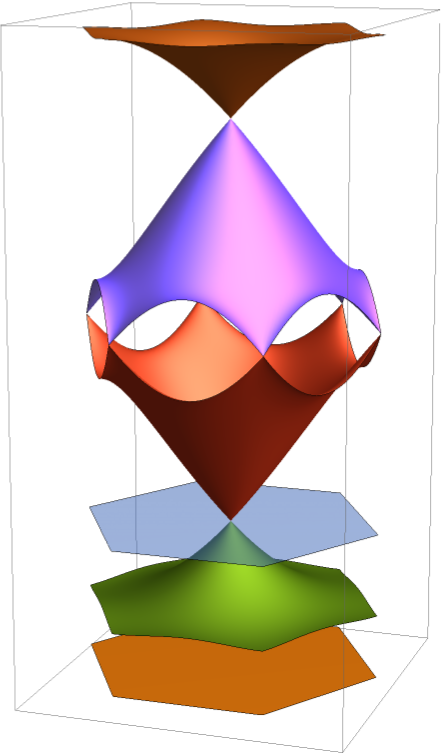}}
\caption{Honeycomb lattice with $\pi$-flux. (Left) Unit cell consisting of 8 sites adopted for the analysis. Fermions hop to nearest neighbors where red links shown have a hopping amplitude with a negative sign. (Right) Band structure showing four bands, each of which is two-fold degenerate. The light blue plane shows the quarter-filled chemical potential. }
\label{fig:piFluxUnitCell}
\end{figure*}

In this section, we revisit the physics of spinless fermions hopping on a honeycomb lattice with a $\pi$-flux through each of the honeycomb plaquettes. We adopt the unit cell shown in \figref{fig:piFluxUnitCell}, and choose a more convenient gauge for obtaining the $\pi$-flux (this enables an efficient implementation of lattice symmetries). We obtain four bands, each of which is two-fold degenerate. At quarter filling, we obtain two Dirac cones located at the $\Gamma$ point of the hexagonal Brillouin zone as shown in \figref{fig:globalBZBands}. The low energy physics is described by a four-component spinor $\chi_\pi$ with a Lagrangian density similar to \eqnref{eqn:DiracGlobal} with,
\beq\label{eqw:PiFluxGamma}
\gamma^0_\pi = M_{03}, \gamma^1_{\pi} = -M_{02}, \gamma^2_\pi = M_{01}
\eeq
where
\beq
M_{\mu \nu} = \sigma_\mu \sigma_\nu.
\eeq
where $\sigma_\mu$ are Pauli matrices defined just below \eqnref{eqn:Mdef}. The system has an emergent global SU(2) symmetry generated by $M_{i0}, i \in \{1,2,3\}$ which is the analog of the chiral symmetry discussed near \eqnref{eq_zetasu2}.

A fermion bilinear of the form
\beq\label{eqn:MassPiFlux}
-i \bar{\chi}_\pi \gamma^0_\pi M_{\mu3} \chi_\pi
\eeq
described by a mass matrix $M_{\mu3}, \mu \in \{0,1,2,3\}$ which anticommutes with $\gamma^1_\pi$ and $\gamma^2_\pi$, gaps out the Dirac fermions. The space of these mass matrices can be reduced in the irreps of the space group, resulting the phases described below.

\subsection{Integer Chern insulator}
This mass transforms as a one-dimensional irrep, breaking time reversal and reflection symmetries of the lattice while preserving all the proper rotational symmetries of the hexagonal lattice
\medskip

\centerline{
\begin{tabular}{|c|c|}
		\hline 
		$\mathcal{A}_{2g}{}^o$&  \\
		\hline 
		1&$M_{03}$ \\ 
		\hline \end{tabular}  
}
\noindent 
resulting in a mass term
\beq
\Delta_{\textup{ICI}} = -i \langle \bar{\chi}_\pi \chi_\pi \rangle.
\eeq
It is clear that the SU(4) symmetric ICI found in \eqnref{eq_chern_insulator} is a ``larger dimensional'' realization of such a phase with a larger value of the Chern-Simons level.

\subsection{Stripy density waves}
The remainder of the three masses organize as a triplet under the space group symmetries, preserving time reversal but breaking rotational and translational symmetries.
\medskip
\centerline{
	\begin{tabular}{|c|c|}
		\hline 
		$\mathcal{T}_{1g}{}^{e}$&  \\
		\hline 
		1&$\frac{1}{\sqrt{2}} M_{13} + \frac{1}{\sqrt{6}} M_{23} - \frac{1}{\sqrt{3}} M_{33}$ \\ 
		\hline 2&$ -\frac{1}{\sqrt{2}} M_{13} + \frac{1}{\sqrt{6}} M_{23} - \frac{1}{\sqrt{3}} M_{33} $ \\ 
		\hline 3&$ \sqrt{\frac{2}{3}} M_{23} + \frac{1}{\sqrt{3}} M_{33}  $ \\ 
		\hline \end{tabular} 
} 
The three components of the masses correspond to the fermion bilinears
\beq
\begin{split}
-i \langle \bar{\chi}_\pi \left( \frac{1}{\sqrt{2}} M_{10} + \frac{1}{\sqrt{6}} M_{20} - \frac{1}{\sqrt{3}} M_{30}  
 \right)\chi_\pi\rangle, \\
 -i \langle \bar{\chi}_\pi \left(   -\frac{1}{\sqrt{2}} M_{10} + \frac{1}{\sqrt{6}} M_{20} - \frac{1}{\sqrt{3}} M_{30}
 \right)\chi_\pi\rangle, \\
  -i \langle \bar{\chi}_\pi \left( \sqrt{\frac{2}{3}} M_{20} + \frac{1}{\sqrt{3}} M_{30} 
 \right)\chi_\pi\rangle
 \end{split}
\eeq
any one of which produces a stripy density wave  similar to that shown in \figref{fig:stripy_CDW}. The mass matrices that appear here are orthogonal linear combinations of the chiral symmetry generators discussed just below \eqnref{eqw:PiFluxGamma}. Indeed, it is evident that the chiral masses shown in \eqnref{eq_su4_cdw} correspond to this case.


{
\section{A model with $j=1/2$ spins}
\label{appen:jHalfModel}

In this section, we construct a model on a honeycomb lattice where the spin-orbit coupling is realized in a $j=1/2$ system, i.~e., as a system with spin-$1/2$ degrees of freedom. Although this model is not directly motivated by a material system, it is nevertheless useful to study, in a simpler setting, the conceptual underpinnings of how spin-orbit coupling produces interesting new phenomena. The model is defined using \figref{fig:2-point honeycomb} where each lattice site has two $j=1/2$ orbitals. The hopping Hamiltonian is same as that in \eqnref{eq:hamiltonian in j=3/2 orbitals}, with the key difference that $U_{\mathbf{r} \mathbf{r}'} = \{ \tau_x,\tau_y,\tau_z \}$  ($\tau_i$ here are the Pauli matrices acting on the $j$-1/2 space) respectively when $\mathbf{r} \mathbf{r}'$ is the $x,y,z$ type link shown in \figref{fig:2-point honeycomb}. This system has the following microscopic symmetries among those listed in table.~\ref{tab:lat_sym} and time reversal,
\begin{enumerate}
	\item Lattice translations as in ${\bf T}_1$ and ${\bf T}_2$
 \item ${\bf C}_3$ rotations 
	\item $\boldsymbol{\sigma}_d$ dihedral reflection
	\item Time reversal $\mathbb{T}$, with $\mathbb{T}^2 = -1$
\end{enumerate}
The interesting aspect of this model is that by carrying out transformations similar to those discussed in appendix \ref{sec:appen_g_matrices}, one can arrive at a system with a $\pi$-flux through each honeycomb plaquette, and a {\em global} SU(2) flavor symmetry. In other words, this model is the ``SU(2) version'' of the SU(4) model discussed in the main text.

We continue to discuss this model in the {\em global} basis. The band structure of this model is identical to that shown in \figref{fig:globalBZBands}, the  difference being that each band is non-degenerate. At quarter filling, the physics can be described by four Dirac cones, one each located at $\Gamma, \textup{M}_1, \textup{M}_2, \textup{M}_3$. We get a Dirac action similar to \eqnref{eqn:DiracGlobal}, with $\chi_g$ as in \eqnref{eqn:globalChi} where $\chi_{g\nu}$, $\nu \in \{\Gamma, \textup{M}_1, \textup{M}_2, \textup{M}_3 \}$  are 2-component spinors. The Dirac gamma matrices are
\beq \label{eqn:su2Vl}
\gamma^0_g = M_{003}, \gamma^1_g = M_{002}, \gamma^{2}_g = - M_{001} 
\eeq
and
\beq
M_{\mu\nu\rho} = \sigma_\mu \sigma_\nu \sigma_\tau 
\eeq
where Pauli matrices $\sigma_\mu$ are described just below \eqnref{eqn:Mdef}

The symmetry transformations are described by
\beq
\begin{split}
 \Omega_{{\bf C}_3}=&\frac{1}{8} \left(M_{000}+i \sqrt{3} M_{003}
-M_{010}-i \sqrt{3} M_{013}  \right. \\
& +i M_{020}-\sqrt{3} M_{023}+M_{030}+i \sqrt{3} M_{033}-M_{100}  \\
&  -i \sqrt{3} M_{103}+M_{110}+i \sqrt{3} M_{113}+i M_{120} \\
& -\sqrt{3} M_{123}+M_{130}+i \sqrt{3} M_{133}-i M_{200}  \\
& +\sqrt{3} M_{203}-i M_{210}+\sqrt{3} M_{213}+M_{220}  \\ 
&  +i \sqrt{3} M_{223} +i M_{230}-\sqrt{3} M_{233}+M_{300} \\
&  +i \sqrt{3} M_{303}+M_{310}+i \sqrt{3} M_{313}-i M_{320}  \\
& \left. +\sqrt{3} M_{323}+M_{330}+i \sqrt{3} M_{333}\right)\\
\Omega_{\boldsymbol{\sigma}_d}=&\displaystyle{\frac{1}{2} \left(-M_{032}+M_{112}+M_{2
22}-M_{302}\right)}\\
\Omega_{{\bf T}_1}=&\displaystyle{M_{300}}\\
\Omega_{{\bf T}_2}=&\displaystyle{M_{030}}\\
\Omega_{\mathbb{T}}=&\displaystyle{i M_{002}}  
\end{split}
\eeq
Again, we see that the spinors $\chi_{g \Gamma}$ (see \eqnref{eqn:GammaProj}) form an invariant subspace under the action of the symmetries above. One, therefore, expects to obtain semimetallic phases when the mass matrices of the type $M_{\mu\nu3}, \mu,\nu \in \{0,1,2,3\}$ that gap out (see \eqnref{eqn:MassGlobal}) the Dirac Lagrangian \eqnref{eqn:DiracGlobal} defined by \eqnref{eqn:su2Vl} are resolved into irreducible components. Below we briefly describe seven irreducible masses and the resulting phases obtained by such an analysis.

\renewcommand{\text}[1]{{#1}} 
\newcommand{\cmark}{\ding{51}} 
\newcommand{\xmark}{\ding{55}} 
\renewcommand{\arraystretch}{1.3} 

\subsection{Chiral masses}
\label{subsec:J2CM}

\subsubsection{Integer Chern insulator}
\label{subsec:QHM}
\medskip
\centerline{
	\begin{tabular}{|c|c|c|c|c|}
		\hline 
		\hline 
		${\bf C}_3$&$\boldsymbol{\sigma}_d$&${\bf T}_1$&${\bf T}_2$&$\mathbb{T}$\\ \hline 
		\cmark&\xmark&\cmark&\cmark&\xmark\\ \hline 
		\hline 
	\end{tabular} 
~~~ \begin{tabular}{|c|c|}
		\hline 
		$\mathcal{A}^o$&  \\
		\hline 
		1&$M_{003}$ \\ 
		\hline \end{tabular}  
}
\medskip
This mass is SU(2) symmetric and produces a fully gapped state. Viewed on the lattice, it produces spin-independent second neighbour hoppings akin to the Haldane honeycomb model as in \figref{fig:chern_insulator_hopping} with an effective Chern-Simons action described by \eqnref{eq_icilag}.

\subsubsection{Stripy Density Wave Phase}
\label{subsec:StDWM}
\medskip
\centerline{
	\begin{tabular}{|c|c|c|c|c|}
		\hline 
		\hline 
  ${\bf C}_3$&$\boldsymbol{\sigma}_d$&${\bf T}_1$&${\bf T}_2$&$\mathbb{T}$\\ \hline 
		\xmark&\xmark&\xmark&\xmark&\cmark\\ \hline 
		\hline 
	\end{tabular} 
~~~~
	\begin{tabular}{|c|c|}
		\hline 
		$\mathcal{T}^{\text{e}}$&  \\
		\hline 
		1&$M_{123}$ \\ 
		\hline 2&$-M_{203}$ \\ 
		\hline 3&$-M_{323}$ \\ 
		\hline \end{tabular} 
} 
\medskip
This is again an SU(2) invariant mass that results in a stripy density wave similar to the SU(4) invariant case found in \eqnref{eq_su4_cdw}.

The two cases described above exhaust the chiral masses.

\subsection{SU(2) Flavor masses}
\label{subsec:J2FM}

\subsubsection{Quantum dipolar Hall mass}
\label{subsec:QDHM}
\medskip
\centerline{
	\begin{tabular}{|c|c|c|c|c|}
		\hline 
		\hline 
  ${\bf C}_3$&$\boldsymbol{\sigma}_d$&${\bf T}_1$&${\bf T}_2$&$\mathbb{T}$\\ \hline 
		\xmark&\xmark&\xmark&\xmark&\cmark\\ \hline 
		\hline 
	\end{tabular} 
~~~~
	\begin{tabular}{|c|c|}
		\hline 
		$\mathcal{T}^{e}$&  \\
		\hline 
		1&$M_{213}$ \\ 
		\hline 2&$-M_{233}$ \\ 
		\hline 3&$-M_{023}$ \\ 
		\hline \end{tabular} 
} 
\medskip
This mass produces spin-dependent second neighbour hopping that produces a uniform SU(2) flux in a second-neighbor triangle, gapping out the system. The phase with this mass is described by a mutual Chern-Simons theory like \eqnref{eq:octupolar_hall_action} resulting in dipole-filtered edge states and is analogous to the phase discussed in section.~\ref{subsec_te1gsop}.

\subsection{Mixed masses}
\label{subsec:J2MM}

\subsubsection{Ferromagnetic insulator}
\label{subsec:UFM}
\medskip
\centerline{
	\begin{tabular}{|c|c|c|c|c|}
		\hline 
		\hline 
  ${\bf C}_3$&$\boldsymbol{\sigma}_d$&${\bf T}_1$&${\bf T}_2$&$\mathbb{T}$\\ \hline 
		\cmark&\xmark&\cmark&\cmark&\xmark\\ \hline 
		\hline 
	\end{tabular} 
~~~~
	\begin{tabular}{|c|c|}
		\hline 
		$\mathcal{A}^o$&  \\
		\hline 
		1&$\frac{M_{033}}{\sqrt{3}}+\frac{M_{303}}{\sqrt{3}}+\frac{M_{333}}{\sqrt{3}}$ \\ 
		\hline \end{tabular} 
} 
\medskip
This mass manifests as a uniform magnetic field in the direction perpendicular to the honeycomb and  fully gap out the Dirac fermions. The analogous state correponding to this in the SU(4) case is discussed in section.~\ref{subsubsec:IsingFerroSpinOctopolar}.

\subsubsection{Ferromagnetic semimetal -- $\Gamma$-Dirac Semimetal}
\label{subsec:FDMIDCSM}
\medskip
\centerline{
	\begin{tabular}{|c|c|c|c|c|}
		\hline 
		\hline 
  ${\bf C}_3$&$\boldsymbol{\sigma}_d$&${\bf T}_1$&${\bf T}_2$&$\mathbb{T}$\\ \hline 
		\xmark&\xmark&\cmark&\cmark&\xmark\\ \hline 
		\hline 
	\end{tabular} 
~~~~
	\begin{tabular}{|c|c|}
		\hline 
		$\mathcal{E}^{{o}}$&  \\
		\hline 
		1&$\frac{M_{033}}{\sqrt{2}}-\frac{M_{333}}{\sqrt{2}}$ \\ 
		\hline 2&$\frac{M_{033}}{\sqrt{6}}-\sqrt{\frac{2}{3}} M_{303}+\frac{M_{333}}{\sqrt{6}}$ \\ 
		\hline \end{tabular} 
} 
\medskip
The doublet mass produces a local magnetic field in the plane of the honeycomb lattice and the components rotate into each other under lattice symmetries; this phase is analogous to the ferro spin-octupolar semimetallic phase discussed in section.~\ref{subsubsec:FSOPSWSM}. For any generic linear combination of the masses, the Dirac cone at the $\Gamma$ point remains ungapped, while for special linear combinations of the two masses, there is one additional gapless mode as discussed in section.~\ref{subsubsec:FSOPSWSM}.   This is a semimetallic phase of $\Gamma$-DSM type.

\subsubsection{Stripy spin density wave semimetal (M-Dirac Semimetal)}
\label{subsec:StSDWMSLDSM}
\medskip
\centerline{
	\begin{tabular}{|c|c|c|c|c|}
		\hline 
		\hline 
  ${\bf C}_3$&$\boldsymbol{\sigma}_d$&${\bf T}_1$&${\bf T}_2$&$\mathbb{T}$\\ \hline 
		\xmark&\xmark&\xmark&\xmark&\xmark\\ \hline 
		\hline 
	\end{tabular} 
~~~~
	\begin{tabular}{|c|c|}
		\hline 
		$\mathcal{T}^{{o}}$&  \\
		\hline 
		1&$\frac{M_{113}}{\sqrt{2}}-\frac{M_{223}}{\sqrt{2}}$ \\ 
		\hline 2&$-\frac{M_{103}}{\sqrt{2}}-\frac{M_{133}}{\sqrt{2}}$ \\ 
		\hline 3&$-\frac{M_{013}}{\sqrt{2}}-\frac{M_{313}}{\sqrt{2}}$ \\ 
		\hline \end{tabular} 
} 
\medskip
This mass produces a spin density wave of the stripy kind similar to that discussed in \eqnref{eq_t01gtriplet}. Interestingly, this produces a semimetallic phase of the M-DSM kind, precisely as discussed for the SU(4) case in \eqnref{eq_t01gtriplet}.

\subsubsection{Stripy spin density wave semimetal -- ($\Gamma$-Dirac Semimetal)}
\label{subsec:StSDWMIDCSM}
\medskip
\centerline{
	\begin{tabular}{|c|c|c|c|c|}
		\hline 
		\hline 
  ${\bf C}_3$&$\boldsymbol{\sigma}_d$&${\bf T}_1$&${\bf T}_2$&$\mathbb{T}$\\ \hline 
		\xmark&\xmark&\xmark&\xmark&\xmark\\ \hline 
		\hline 
	\end{tabular} 
~~~~
	\begin{tabular}{|c|c|}
		\hline 
		$\text{T}_{\text{u}}{}^{\text{o}}$&  \\
		\hline 
		1&$\frac{M_{113}}{\sqrt{2}}+\frac{M_{223}}{\sqrt{2}}$ \\ 
		\hline 2&$\frac{M_{133}}{\sqrt{2}}-\frac{M_{103}}{\sqrt{2}}$ \\ 
		\hline 3&$\frac{M_{313}}{\sqrt{2}}-\frac{M_{013}}{\sqrt{2}}$ \\ 
		\hline \end{tabular} 
} 
\medskip
This is a triplet mass that produces a stripy magnetic field; the key difference between the one just discussed above, is that this possesses an isolated Dirac cone, where a single Dirac cone at $\Gamma$ is always left ungapped. This is similar to the SU(4) case discussed in \eqnref{eq_gapless_t2go} that produces a stripy spin-octupolar density wave. For the SU(2) case one has a stripy density wave. Again, just as in the case discussed in \eqnref{eq_gapless_t2go}, there are special linear combinations of the masses that obtain additional gapless modes.

}


\bibliography{ref}
\end{document}